\documentclass[longauth]{aa}  

\usepackage{graphicx}
\usepackage{txfonts}
\usepackage{natbib}
\bibpunct{(}{)}{;}{a}{}{,} 

\usepackage{wrapfig}
\usepackage{color}
\usepackage{placeins}

\newcommand{\hi}{H\,{\sc i}\xspace}

\newcommand{\HI}{H\,{\sc i}\xspace}
\newcommand{\kms}{km~s$^{-1}$\xspace}
\newcommand{\msun}{\ensuremath{M_{\odot}}\xspace}
\newcommand{\msunpc}{\ensuremath{M_{\odot}\,{\rm pc}^{-2}}\xspace}
\newcommand{\degree}{\ensuremath{^\circ\xspace}}
\newcommand{\wsclean}{{\tt wsclean}}
\newcommand{\matHI}{{\rm H{\hskip 0.02cm\tt I}}}

\begin{document} 

\title{MHONGOOSE --- A MeerKAT nearby galaxy \HI survey}

\titlerunning{MHONGOOSE}
\authorrunning{de Blok et al.}

   \author{W.J.G. de Blok
          \inst{1,2,3}
          \and
          J. Healy\inst{1}
          \and
          F.M. Maccagni\inst{1,4}
          \and
D.J. Pisano\inst{2,5}
\and
A. Bosma\inst{6}
\and
J. English\inst{7}
\and
T. Jarrett\inst{2}
\and
A. Marasco\inst{8}
\and
G.R. Meurer\inst{9}
\and
S. Veronese\inst{1,3}
\and
F. Bigiel\inst{10}
\and
L. Chemin\inst{11}
\and
F. Fraternali\inst{3}
\and
B.W. Holwerda\inst{12}
\and
P. Kamphuis\inst{13}
\and
H.R. Kl\"ockner\inst{14}
\and
D. Kleiner\inst{1,4}
\and
A.K. Leroy\inst{15,16}
\and
M. Mogotsi\inst{17,18}
\and
K.A. Oman\inst{19,20}
\and
E. Schinnerer\inst{21}
\and
L. Verdes-Montenegro\inst{22}
\and
T. Westmeier\inst{9}
\and
O.I. Wong\inst{23,9}
\and
N. Zabel\inst{2}
\and
P. Amram\inst{6}
\and
C. Carignan\inst{24,2}
\and
F. Combes\inst{25}
\and
E. Brinks\inst{26}
\and
R.J. Dettmar\inst{13}
\and
B.K. Gibson\inst{27}
\and
G.I.G. Jozsa\inst{14,28}
\and
B.S. Koribalski\inst{29,30}
\and
S.S. McGaugh\inst{31}
\and
T.A. Oosterloo\inst{1,3}
\and
K. Spekkens\inst{32,33}
\and
A.C. Schr\"oder\inst{34}
\and
E.A.K. Adams\inst{1,3}
\and
E. Athanassoula\inst{6}
\and
M.A. Bershady\inst{35,17,2}
\and
R.J. Beswick\inst{36}
\and
S. Blyth\inst{2}
\and
E.C. Elson\inst{37}
\and
B.S. Frank\inst{2,17,38}
\and
G. Heald\inst{23}
\and
P.A. Henning\inst{39,40}
\and
S. Kurapati\inst{2}
\and
S.I. Loubser\inst{41}
\and
D. Lucero\inst{42}
\and
M. Meyer\inst{9}
\and
B. Namumba\inst{43,22}
\and
S.-H. Oh\inst{44}
\and
A. Sardone\inst{15}
\and
K. Sheth\inst{45}
\and
M.W.L. Smith\inst{46}
\and
A. Sorgho\inst{22}
\and
F. Walter\inst{21}
\and
T. Williams\inst{47}
\and
P.A. Woudt\inst{2}
\and
A. Zijlstra\inst{48,49}
}

\institute{Netherlands Institute for Radio Astronomy (ASTRON), Oude Hoogeveensedijk 4, 7991 PD Dwingeloo, the Netherlands
\and
             Dept.\ of Astronomy, Univ.\ of Cape Town, Private Bag X3, Rondebosch 7701, South Africa
\and
             Kapteyn Astronomical Institute, University of Groningen, PO Box 800, 9700 AV Groningen, The Netherlands
\and
             INAF -- Osservatorio Astronomico di Cagliari, Via della Scienza 5, 09047, Selargius, CA, Italy
\and
              Adjunct Astronomer, Green Bank Observatory, 155 Observatory Road, Green Bank, WV 24944, USA
\and            
             Aix Marseille Univ, CNRS, CNES, LAM, Marseille, France
\and
             Department of Physics and Astronomy, University of Manitoba, Winnipeg, Manitoba, Canada, R3T 2N2, Canada
\and
            INAF – Padova Astronomical Observatory, Vicolo dell’Osservatorio 5, I-35122 Padova, Italy
\and
            International Centre for Radio Astronomy Research, The University of Western Australia, 35 Stirling Highway, Crawley, WA 6009, Australia
\and
            Argelander-Institut f\"ur Astronomie, Auf dem H\"ugel 71, 53121, Bonn, Germany
\and 
             Instituto de Astrof\'isica, Departamento de Ciencias F\'isicas, Universidad Andr\'es Bello, Fernandez Concha 700, Las Condes, Santiago, Chile
\and
                Department of Physics and Astronomy, 102 Natural Science Building, University of Louisville, Louisville, KY, 40292, USA
                \and
             Ruhr University Bochum, Faculty of Physics and Astronomy, Astronomical Institute (AIRUB), Universit\"atsstrasse 150, 44801 Bochum, Germany
\and
             Max-Planck-Institut f\"ur Radioastronomie, Auf dem H\"ugel 69, 53121, Bonn, Germany
\and             
            Department of Astronomy, The Ohio State University, 140 West 18th Avenue, Columbus, OH 43210, USA
\and
            Center for Cosmology and Astroparticle Physics, 191 West Woodruff Avenue, Columbus, OH 43210, USA
\and
South African Astronomical Observatory (SAAO), P.O. Box 9, Observatory 7935, Cape Town, South Africa
\and
Southern African Larger Telescope (SALT), P.O. Box 9, Observatory 7935, Cape Town, South Africa
\and
Institute for Computational Cosmology, Department of Physics, Durham University, South Road, Durham DH1 3LE, UK
\and
Centre for Extragalactic Astronomy, Department of Physics, Durham University, South Road, Durham DH1 3LE, UK
\and
             Max Planck Institute for Astronomy, K\"onigstuhl 17, 69117 Heidelberg, Germany
\and
Instituto de Astrof\'isica de Andaluc\'ia-CSIC, Glorieta de la Astronom\'ia s/n, E-18008 Granada, Spain
\and
                          CSIRO, Space \& Astronomy, PO Box 1130, Bentley WA 6102, Australia
\and
Laboratoire de Physique et de Chimie de l’Environnement, Observatoire d’Astrophysique de l’Universit\'e Ouaga I Pr Joseph KiZerbo (ODAUO), 03 BP 7021, Ouaga 03, Burkina Faso
\and
             Observatoire de Paris, Coll\`ege de France, Universit\'e PSL, Sorbonne Universit\'e, CNRS, LERMA, Paris, France
\and
             Centre for Astrophysics Research, University of Hertfordshire, College Lane, Hatfield, AL10 9AB, UK   
\and
              E.A. Milne Centre for Astrophysics, University of Hull, Hull, HU6 7RX, United Kingdom
\and
Department of Physics and Electronics, Rhodes University, PO Box 94, Makhanda, 6140, South Africa
\and
Australia Telescope National Facility, CSIRO Astronomy and Space Science, P.O. Box 76, Epping, NSW 1710, Australia
     \and 
Western Sydney University, Locked Bag 1797, Penrith South, NSW 1797, Australia
\and
Department of Astronomy, Case Western Reserve University, 10900 Euclid Avenue, Cleveland, OH 44106, USA
\and
Department of Physics and Space Science, Royal Military College of Canada P.O. Box 17000, Station Forces Kingston, ON K7K 7B4, Canada
\and
Department of Physics, Engineering Physics and Astronomy, Queen’s University, Kingston, ON K7L 3N6, Canada
\and
Max Planck Institute for Extraterrestrial Physics, Gie{\ss}enbachstra{\ss}e 1, 85748 Garching, Germany
\and
University of Wisconsin-Madison, Department of Astronomy, 475 N. Charter Street, Madison, WI 53706-1582, USA
\and
Jodrell Bank Centre for Astrophysics, School of Physics and Astronomy, University of Manchester, Oxford Road, Manchester M13 9PL, UK
\and
Department of Physics and Astronomy, University of the Western Cape, Robert Sobukwe Rd, Bellville 7535, South Africa
\and
The Inter-University Institute for Data Intensive Astronomy (IDIA), University of Cape Town, Private Bag X3, Rondebosch 7701, South Africa
\and
National Radio Astronomy Observatory, P.O. Box O, Socorro, NM 87801, USA
\and
Department of Physics and Astronomy, MSC07 4220, 1 University of New Mexico, Albuquerque NM 87131, USA
\and
Centre for Space Research, North-West University, Potchefstroom 2520, South Africa
\and
Department of Physics, Virginia Polytechnic Institute and State University, 50 West Campus Drive, Blacksburg, VA 24061, USA
\and
Wits Centre for Astrophysics, School of Physics, University of the Witwatersrand, 1 Jan Smuts Avenue, 2000, South Africa 
\and
Department of Physics and Astronomy, Sejong University, Seoul 05006, South-Korea
\and
NASA Headquarters, 300 Hidden Figures Way, SE, Mary W. Jackson NASA HQ Building, Washington, DC 20546, USA
\and
School of Physics and Astronomy, Cardiff University, Queens Building, The Parade, Cardiff CF24 3AA, UK
\and
Department of Physics and Astronomy, Rutgers, The State University of New Jersey, 136 Frelinghuysen Road, Piscataway, NJ, 08854-8019, USA
\and
Department of Physics and Astronomy, The University of Manchester, Manchester M13 9PL, UK
\and
School of Mathematical and Physical Sciences, Macquarie University, Balaclava Road, North Ryde, Sydney, NSW 2109, Australia
}

   \date{Received --, 2023; accepted --, 2023}

 
  \abstract{The MHONGOOSE (MeerKAT \HI Observations of Nearby
    Galactic Objects: Observing Southern Emitters) survey maps the
    distribution and kinematics of the neutral atomic hydrogen (\HI) gas in and
    around 30 nearby star-forming spiral and dwarf galaxies to extremely 
    low \HI column densities.   The \HI 
    column density sensitivity (3$\sigma$ over 16 \kms) ranges from $\sim 5\cdot 10^{17}$ cm$^{-2}$ at 
    $90''$ resolution to $\sim 4 \cdot 10^{19}$ cm$^{-2}$ at the highest resolution of $7''$. 
    The \HI mass sensitivity 
    ($3\sigma$ over 50 \kms) is $\sim 5.5 \cdot 10^5$ \msun at a distance of 10 Mpc (the median distance of the sample galaxies). The velocity resolution of the data is 1.4 \kms.  One of the main science goals of the survey is the 
    detection of cold accreting gas in the outskirts of the sample galaxies.
    The sample was selected to cover a range in \HI masses from $10^7$ \msun to almost $10^{11}$ \msun\ in order 
    to optimally sample possible accretion scenarios and environments. The distance to the sample 
    galaxies ranges from 3 to 23 Mpc. In this paper, we present the sample selection, survey design, and 
    observation and reduction procedures.  We compared the integrated \HI fluxes based on the MeerKAT 
    data with those derived from single-dish measurement and find good agreement, indicating that 
    our MeerKAT observations are  recovering all flux. We present \HI moment maps of the entire sample based on the first 
    ten percent of the survey data, and find that a comparison of the zeroth- and second-moment values shows a 
    clear separation in the physical properties of the \HI between areas with star formation and areas 
    without related to the formation of a cold neutral medium. Finally, we give an overview of the \HI-detected
    companion and satellite galaxies in the 30 fields, five of which have not previously been cataloged. We find a clear relation between the number of companion galaxies and the mass of the main 
    target galaxy. }


\maketitle
%

\section{Introduction}

The evolution of the baryonic matter in the Universe can to a large
degree be described as the gradual transformation of primordial
atomic hydrogen into galaxies over cosmic time.  This
transformation involves physical processes on many scales, from the
size of galaxy clusters to those of individual gas clouds within a
galaxy. These processes, such as gas infall, collapse of
clouds, formation of stars, feedback due to stellar winds, and
supernovae, form part of the baryon cycle. Although many of them act on subgalactic
scales, they nevertheless affect the evolution of galaxies as a whole.

The only place where a comprehensive detailed survey of these
processes can be made is the nearby Universe: locally we can
study all aspects of the baryon cycle in detail. Resolved
observations of the cold gas, and, specifically of neutral hydrogen (\HI)
can contribute to determining how galaxies acquire their gas, 
how star formation is sustained, and ultimately how the dark and visible
matter interact to determine and regulate the evolution of galaxies.

We concentrate here on the first issue, that is, the origin of the gas in
galaxies.  This question first arose from the observation that in the 
inner regions of nearby spiral galaxies, timescales for the consumption of 
gas by star formation are much shorter than a Hubble time, even though the star formation rate (SFR) 
has been approximately constant over most of the lifetime of the
galaxies (e.g., \citealt{Kennicutt.1998,Bigiel.2011,Leroy.2013,Fraternali.2012}).
If galaxies are to continue forming stars at their current rate beyond
the current epoch, there must be a gas supply external to galaxies. A similar
argument can also be made based on observations of high-redshift galaxies (e.g., \citealt{Saintonge.2013, Tacconi.2018}).  These
form stars at a much higher rate than that observed in the local
Universe. This would imply a strong decrease in the gas content of galaxies
between then and now; yet we observe an almost constant density, again
implying there is an external supply of gas (see, e.g., \citealt{Walter.2020}).
In other words, we observe that the gas depletion time for high-redshift galaxies is $\sim 1$ Gyr or less \citep{Saintonge.2013, Tacconi.2018} and yet most galaxies are not quenched and keep forming stars for a much longer time until today.

Galactic disks can, in principle, be replenished by accreting gas-rich
companion galaxies, but the slope of the \HI mass function is not
steep enough for small companions to supply larger galaxies with a
substantial amount of gas for a sufficiently long time
\citep{Sancisi.2008}. In addition, observations suggest that gas-rich mergers can, at most, 
provide~20\% of the gas required to maintain the SFR in nearby late-type 
galaxies \citep{DiTeodoro.2015b}. This implies that spirals have to accrete directly from 
the intergalactic medium (IGM).

The presence of cold gas in the halos of our Milky Way and other
galaxies has been known for several decades (see, e.g.,
\citealt{Wakker.1997, Oosterloo.2007, Sancisi.2008, Heald.2011} and
the reviews by \citealt{Putman.2012} and
\citealt{Tumlinson.2017}). The \HI in some halos may be part of a
star-formation-driven ``galactic fountain'' \citep{Shapiro.1976}. This
galactic fountain has been proposed to lead to an ``indirect''
cold-gas-accretion mechanism, where the gas that is expelled from the
disk, drags additional halo gas along as it returns to the disk
\citep{Marasco.2012, Fraternali.2017,Marasco.2019}.

This is suggested, for example, for the galaxy NGC 2403 by the
observation that most of its extraplanar \HI has a similar projected
radial distribution to the star formation in the disk and that it has
disk-like kinematics: rotating but lagging behind the main disk (see,
e.g., \citealt{Fraternali.2001}). Similar conclusions were reached for
nearby galaxy NGC 253 \citep{Lucero.2015}. However, we note that in
NGC 2403, we also observe \HI in the halo that likely has an external
origin \citep{Veronese.2023}, and indeed some of the \HI complexes
found outside of the main \HI disks of galaxies are counter-rotating
with respect to the disk, confirming that they cannot have originated
from it.  Numerical simulations (e.g.,
\citealt{Keres.2005,Ramesh.2023}) predict that filaments of cooler gas
from the IGM can penetrate the hot halos surrounding galaxies and
deposit gas onto the disk. This process is called cold accretion.
These filaments are a prediction of high-resolution cosmological
hydrodynamical simulations of structure formation (e.g.,
\citealt{Dave.1999,Crain.2016,Ramesh.2023}), which suggest that most
of the baryons at low redshift are in a warm-hot ($T\sim 10^5 - 10^7$
K) intergalactic medium and most of the gas in the cosmic web is
therefore ionized and difficult to observe directly. To detect the
smaller fraction of cooler ($T<10^4$ K) baryons in the cosmic web, an
\HI column density sensitivity of $\sim 10^{17-18}$ cm$^{-2}$ is
required \citep{Popping.2009}.

It is in the context of this cold accretion that the study of \HI halos
of galaxies is relevant: it could provide direct observations of the
accretion of gas onto galaxies as well as a strong observational test
for models of galaxy evolution.
An extensive study of the \HI halos of nearby spiral galaxies was made
by the Hydrogen Accretion in LOcal GAlaxieS (HALOGAS) 
project on the Westerbork Synthesis Radio Telescope (WSRT;  \citealt{Heald.2011}). To this end, 22 disk
galaxies were mapped down to an \HI  column density limit of $\sim 10^{19}$
cm$^{-2}$, which is an order of magnitude lower than the surface densities
typically found in  star-forming \HI disks.  The results of HALOGAS
indicate that some galaxies have more extended low-column-density
emission, while others do not: extended \HI distributions have been
detected in about 12 of the 22 galaxies observed. It is possible that
some of this gas is related to star formation and galactic fountain
processes \citep{Marasco.2019}, but  external accretion cannot be excluded. The average 
rate at which cold neutral hydrogen gas is accreted  by the HALOGAS galaxies is between 0.05 and 0.09~\msun~yr$^{-1}$ \citep{Kamphuis.2022}, with the exact value depending on the treatment of the amount of \HI detected by the Green
Bank Telescope (GBT) but not by the WSRT.
  This accretion rate is generally lower than the SFR (cf.\ Table 4 in \citealt{Kamphuis.2022}).
If in the HALOGAS sample, the \HI accretion rate balances the star formation rate, direct accretion must occur at much lower neutral gas column densities than detected by HALOGAS and on spatial scales not resolved by the GBT.

Very deep single-dish \HI observations show the presence of \HI at
these very low-column densities.  \citet{Braun.2004}, for example,
observed low-column-density features around and between M31 and M33
and, using the WSRT as a single dish, reached a $3\sigma$ limit over
16 \kms of $1.1 \cdot 10^{17}$ cm$^{-2}$, but with an angular
resolution of $\sim 49'$. Further deep observations with the GBT of
the HALOGAS \citep{Heald.2011} and The \HI Nearby Galaxy Survey
(THINGS; \citealt{Walter.2008}) galaxies are described in
\citet{Pisano.2014}, \citet{deBlok.2014}, and
\citet{Pingel.2018}. These reach a $3\sigma$ over 16 \kms sensitivity
of $\sim 6 \cdot 10^{17}$ cm$^{-2}$ at an angular resolution of about
$9'$ or $\sim$ 14 kpc and $\sim 24$ kpc at the median distances of the
THINGS and HALOGAS galaxies, respectively. However, sheer column
density sensitivity is not enough, as these relatively coarse spatial
resolutions already suggest.  \citet{Wolfe.2013, Wolfe.2016} show that
the diffuse $\sim 10^{17}$ cm$^{-2}$ low-column-density gas between
M31 and M33 observed by \citet{Braun.2004} is resolved in several
kiloparsec-sized clouds with peak column densities of a few times
$10^{18}$ cm$^{-2}$ when observed at higher spatial resolutions.

To properly detect and characterize accretion features both a kpc-scale (or better)
spatial resolution and a column density sensitivity of $\sim 10^{18}$ cm$^{-2}$ (or better) are thus
needed.  Surveys such as THINGS
\citep{Walter.2008} and HALOGAS \citep{Heald.2011} have concentrated
on either obtaining a high spatial resolution or a high column-density
sensitivity. Optimising both simultaneously is rarely an option; this
has so far limited our knowledge of how any low-column-density gas is
connected with the cosmic web and where accretion occurs.

\section{The MHONGOOSE survey}

The MeerKAT radio telescope \citep{Jonas.2018, Camilo.2018, Mauch.2020} is
making it possible to take this large step forward. MeerKAT consists
of 64 dishes each with an effective diameter of 13.5m, located in the
Karoo semi-desert in South Africa.  The array has a compact and dense core
(the shortest baseline being 29m), with 70\% of the collecting area
located close to the centre of the array, and with longest core baselines of 1~km.  The rest of the array is distributed at larger distances, to a
maximum baseline of 7.7 km.  This combination of a dense core and long
baselines means the telescope has the ability to produce
high-resolution imaging, while retaining a good column density
sensitivity. A third factor contributing to the high sensitivity is
the low system temperature of the receivers. At 1.4 GHz, the effective
system temperature is $T_{\rm sys}/\eta \simeq 20.5$K (where $\eta$ is the antenna efficiency).

\begin{figure}
  \resizebox{\hsize}{!}{\includegraphics{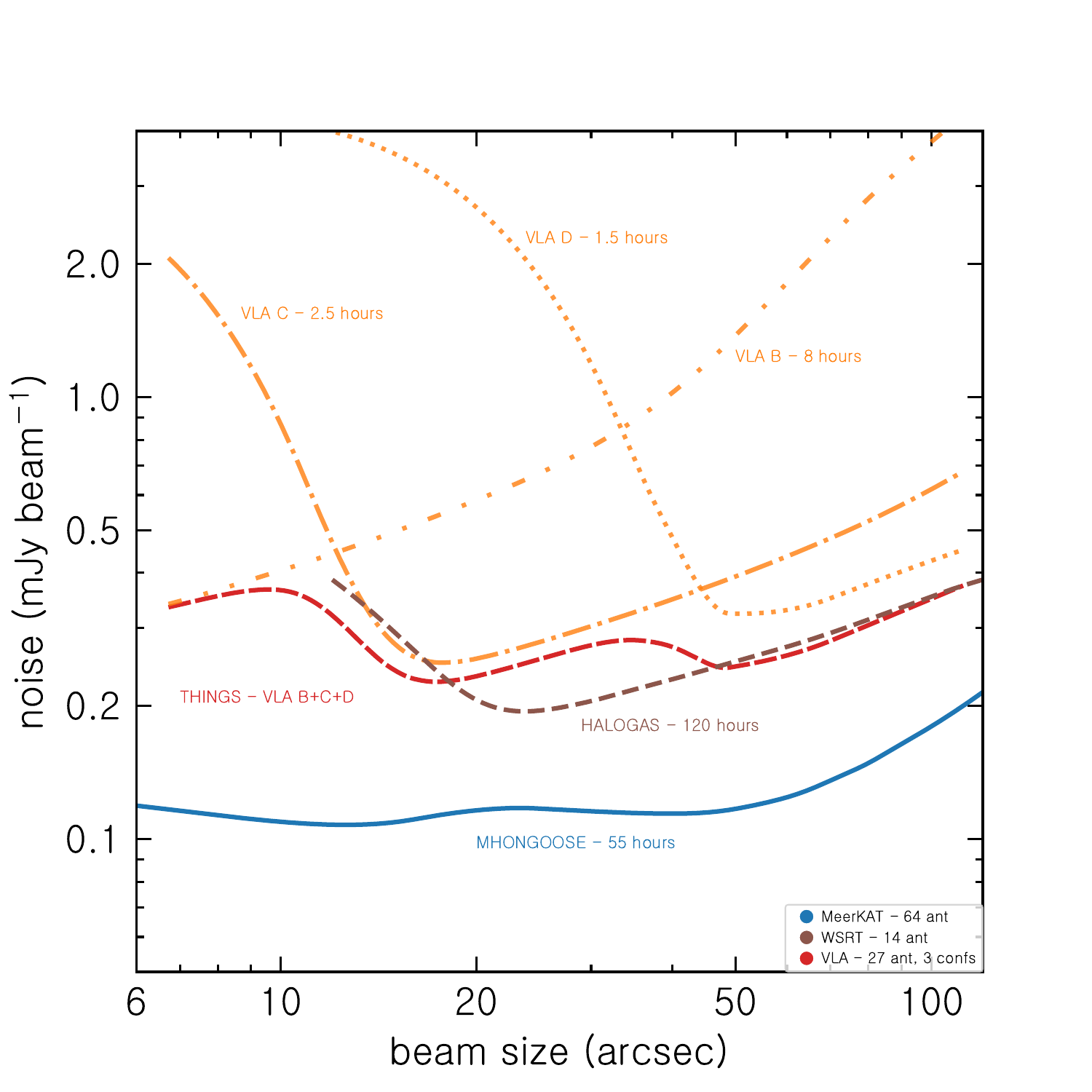}}
  \caption{Noise level as a function of resolution for the HALOGAS,
    THINGS, and MHONGOOSE surveys. The blue curve shows that the
    MHONGOOSE survey has a flat noise distribution, meaning
    equally sensitive imaging can be obtained over a large range of
    resolutions between $\sim 6''$ and $\sim 90''$. The dashed brown curve
    shows the HALOGAS survey, with an optimal sensitivity at around
    $\sim 20''$. The dashed red curve represents the THINGS survey. This is a combination of three
    separate VLA array configurations (also shown at the top of the
    plot calculated using the correct relative observing times). THINGS has optimal sensitivity at around $\sim 15''$ and 
    $\sim 50''$.  All noise values are calculated over a 5 \kms channel.}
  \label{fig:flat}  
\end{figure}

The array design of MeerKAT is such that the noise level for observations producing beam sizes between
$\sim 6''$ and $\sim 90''$ is approximately flat (see
Fig.~\ref{fig:flat}).  MeerKAT is therefore not optimized for one
particular resolution, but allows high-quality imaging over a large
range in resolutions using weighting or tapering.

Thanks to the combination of exquisite column density sensitivity,
high spatial resolution (down to $\sim 7''$ for \HI) and a large field of view with a primary beam full width at half-maximum (FWHM) diameter of 1$\degree$, we can study
nearby galaxies in \HI at the required quality to characterize any low-column density \HI that may be accreting onto a galaxy.

MHONGOOSE (MeerKAT \HI Observations of Nearby Galactic Objects:
Observing Southern Emitters) is a MeerKAT Large Survey Project designed
to produce ultra-deep \HI observations of 30 nearby gas-rich spiral and
dwarf galaxies in order to detect and characterize any low-column
density, potentially infalling, atomic gas, and to probe its link to star
formation (see also \citealt{deBlok.2016, deBlok.2020}). These deep observations can  provide information on gas
flows into and out of the galaxy disks, accretion from the IGM, the fuelling of
star formation, the connection with the cosmic web and even the
possible existence of low-mass cold dark matter (CDM) halos. The
relation between dark and baryonic matter and the distribution of dark
matter within galaxies can also be comprehensively studied due to the
high resolution combined with high sensitivity. The large field of view of 
MeerKAT means a significant fraction of the virial volumes of the target galaxies can be observed.

For the MHONGOOSE survey, each sample galaxy is observed for 55 hours in order to reach a $3\sigma$ over 16 \kms sensitivity limit of
$5 \cdot 10^{17}$ cm$^{-2}$ to detect the low-column density component discussed above.  
Such column densities are close to
three orders of magnitude below those typically found in the
star-forming disks of galaxies and equal those of the accreting cool neutral
gas as predicted by many numerical simulations \citep{Keres.2005,Popping.2009, Ramesh.2023}

\begin{figure*}
  \centering
  \sidecaption
  \resizebox{12cm}{!}{\includegraphics{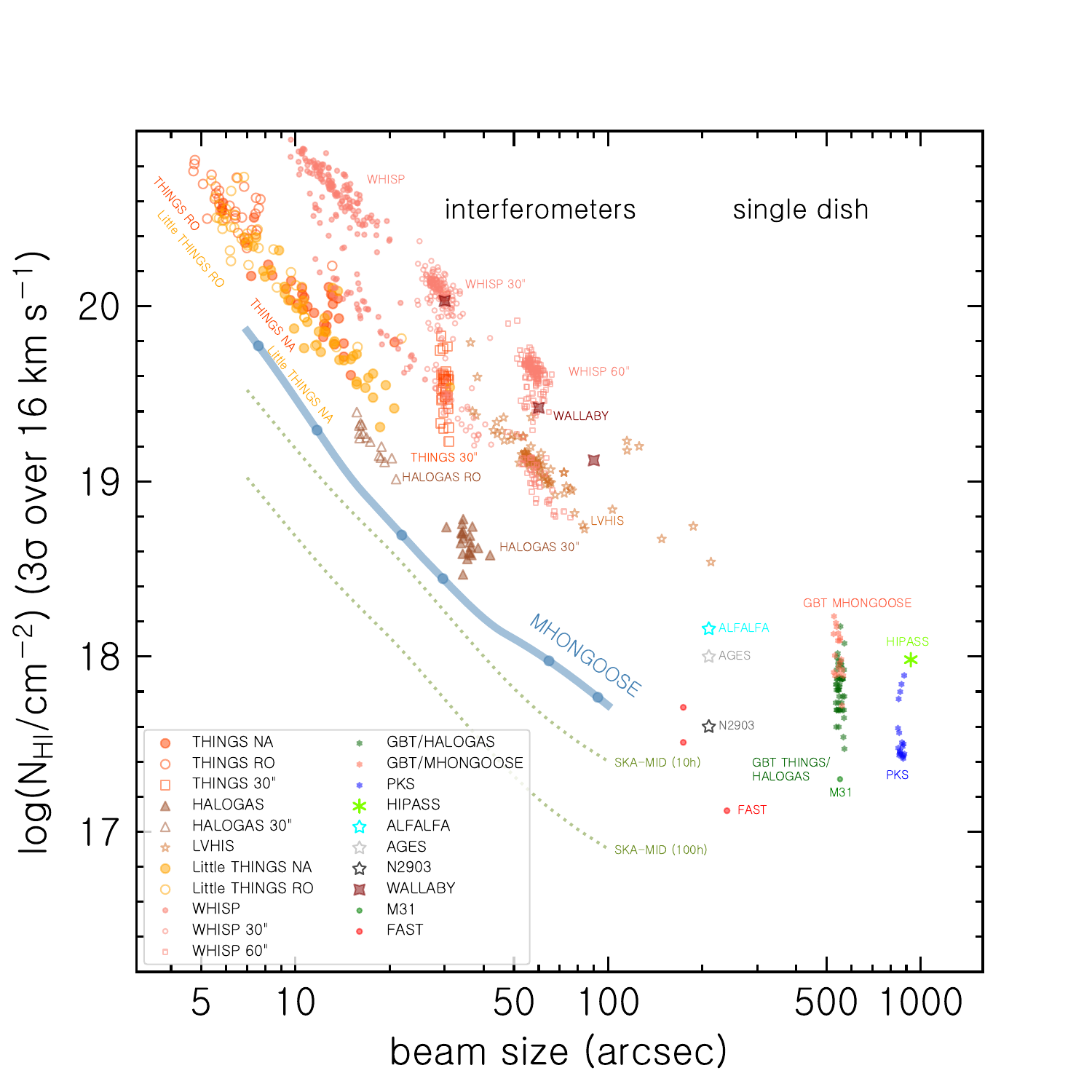}}
  \caption{Sensitivity versus resolution in \HI surveys. Colored
    symbols show the $3\sigma$ column density sensitivity over 16 \kms
    for various interferometric and single-dish surveys, as indicated
    by the labels and legend and listed in the text. The thick blue
    line shows the observed MHONGOOSE sensitivities. MHONGOOSE reaches
    single-dish sensitivities but at a 10--50 times better angular
    resolution. To give an indication of the physical scales: at 10.3
    Mpc (the median distance of the MHONGOOSE sample), $10''$
    corresponds to 0.5 kpc. Galaxies that are part of a sample with a
    fixed angular resolution (THINGS $30''$, GBT, and PKS) were given
    small, random horizontal offsets for clarity. References are given
    in the text.}
  \label{fig:surveys}
\end{figure*}

Figure \ref{fig:surveys} compares the MHONGOOSE column density
sensitivities with those from previous \HI surveys using both
interferometers and single-dish telescopes. Interferometric
targeted surveys shown here are THINGS \citep{Walter.2008},
HALOGAS \citep{Heald.2011}, the Westerbork \HI Survey Project (WHISP;
van der Hulst et al.\ 2001), the Local Irregulars That Trace
Luminosity Extremes, The HI Nearby Galaxy Survey (LITTLE THINGS;
\citealt{Hunter.2012}) and the Local Volume \HI Survey (LVHIS;
\citealt{Koribalski.2018}). Also shown is the untargeted
Widefield ASKAP L-band Legacy All-sky Blind surveY (WALLABY;
\citealt{Koribalski.2020}). For the single-dish observations we show
the column density sensitivities for the \HI Parkes All Sky Survey
(HIPASS; \citealt{Barnes.2001, Meyer.2004}), the Arecibo Legacy Fast
ALFA (ALFALFA; \citealt{Haynes.2018}), the Arecibo Galaxy Environment
Survey (AGES; \citealt{Auld.2006}), deep observations of M31
\citep{Wolfe.2016} and NGC 2903 \citep{Irwin.2009}, as well as a
number of deep GBT and Parkes observations of nearby galaxies
(\citealt{Sorgho.2019}, \citealt{Sardone.2021}, \citealt{Pingel.2018},
D.J.\ Pisano, priv.\ comm.). We also show a number of representative
deep \HI observations taken with the Five-hundred-meter Aperture
Spherical Telescope (FAST) from \citet{Xu.2022} and \citet{Liu.2023}.

To ensure a proper comparison we have taken the noise per channel and
the channel widths from the source papers or the corresponding
publicly available data and homogenized these quantities to a common
channel width of 16 \kms, assuming square-root scaling of noise with
channel-width.  A 16 \kms channel width corresponds approximately to
the FWHM of an \HI line with a velocity dispersion of 7 \kms, which is
comparable to the lowest values seen in previous \HI observations of
nearby galaxies \citep{Ianjamasimanana.2017}.

In Fig.~\ref{fig:surveys} we show the sensitivities of
these  \HI surveys and compare these with the observed sensitivity of
MHONGOOSE.  It is clear that at all resolutions the MHONGOOSE data are
significantly more sensitive than previous interferometric observations.  Note that
the sensitivity does not change simply as the inverse of the beam size
squared (which would be a straight line with a slope of $-2$) as one might
expect. This is due to the different weightings and taperings used to
produce the data at the various resolutions. We return to this in
Sec.~\ref{sec:sens}.

For comparison, we also show estimated sensitivities for \HI surveys
on SKA-MID. These are simply calculated by scaling up the MeerKAT
collecting area to the SKA-MID ``baseline design'' collecting area
(i.e., the equivalent of 64 MeerKAT dishes with 13.5m diameter and 133
SKA dishes with 15m diameter) and should therefore only be regarded as
indicative, as they do not take into account potentially different
baseline distributions, dish designs, or system temperatures.

A striking aspect of the comparison is that between $\sim 30''$ and $\sim 100''$,
MHONGOOSE is probing unexplored territory, with observations that have
the column density sensitivity of deep single-dish \HI observations of
the Local Universe, but at an angular resolution that is more than an
order of magnitude better.  With these observations it will therefore be possible to investigate
the low-column density \HI,  from the outskirts of the star-forming
disks out into the far reaches of the dark matter halo.

\section{Sample selection}

One of the main goals of MHONGOOSE is to trace accretion and star
formation processes over a large range in galaxy properties. It is
thus important to ensure that within the limited observing time
available a representative range of \HI masses, stellar masses, star
formation rates and rotation velocities (and hence, halo masses) is
sampled. Furthermore, to isolate accretion features from tidal and
interaction features as much as possible, strongly interacting galaxies
and dense environments are to be avoided.

Primary criteria for the sample selection are that the galaxies
should have been detected in \HI and located in the southern hemisphere. This makes the HIPASS
catalog \citep{Meyer.2004} a natural starting point. To ensure the
availability of a significant set of multi-wavelength data, we limited
our selection to HIPASS galaxies that are part of
the Survey of Ionization in Neutral Gas Galaxies (SINGG; \citealt{Meurer.2006}) and  the Survey of 
Ultraviolet emission in Nearby Galaxies (SUNGG; \citealt{Wong.2016}).
These surveys collected H$\alpha$, optical, infrared and ultraviolet data for a large
number of HIPASS-detected nearby galaxies.

The SINGG/SUNGG criteria
were as follows:
\begin{enumerate}[\itshape(i)]
\item HIPASS peak flux density $> 50$ mJy ($3.8\sigma$ in HIPASS),
\item galactic latitude $|b| > 30^{\circ}$,
\item projected distance from the center of the Large (Small) Magellanic Cloud $> 10^\circ$ ($> 5^\circ$),
\item Galactic standard of rest velocity $> 200$ km s$^{-1}$.
\end{enumerate}
 
This SINGG/SUNGG protosample was divided in bins of 0.2 dex in
$\log(M_{\matHI})$ and in each bin the closest 30 to 40 galaxies were
selected. This gave a flat number distribution in the range $8.5 <
\log(M_{\matHI}/M_{\rm \odot}) < 10.5$, with an additional, small number
of galaxies around $\log(M_{\matHI}/M_{\rm \odot}) \sim 7.0$ and
$\log(M_{\matHI}/M_{\rm \odot}) \sim 11$. Most of these selected galaxies
have a radial velocity $ < 2000$~\kms with a median velocity  of 1300
\kms.  These criteria resulted in the final SINGG sample of 468 galaxies as published in \citet{Meurer.2006}.
The key characteristic of this selection is that it was done
uniformly in $\log(M_{\matHI})$ in order to guarantee the broadest possible range in \HI masses.

For MHONGOOSE we further narrowed down the SINGG sample by requiring
that for a given galaxy, a complete set of H$\alpha$, $R$-band and
GALEX ultra-violet data was available. This reduced the number of
galaxies to 151.  To retain sufficient spatial resolution, we
furthermore removed all galaxies with a distance $D > 30$ Mpc
(heliocentric velocity $v_{\rm hel} > 2100$ \kms). In addition, we
only selected galaxies with a southern declination, and we excluded
the survey area of the MeerKAT Fornax Survey \citep{Serra.2023}.  We
additionally checked that the potential sample galaxies were not
located in the densest environments (i.e., inner parts of galaxy
clusters or major galaxy groups).  Galaxies that were located in the
outer parts of (smaller) groups were retained (as were, obviously,
isolated galaxies). This enables quantifying the effects of
environment on accretion processes in medium- to low-density
environments, without the target galaxies themselves being majorly
affected by the environment.  Studies of isolated galaxies have shown
that the gas captured from companion galaxies and galactic fountain
processes (due to SF and Active Galactic Nuclei [AGN]) are minimized
there
\citep{Jones.2018,Espada.2011,Espada.2011a,Leon.2008,Sabater.2012,Lisenfeld.2007}.
These criteria resulted in a sample of 88 galaxies.

As the goal of MHONGOOSE is to characterize the low-column density \HI
over a large range in \HI masses, we aimed for
a flat number distribution in $\log(M_{\matHI})$ over the available mass
range $6.0 < \log(M_{\rm \matHI}/M_{\rm \odot})< 10.5$, analogous to SINGG.  To compensate for
the smaller number of galaxies, we used six bins each with a width of
0.5 dex, except for the lowest-mass bin, which we defined as
$\log(M_{\matHI}/M_{\odot}) <8.0$.

The final sample size of 30 galaxies was set by the assigned observing time and the desired column density sentivity.
This total number implies  a total of  five
galaxies per mass bin. To optimize the selection for the science
objectives of the survey, we required that galaxies be in one of three
well-defined inclination categories as follows:
{\it i)} (close to) face-on, {\it ii}) (close to) edge-on and {\it iii)}
inclination (close to) 60 degrees. Face-on allows the best
characterization of the morphology of the interstellar medium (ISM), as well as
determination of vertical motions. Edge-on allows an unambiguous
characterization of the vertical structure of the ISM and possibly the dark matter distribution out of the disk plane. The
intermediate inclination range allows (following a limited amount
of modeling) determination of both, and can be used to tie results
from edge-on and face-on classes together. It is also an optimal inclination for rotation curve measurements and mass-modeling of the galaxy. 

Selection of the final 30 galaxies was done in two steps. In the
first step, optical SINGG and Digital Sky Survey images were
examined and galaxies that were affected by bright foreground stars
were rejected.  In addition, we rejected cases where the galaxy was
too big to comfortably fit in the $1\degree$ FWHM MeerKAT primary beam. This to avoid
reduced sensitivity to \HI in the outer parts of the galaxies (due to the primary beam
attenuation) and to ensure mosaicking was not needed.  Generally we
insisted that the optical diameter was smaller than $15'$ (though few
galaxies actually reached that size). In SINGG, the optical diameter is 
determined from optical surface brightness profiles and corresponds to 
twice the radius beyond which no detectable R-band, H$\alpha$, or UV emission is found. 
\citet{Meurer.2013} show that this radius appears to correspond to the edge of the stellar disk. 

We also did not include galaxies
that were clearly strongly interacting. Finally, we checked that no extremely
bright radio continuum sources were present within or close to the galaxy
positions. All of this resulted in the rejection of 11 galaxies.

We then stepped through each of the mass bins and used the
SINGG imaging to select in each bin the most optimal edge-on, face-on,
and intermediate-inclination galaxies.  Factors that went into this
were how close the galaxies were to the preferred inclinations (as
judged from the apparent major to minor axis ratios estimated from the SINGG $R$-band images), their angular sizes, and their distances (where
the nearest galaxies were preferred). A secondary condition was to ensure that a range in star formation rates (as judged
from the SINGG H$\alpha$ SFR measurements) was covered for each bin.

This resulted in the final sample of 30 galaxies. These are listed in
Table \ref{tab:sample} along with some fundamental properties. We note
that some properties listed here are different from those given in the
earlier sample table in \citet{deBlok.2020}. That table lists the
sample with the galaxy parameters originally used for the sample
selection as based on parameters and distances from SINGG/SUNGG.  In
the current paper, we have adopted more recent distance estimates for
the galaxies (including Tip of the Red Giant Branch [TRGB]
measurements). These revised distances resulted in small changes in
distance-dependent properties (such as the \HI mass), sometimes
resulting in a galaxy moving to a different \HI mass bin. These
  changes have no significant impact on the final science goals. The
  sample still covers the desired large range in \HI mass as well as a
  representative range in (gas-rich) galaxy properties.  The current
Table lists the galaxies in order of \HI mass assuming the revised
distances. For the median distance of the galaxies in the MHONGOOSE
sample (10.3 Mpc), $10''$ corresponds to 0.5 kpc.  \HI-related
parameters listed in the Table are based on the observations presented
in this paper. The inclination values are based on the optical axis
ratios and should be regarded as indicative only.

\begin{figure}
 \centering
  \resizebox{0.95\hsize}{!}{\includegraphics{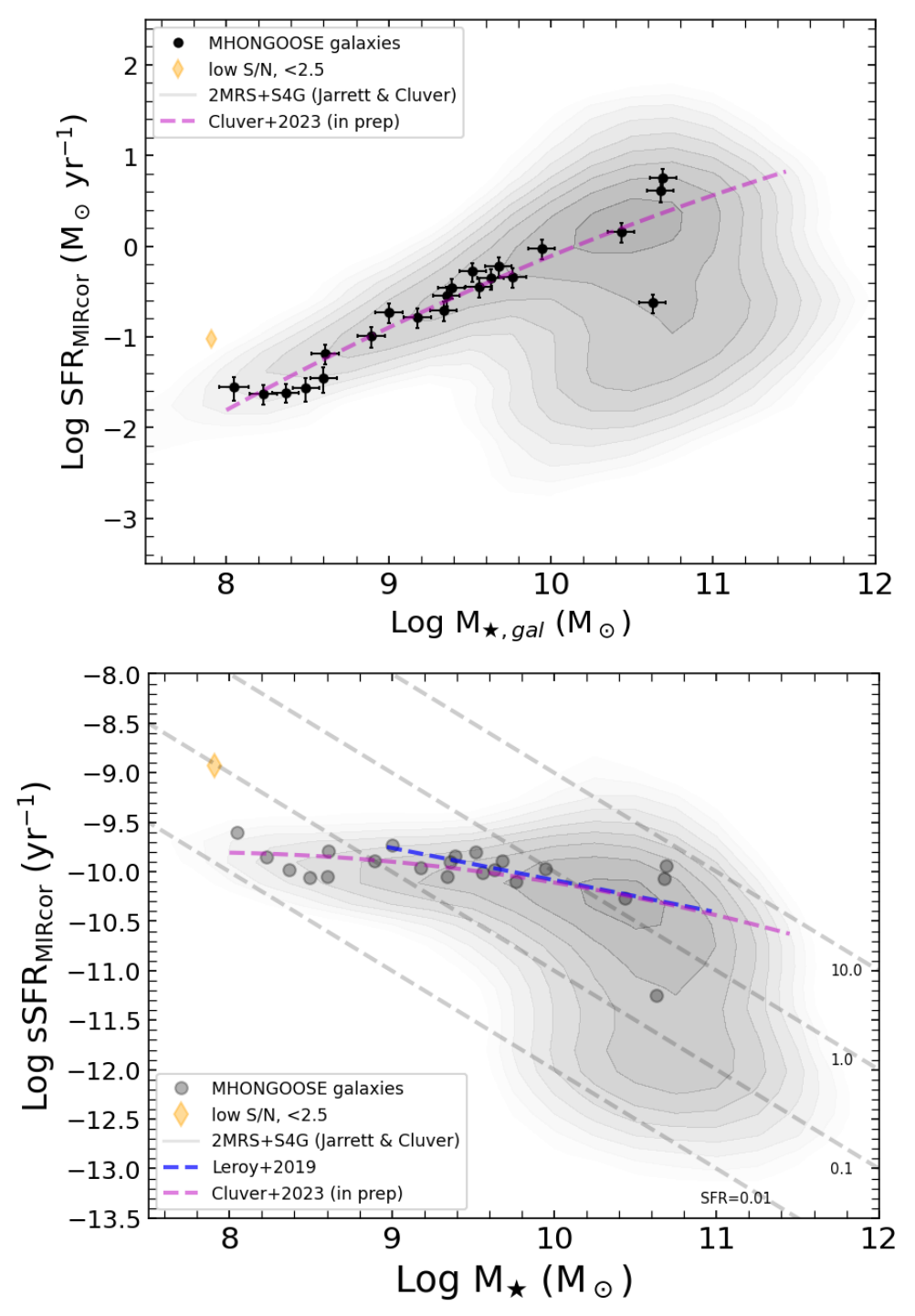}}
  \caption{Star formation rates of the MHONGOOSE galaxies plotted
    against their stellar masses. Top panel: the
    SFR as derived using the method described in Cluver et al.\ (in
    prep.) based on mid-IR and UV SFRs. The stellar masses are derived
    from WISE W3 and W4 luminosities as described in
    \citet{Jarrett.2023}. Background grayscale and gray contours show
    the distribution of galaxies in the local Universe based on mid-IR
    and UV data (Jarrett \& Cluver, in prep.). The MHONGOOSE
      sample was selected to be representative of the SFR--$M_\star$
      main sequence (the upper ridge of the distributions) and almost
      all galaxies are indeed on this sequence. Only NGC 1371
    (J0335--24) has a low SFR for its mass. Galaxies with uncertain
    SFR values are not plotted (see Tab.~\ref{tab:sample}). UGCA320
    (J1303--17b) has a low S/N in the WISE observations as indicated
    by the diamond symbol. The purple dashed curve shows the average
    trend derived in Cluver et al.\ (in prep.). Bottom panel: this shows the specific
    SFR of the MHONGOOSE galaxies plotted against stellar
    mass. Symbols and curves as in top panel. In addition, the blue
    curve shows a fit to the data presented in \citet{Leroy.2019}. The
    diagonal gray dashed lines indicate corresponding SFR values in
    \msun yr$^{-1}$. }
  \label{fig:sfr_mir}
\end{figure}

In Table \ref{tab:sample} we also give the stellar masses and star
formation rates based on WISE infrared measurements. The stellar
masses were derived using the new GAMA Stellar Masses calibration and
(light and colors) method of \citet{Jarrett.2023}. The star formation
rates are a combination mid-IR and UV SFR based on the method
described in Cluver et al.\ (in prep.).  Figure~\ref{fig:sfr_mir}
shows the distribution of the MHONGOOSE galaxies along the
SFR-$M_\star$ main sequence (defined by the upper ridges in the
diagrams). As the sample was selected to be representative of
gas-rich, star-forming galaxies, it therefore covers the main
sequence well.  The only galaxy not on it is NGC 1371 (J0335--24),
which has a low SFR for its stellar mass.

The MHONGOOSE sample galaxies span a range of galaxy environments, although, as described above, the selection was 
intentionally biased against those galaxies residing in the densest environments.  
Using the group catalog from \citet{Kourkchi.2017}, we quantify the group environment for all 30 galaxies.  
Of our sample galaxies, 13 are ``isolated''; that is, not in a group with another galaxy although they are 
still in larger associations.  The remaining 17 galaxies reside in groups with anywhere from 2--31 members. Their dynamical group masses range from those of dwarf galaxy associations ($\lesssim 10^{12}$~\msun) or 
single, massive galaxies up to -- for a few galaxies -- those of small groups 
($\sim10^{13}$~\msun).  

The identification and masses of all groups and associations as listed
in the \citet{Kourkchi.2017} catalog are given in
Table~\ref{tab:environment}. A number of galaxies are found to be
  part of the same groups or associations as indicated in
  Table~\ref{tab:environment}.
 
In addition, we use the stellar masses and distances given in
Table~\ref{tab:sample} to calculate the halo mass $M_{200}$ using the
stellar mass-halo mass relation given in \citet{Moster.2013}. Using
standard cosmology ($H_0 = 69$ \kms\ Mpc$^{-1}$,
$\Omega_{\Lambda}=0.27$) we also derive the virial radius $R_{200}$
for the target galaxies. We also list the ratio of a $1.5\degree$
  field of view (cf.\ Sect.~\ref{sec:fullcube}) in kpc at the distance
  of the galaxy, and the virial diameter $D_{200} = 2 R_{200}$. Note
that we have made no attempt to homogenize the \citet{Kourkchi.2017}
and \citet{Moster.2013} numbers, so some differences may exist in halo
or group masses due to different methods, assumptions or input data.

\begin{table*}
    \small
    \caption[]{Properties of galaxies in the MHONGOOSE sample}
    \label{tab:sample}
    \centering
    \begin{tabular}{l l l l r r  r r r r r}
      \hline
      \hline
      \noalign{\vspace{2pt}}

      HIPASS &  Name  &      $\alpha$ (J2000.0)& $\delta$ (J2000.0)   &     $D$   &  $V_{\rm cen}$ & $\log(M_{\rm HI})$ & $i$ &  $W_{20}$ &  $\log(M_{\rm \star})$ & log(SFR)\rlap{\tablefootmark{g}}\\
             &        &      $(^h\ ^m\ ^s)$   &   $(^\circ\ '\ '')$    &     (Mpc)   &  (\kms)  &  [\msun] & $({}^{\circ})$& (\kms) & [\msun] & [\msun\ yr$^{-1}$] \\
      (1)& (2)& (3)& (4)& (5)& (6) & (7)& (8) & (9) & (10) & (11) \\
      \hline
      \noalign{\vspace{2pt}}
      J0049--20   & UGCA015    & 00 49 49.20  &--21 00 54.0  &3.4  &293.8                          &               6.99  & 66 & 32  & 6.50 & (-2.52\rlap{)}\\
      J0008--34   & ESO349-G031& 00 08 13.36  &--34 34 42.0  &3.3\rlap{\tablefootmark{b}}  &219.6  &               7.11  & 35 & 39  & 6.13 & (-2.52\rlap{)}\\
      J1321--31   & KK98-195   & 13 21 08.20  &--31 31 45.0  &5.2\rlap{\tablefootmark{d}}  &570.8  &               7.60  & 63 & 38  & 4.71 & (-2.29\rlap{)}\\
      J0310--39   & ESO300-G016& 03 10 10.48  &--40 00 10.5  &8.0\rlap{\tablefootmark{c}}  &707.6  &               7.80  & 36 & 42  & 7.22 & (-2.52\rlap{)}\\
      J0031--22   & ESO473-G024& 00 31 22.51  &--22 45 57.5  &7.2\rlap{\tablefootmark{c}}  &538.1  &               7.95  & 57 & 50  & 6.84 & -2.22\\
      \hline
      \noalign{\vspace{2pt}}
      J1337--28   & ESO444-G084& 13 37 19.99  &--28 02 42.0  &4.6\rlap{\tablefootmark{e}}  &586.7  &               8.02  & 40 & 74  & 6.69 & (-2.52\rlap{)}\\
      J0320--52   & NGC1311    & 03 20 06.96  &--52 11 07.9  &5.6  &569.5                          &               8.05  & 74 & 101  & 8.23 & -1.62\\
      J0454--53   & NGC1705    & 04 54 13.50  &--53 21 39.8  &5.7   &631.6                         &               8.07  & 42 & 155 & 8.37 & -1.61\\
      J0429--27   & NGC1592    & 04 29 40.13  &--27 24 30.7  &10.3  &943.2                         &               8.10  & 60 & 80  & 8.49 & -1.57\\
      J0135--41   & NGC0625    & 01 35 04.63  &--41 26 10.3  &4.0  &393.4                          &               8.12  & 71 & 95  & 8.61 & -1.28\\
      \hline
      \noalign{\vspace{2pt}}
      J1106--14   & KKS2000-23 & 11 06 12.00  &--14 24 25.7  &13.9\rlap{\tablefootmark{c}} &1035.9 &               8.74  & 70 & 92  & 7.51 &( -1.82\rlap{)}\\
      J1253--12   & UGCA307    & 12 53 57.29  &--12 06 21.0  &11.0  &821.7                         &               8.88  & 60 & 89  & 8.05 & -1.56\\
      J0309--41   & ESO300-G014& 03 09 37.87  &--41 01 49.7  &10.9 &953.6                          &               8.89  & 59 & 140 & 8.90 & -0.99\\
      J0331--51   & IC1954     & 03 31 31.39  &--51 54 17.4  &12.8\rlap{\tablefootmark{f}} &1057.1 &               8.90  & 62 & 231 & 9.63 & -0.35\\
      J0351--38   & ESO302-G014& 03 51 40.90  &--38 27 08.0  &16.8 &869.6                          &               8.90  & 28 & 82  & 8.19 & (-2.52\rlap{)}\\
      J2009--61   & IC4951     & 20 09 31.77  &--61 51 01.7  &11.6 &811.4                          &               8.93  & 80 & 131 & 8.60 & -1.45\\
      J2357--32   & NGC7793    & 23 57 49.83  &--32 35 27.7  &3.6\rlap{\tablefootmark{f}}  &225.7  &               8.96  & 47 & 189 & 9.34 & -0.71\\
      J1303--17b  & UGCA320    & 13 03 16.74  &--17 25 22.9  &6.0  &739.9                          &               8.97  & 83 & 125 & 7.91 & -1.02\\
      \hline
      \noalign{\vspace{2pt}}
      J1318--21   & NGC5068    & 13 18 54.81  &--21 02 20.8  &5.2\rlap{\tablefootmark{f}}  &668.9  &               9.01  & 28 & 107  & 9.36 & -0.54\\
      J0546--52   & NGC2101    & 05 46 24.17  &--52 05 18.7  &16.4 &1186.8                         &               9.22  & 47 & 117 & 9.00 & -0.73\\
      J1254--10a  & NGC4781    & 12 54 27.00  &--10 30 30.0  &11.3\rlap{\tablefootmark{f}} &1255.9 &               9.22  & 65 & 250 & 9.52 & -0.28\\
      \hline
      \noalign{\vspace{2pt}}
      J0516--37   & ESO362-G011& 05 16 38.80  &--37 06 09.1  &15.6  &1337.0                        &               9.57  & 81 & 289 & 9.68  & -0.22\\
      J0459--26   & NGC1744    & 04 59 57.80  &--26 01 20.0  &9.3 &739.9                           &               9.54  & 57 & 207 & 9.18  & -0.78\\
      J1103--23\tablefootmark{a}& NGC3511  & 11 03 23.77  &--23 05 12.4  &13.9\rlap{\tablefootmark{f}} &1099.1   & 9.54  & 70 & 307 & 9.94  & -0.03 \\
      J2257--41   & NGC7424    & 22 57 18.37  &--41 04 14.1  &7.9 &936.6  &                                        9.60  & 32 & 170 & 9.56  & -0.46\\
      J1153--28   & UGCA250    & 11 53 24.06  &--28 33 11.4  &20.2 &1696.1 &                                       9.84  & 82 & 285 & 9.76  & -0.34 \\
      J0335--24   & NGC1371    & 03 35 01.34  &--24 55 59.6  &22.7 &1456.4                         &               9.97  & 46 & 398 & 10.63 & -0.62\\
      \hline
      \noalign{\vspace{2pt}}
      J0052--31   & NGC0289    & 00 52 42.36  &--31 12 21.0  &21.5  &1620.1 &                                      10.35 & 45 & 292 & 10.43 & 0.16\\
      J0419--54   & NGC1566    & 04 20 00.42  &--54 56 16.1  &17.7\rlap{\tablefootmark{f}} &1496.0 &               10.08 & 37 & 219 & 10.68 & 0.61\\
      J0445--59   & NGC1672    & 04 45 42.50  &--59 14 49.9  &19.4\rlap{\tablefootmark{f}} &1327.0 &               10.29 & 34 & 298 & 10.69 & 0.75\\
        \hline
    \end{tabular}
    \tablefoot{ \tablefoottext{a}{pair with NGC 3513.} Columns: (1)
      HIPASS identification. (2) Common name. (3) Right ascension
      (J2000.0). (4) Declination (J2000.0). (5) Distance. Distances taken from \citet{Leroy.2019}, unless specified otherwise in
      the notes below. (6) Central velocity (this
      work). (7) Logarithmic \HI mass (this work). (8) Indicative inclination (NED). (9) $W_{20}$ \hi
      velocity width (this work). (10) WISE-derived logarithmic stellar mass (11). WISE-derived logarithmic SFR.\\ 
      \emph{Distance references:}
      \tablefoottext{b}{\citet{Tully.2013}}, \tablefoottext{c}{\citet{Tully.2016}},
      \tablefoottext{d}{\citet{Pritzl.2003}},
      \tablefoottext{e}{\citet{Karachentsev.2002}}, \tablefoottext{f}{\citet{Anand.2021}}.\\
      \tablefoottext{g}{Values in parentheses are uncertain due to faintness of the mid-IR emission and should be regarded as indicative only.}}
\end{table*}

\begin{table*}
    \small
    \caption[]{Group membership and halo properties of the galaxies in the MHONGOOSE sample}
    \label{tab:environment}
    \centering
    \begin{tabular}{l l r r r r  r r r r r}
      \hline
      \hline
      \noalign{\vspace{2pt}}
      HIPASS &  Name  &      PGC & PGC1   &  PGC1+  &  $N$ & $\log(M_{K})$ & $\log(M_d)$ & $\log(M_{200})$  &  $R_{200}$ & $f_{D200}$\\
             &        &          &        &         &      &  [\msun] & [\msun]& [\msun]& (kpc)&  \\
      (1)& (2)& (3)& (4)& (5)& (6) & (7)& (8) & (9) & (10) & (11) \\
      \hline
      \noalign{\vspace{2pt}}
      J0049--20   & UGCA015     &  2902 &    2789   & 2789\rlap{\tablefootmark{A}}      &                  4 & 12.78 &12.71          &9.93   &67 & 0.67 \\
      J0008--34   & ESO349-G031 &   621 &    621   & 2789\rlap{\tablefootmark{A}}       &                  1 & 10.02& --        & 9.77 & 59 & 0.73 \\
      J1321--31   & KK98-195    &166163 &   48082  & 46957\rlap{\tablefootmark{a}}      &                  13& 12.25 &12.59      &9.20   &38 & 1.79\\
      J0310--39   & ESO300-G016 & 11842 &   12209  & 12209    &                         3& 12.53 &12.43      &10.24  &85 & 1.23\\
      J0031--22   & ESO473-G024 &  1920 &    2789  &  2789\rlap{\tablefootmark{A}}      &                   4& 12.78 &12.71      &10.07  &74 & 1.28\\
      \hline
      \noalign{\vspace{2pt}}
      J1337--28   & ESO444-G084 & 48111 &   48082  & 46957\rlap{\tablefootmark{a}}      &                  13& 12.25 &12.59      &10.07  &74 & 0.81\\
      J0320--52   & NGC1311     & 12460 &   12460  & 12460     &                   1& 10.61 &--          &10.66  &116 & 0.63\\
      J0454--53   & NGC1705     & 16282 &   16282  & 16282    &                   1& 10.72 &--          &10.70  &121 & 0.62 \\
      J0429--27   & NGC1592     & 15292 &   15292  & 15292    &                   1& 10.65 &--          &10.78  &128 & 1.05\\
      J0135--41   & NGC0625     &  5896 &    5896  &  2789\rlap{\tablefootmark{A}}    &                   1& 10.91 &--          &10.83  &133 & 0.39\\
      \hline
      \noalign{\vspace{2pt}}
      J1106--14   & KKS2000-23  &3097702 &  3097702 & 3097702  &                  1 &10.54 &--          &10.26  &86 &2.12 \\
      J1253--12   & UGCA307     &  43851 &   43851  & 41220\rlap{\tablefootmark{C}}   &                   1 &10.73 &--          &10.53  &106 & 1.36\\
      J0309--41   & ESO300-G014 &  11812 &   12209  & 12209   &                   3 &12.53 &12.43      &11.01  &153 & 0.93 \\
      J0331--51   & IC1954      &  13090 &   13090  & 14391   &                   2 &11.52 &10.70      &11.29  &190 & 0.88\\
      J0351--38   & ESO302-G014 &  13985 &   13985  & 12209   &                   1 &10.59 &--          &10.61  &113 & 1.95 \\
      J2009--61   & IC4951      &  64181 &   64181  & 62836   &                   1 &10.92 &--          &10.82  &132 & 1.15\\
      J2357--32   & NGC7793     &  73049 &   73049  &  2789   &                   1 &11.23 &--          &11.15  &170 & 0.28\\
      J1303--17b  & UGCA320     &  45084 &   45084  & 46400\rlap{\tablefootmark{B}}   &                   2 &10.99 & 9.00      &10.65  &116 & 0.68\\
      \hline
      \noalign{\vspace{2pt}}
      J1318--21   & NGC5068     &  46400 &   46400  & 46400\rlap{\tablefootmark{B}}   &                   1 &11.43 &--          &11.17  &173 & 0.39\\
      J0546--52   & NGC2101     &  17793 &   17822  & 17822   &                   2 &11.18 &10.67      &11.00  &152 & 1.41 \\
      J1254--10a  & NGC4781     &  43902 &   42407  & 41220\rlap{\tablefootmark{C}}   &                  17 &13.37 &13.96      &11.24  &183& 0.81 \\
      \hline        
      \noalign{\vspace{2pt}}
      J0516--37   & ESO362-G011 &  17027 &   17027  & 13418\rlap{\tablefootmark{D}}   &                   2 &11.59 & 7.83      &11.31  &193 & 1.06\\
      J0459--26   & NGC1744     &  16517 &   16517  & 16517   &                   1 &10.97 &--          &11.07  &160 & 0.76\\
      J1103--23   & NGC3511     &  33385 &   33385  & 33385   &                   2 &11.86 &10.73      &11.46  &216 & 0.84 \\
      J2257--41   & NGC7424     &  70096 &   70096  & 70096   &                   1 &11.35 &--          &11.19  &175 & 0.59 \\
      J1153--28   & UGCA250     &  37271 &   37061  & 37061   &                  25 &13.32 &13.17      &11.36  &200 & 1.32 \\
      J0335--24   & NGC1371     &  13255 &   13419  & 13418\rlap{\tablefootmark{D}}   &                  29 &13.33 &13.25      &12.16  &370 & 0.80\\
      \hline
      \noalign{\vspace{2pt}}
      J0052--31   & NGC0289     &   3089 &    3089  &  3089   &                  11 &12.57 &12.84      &11.86  &293 & 0.96 \\
      J0419--54   & NGC1566     &  14897 &   14765  & 13418\rlap{\tablefootmark{D}}   &                  31 &13.45 &13.54      &12.26  &399 & 0.58\\
      J0445--59   & NGC1672     &  15941 &   15941  & 15941   &                   5 &12.31 &12.44      &12.28  &406 & 0.63\\
        \hline
    \end{tabular}
    \tablefoot{ Columns: (1)
      HIPASS identification. (2) Common name. (3) PGC number of galaxy. (4) PGC number of dominant galaxy in group. (5) PGC number of dominant galaxy in association. (6) Number of galaxies in group. (7) Logarithmic total mass of group based on luminosity. (8) Logarithmic total mass of group based on velocity dispersion (only for $>1$ group members). Columns (3) to (8) from \citet{Kourkchi.2017}. (9) Logarithmic total halo mass $M_{200}$ derived from stellar mass given in Table~\ref{tab:sample} and the stellar mass-halo mass relation in \citet{Moster.2013}. 
      (10) $R_{200}$ (virial) radius based on column (9) and standard cosmology. (11) Ratio of the diameter of a $1.5\degree$ field of view in kpc at the distance of the galaxy and the $D_{200}$ diameter, where $D_{200} = 2 R_{200}$.
      \tablefoottext{a}{Part of same group and association.}
      \tablefoottext{A,B,C,D}{Part of same association as indicated by footnote label.}}
\end{table*}

\section{Observations}

The MeerKAT MHONGOOSE observations started in October 2020.  For ease of
scheduling, the 55h of observing time per galaxy are divided into 10
observations of 5.5h. These consist of five ``rising'' observations
and five ``setting'' ones, each of which is referred to as a ``single track'' in this paper. The complete 55h observation is referred to as the ``full-depth'' data.

The rising observations generally start
sometime in the first 1.5 hours after the source has become visible to
MeerKAT, and end when the source is close to transit. The setting
observations start close to transit and generally end in the last 1.5
hours before it crosses MeerKAT's observing horizon. There is some
overlap between rising and setting tracks close to transit. The
amount of overlap depends on the declination of the galaxy but is
never more than 1.5 hours.

A typical 5.5h observation consists of 10 mins of observing time on one
of the primary calibrators J1939--6342 or J0408--6545. This is then
followed by five cycles of \emph{(i)} two mins on a secondary or phase calibrator and  \emph{(ii)}
a subsequent $\sim 55$ mins on the target galaxy. The duration of the
latter varies slightly from galaxy to galaxy to take into account the
different slewing times while staying within the 5.5h overall
duration.

If J1939--6342 was used as the primary calibrator, two additional
three-minute observations of a polarisation calibrator were inserted
between different cycles. If J0408--6545 was used, a polarization
calibrator was observed for three minutes at the end of every phase
calibrator-target cycle. This increased number of polarization
calibrator observations reflects the less well-characterized
polarisation properties of J0408--6545. However, since we are only
concerned with the Stokes $I$ data, we do not discuss the polarization
aspects here.

A small number of observations deviated from our standard observing 
template. For J0008--34, J0031--22 and J2257--41 no suitable
polarization calibrator was available during the rising observations,
so these were extended to 6.5h. The corresponding setting observations
were reduced to 4.5h.

We used the {\tt c856M4k\_n107M} SKARAB correlator mode, allowing for
a 32k narrow band (primarily used for \HI studies) and a 4k wide band
(primarily used for continuum and polarization studies) to be used
simultaneously. The narrow band has 32768 channels of 3.265~kHz (0.7
\kms at the \HI rest-frequency) each, giving a total bandwidth of 107
MHz. The 4k-mode gives 4096 channels of 208.984 kHz ($\sim 43$ \kms at
the \HI rest-frequency), covering the total bandwidth of MeerKAT of
856 MHz. For the MHONGOOSE observations, the central topocentric
frequency of the 32k narrow-band was always set to 1390 MHz. For the
4k-mode this value was 1284 MHz. Both modes observed four
polarizations, though for the narrowband mode (\HI) we only used the
HH and VV polarisations.  The correlator data were integrated over 8s
in the correlator before being output.

To ensure high-quality data, observations were only started if there
were 58 or more active antennas present in the array.  As the science
critically depends on the detection of diffuse low-column density gas,
additional checks were made prior to observing to ensure that a
sufficient number of short baselines was present in the array, with
the strictest checks for the shortest baselines. An overview of the
required fraction of the number of total baselines as a function of
baseline length is given in Table \ref{tab:baseline}.

A small number of early observations were done partly during daytime, but it
was quickly found that solar radio frequency interference (RFI) had a
severe impact on the shortest baselines. Observations that were too
severely affected were redone, while for less-affected ones,
baselines shorter than 1.5 k$\lambda$ (or 315m for rest-frame 21-cm emission) were flagged. To avoid solar
RFI affecting the short baselines, all
other observations were done at night time.

\begin{table}
\small
\caption{Baseline requirements}
\label{tab:baseline}
\centering
\begin{tabular}{l r r}
  \hline
  \hline
  \noalign{\vspace{2pt}}
Baseline range & number  & required \\
 & in range & fraction \\
 (1) & (2) & (3) \\
\hline
\noalign{\vspace{2pt}}
    $<$ 50m &  5     & 80\% \\ 
    50--100m &  29   & 66\% \\
  100--200m &  118   & 66\% \\
  200--400m &  314   & 66\% \\
  400--1000m &  568   & 66\% \\
 1000--3000m &  563   & 66\% \\
 3000--6000m &  405   & 50\% \\
 $>$ 6000m &  14    & 0\% \\
 \hline
\end{tabular}
\tablefoot{Total number of antennas must be 58 or more. (1) Range in baseline length. (2) Maximum number of baselines in this range. (3) Fraction of baselines in this range that needs to be present before observation can commence.}
\end{table}

\section{Data reduction\label{sec:reduction}}

After an observation was completed and deposited in the SARAO MeerKAT
archive, we extracted the data covering the channel range 16384 to 26383 (10,000 channels) from
the 32k narrowband data set retaining only the HH and VV
polarisations. This covers the topocentric frequency range
1390.0--1422.7 MHz. These data were binned by two channels leading to
a 5000-channel measurement set with a channel width of 6.53 kHz (1.4
\kms).  The data were then processed using the CARACal data reduction
pipeline\footnote{CARACal and its related software packages can be found in the repository collection at \tt https://github.com/caracal-pipeline/}
\citep{Jozsa.2020}.  As the observations cover multiple years, and
MHONGOOSE observations took place simultaneously with CARACal
development, several (pre)releases of CARACal were used.
Where necessary data were rereduced to ensure consistent outcomes.

CARACal is a Python-based, containerized pipeline that uses a
``best-of'' approach to link together  reduction and analysis tasks and
applications from various radio astronomy packages to optimally reduce
the data, passing it  seamlessly between the various reduction
stages using {\tt Stimela} \citep{Makhathini.2018}. These stages consist of:

\begin{enumerate}[\itshape(i)]
\item flagging the calibrators;
\item deriving and applying the cross-calibration to the target;
\item flagging the target;
\item imaging the continuum for several cycles of self-calibration;
\item transferring the self-calibration solutions to the original
measurement set;
\item using the self-calibration sky model to subtract the continuum from the original measurement set;
\item creating and deconvolving the spectral line \HI data cubes.
\end{enumerate}
  
We describe these steps in more detail below. Much of the
reduction procedure was developed in collaboration with the MeerKAT
Fornax Survey \citep{Serra.2023}. The description given here should
therefore be regarded as complementary to that given in
\citet{Serra.2023}. Unless stated otherwise, all software tasks
mentioned below in Sec.~\ref{sec:xcal}--\ref{sec:fullcube} were
executed as part of the CARACal pipeline environment. All reduction was
done on the dedicated MeerGas cluster at ASTRON consisting of four computing nodes with 128 cores and  1~Tb of memory each.

\subsection{Cross calibration\label{sec:xcal}}

The calibrator observations were split off in a separate measurement
set and  flagged for shadowing and
RFI where necessary. For the latter {\tt aoflagger} (as included in CARACal) searched
and flagged in Stokes Q. The frequency range $1419.8$--$1421.3$ MHz
($+125$ to $-190$ \kms) was also flagged to avoid contamination from
Galactic emission.

For each observation the primary calibrator  was used to derive the delay, gain and bandpass
calibration using baselines longer than 150m. To improve the
signal-to-noise ratio (S/N), the bandpass was smoothed using a
9-channel box-car filter. Any gaps in the bandpass (e.g., due to the
flagging of the Galactic emission) were interpolated. A second cycle
of flagging and calibrating then refined these solutions, which were
subsequently applied to the secondary calibrator. The latter was used to track
the gain and phase variations over time. These time-dependent solutions were also
derived twice with an intermediate flagging step.  The final solutions
were then applied to the target observations.

\subsection{Self calibration and continuum subtraction}

The cross-calibrated target scans were checked for shadowing (where the line-of-sight to the source is partially or fully blocked by another antenna) and
RFI and, when found, flagged. The clean RFI environment at MeerKAT, and the targets being located in the protected frequency band, meant that on average only 1-2 percent of the data were flagged.

For the self-calibration procedure
the target measurement set was binned in frequency to a channel width
of 1 MHz. The frequency range covering Galactic emission was
flagged, as well as the frequency range of the target galaxy.  The
latter was done to prevent bright \HI features (present for some
galaxies) from entering the self-calibration procedure.

Three cycles of imaging and self-calibration were then performed using
three spectral solution intervals of $\sim 10$ MHz each. During each
cycle, the data were deconvolved and imaged (using \wsclean; \citealt{Offringa.2014} ) to progressively deeper
limits from 5 times the noise for the first imaging run, to 3.5 times
the noise for final imaging. We imaged an
area of $3 \times 3$ degrees, using a robust weighting value of
$-0.5$, a taper of $10''$ and a pixel size of $4''$. 
The SoFiA source finder \citep{Serra.2015} (as implemented in CARACal) was used to automatically create a mask 
of the sky based on the output cleaned image.  Care was taken that a sufficiently large window for the noise
measurement was chosen such that bright, extended continuum emission
from some of the target galaxies did not affect the measured noise
levels and therefore the self-calibration procedure. 
Using this mask, a new image and sky model were created. 
This sky model was used for 
 the self-calibration, after which a new cycle of image, mask and sky-model calibration was started, followed by further self-calibration. All cycles were phase-calibration only.
 In this way the three cleaning and 
self-calibration cycles to progressively larger depths lead to the final solutions.

The self-calibration solutions were interpolated in frequency and
transferred to the cross-calibrated target measurement set. The most
recent self-calibration sky-model clean components were also
transferred to the target measurement set at this stage, and subsequently subtracted
using the {\tt Crystalball} software in CARACal. This step is
time-consuming and accounts for one-third to one-half of the total
processing time.

A second follow-up continuum subtraction using the CASA task {\tt
  mstransform} within CARACal was used to subtract any residual
continuum by fitting a first or second-order polynomial to the
line-free visibilities. A catalog of known \HI sources was used to
define the line-free channels. This second procedure on rare
occasions produces bright artefacts related to the fitting of
completely flagged channels, and a simple sigma-clipping strategy was
used as part of the pipeline to remove these artefacts.

\subsection{Quality control imaging}

For each observation, two data cubes were created to check for any
remaining RFI or reduction artefacts and for general quality
control.  One cube has high angular resolution
using a robustness parameter of 0.5 and covering the 1000 \kms
velocity range straddling the target galaxy with a channel spacing of
1.4 \kms. A second, lower resolution cube also used a robustness
parameter of 0.5, but with a taper of $25''$ applied. Here we imaged
the entire frequency range above 1390~MHz but at a velocity resolution
of 7 \kms (5-channel binning). 

\subsection{$u=0$ flagging}

\begin{figure}
  \resizebox{\hsize}{!}{\includegraphics{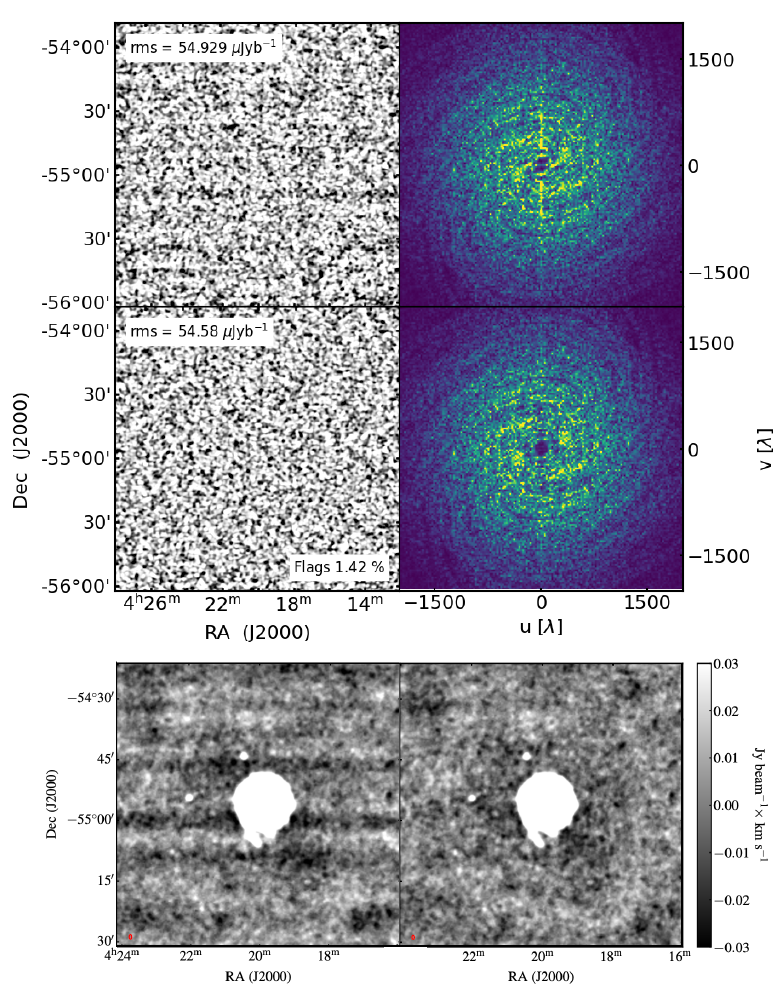}}
  \caption{Overview of the $u=0$ flagging procedure. The top row shows
    a 100-channel binned empty  map with horizontal stripes of one
    target scan of a NGC 1566 observation on the left, and on the
    right the corresponding FFT amplitude image, which shows a prominent
    vertical RFI feature at $u=0$. The middle row shows the same
    plots, but after removal of the $u=0$ stripe using our flagging
    procedure. The bottom row shows two zeroth-moment \HI maps of NGC
    1566, both created using a simple $2\sigma$ intensity cut. The
    bottom-left panel shows the map with the $u=0$ RFI still included. The bottom-right panel 
    shows the same map but with the $u=0$ RFI removed.}
  \label{fig:u0}
\end{figure}

After combining the separate observations of our first completed
galaxies and the subsequent creation of the first full-depth cubes, we
encountered unexpected coherent horizontal stripes in the data cubes.
Investigation showed that these were caused by RFI, which in visibility
space is located at $u=0$ over a small range of $\varv$-values centred symmetrically 
around $\varv=0$ (see
Fig.~\ref{fig:u0}).  Most likely this is due to the low or zero fringe
rate near $u=0$, which allows low-level RFI to accumulate
coherently. Away from $u=0$ this RFI is added incoherently due to the
rapid change in amplitude and phase of the visibilities. These stripes
are an issue as they are at the level of the faint \HI emission that we are
trying to detect and we thus need to remove them from the
data.

The $u=0$ artefacts have been seen before in observations taken with
different arrays (e.g., \citealt{Hess.2015,Lucero.2015,Heald.2016}).  Generally, the
solution has been to simply flag a small range around $u=0$ for
the relevant range in $\varv$. Applying these strategies to the MHONGOOSE data would,
however, result in flagging close to $\sim 10\%$ of the data,
due to the large number of short baselines in the MeerKAT array and
the resulting high filling factor of the inner $u\varv$ plane. Such a solution is clearly not ideal, and we therefore
developed a more sophisticated flagging strategy
\citep{Maccagni.2022} as described below.

We first split the observation into the separate target scans of $\sim 55$ minutes each, and then
perform the following procedure on a scan-by-scan basis. Since in each scan the $u\varv$-coverage 
is different, we calculate
the amplitude image of a Fast Fourier Transform (FFT) of the sum of
100 line-free channels in a custom-made data cube of each scan. We adopt a
pixel size of $20''$ and an image size of $8000''$. This leads to a
pixel size of $25.8\lambda$ in the FFT image. The $u=0$ artefacts are
then visible as a vertical stripe ($u=0$ for a range in $\varv$) in the
amplitude image. We found that using an empty data cube created using
a robust value of 1.5 combined with a $60''$ tapering as the basis for
the FFT image gave the best results.

We determine the median absolute deviation (MAD) of the distribution
of amplitudes $a_i$ in the $u\varv$-plane, where MAD $=$
median$(|a_{i,(u,\varv)} - {\rm median}(a_{i,(u,\varv)})|)$. We then define a
cutoff threshold (median$(a_{i,(u,\varv)}) + M \times {\rm MAD}$), where $M=100, 150, 200, 300, 500$. For each $M$, we
find the pixels in the amplitude image where the value is larger than
the cut-off and we flag the corresponding visibilities in the 
measurement set of that scan.

The empty cube is then reimaged and the noise measured. We adopt the
value of $M$ that gives the lowest noise and identify the
corresponding flagged pixels in the amplitude image. Flags are
extended by 1 pixel in $u$ and 2 in $\varv$ and the corresponding
visibilities are flagged, but this time throughout the entire spectral
range of the measurement set of that scan. Doing this for all scans in
an observation results in removal of the $u=0$ artefacts at the
cost of flagging typically only  a few tenths of a percent of the visibilities. This $u=0$
flagging solution has been implemented in CARACal. We refer to \citet{Maccagni.2022} for further 
information on the flagging fraction and its sky distribution.

\subsection{Time flagging\label{sec:tflag}}

The initial flagging steps removed most of the RFI in the
data. However, we found that in some cases the ``wings'' (i.e.,
ramp-up and ramp-down) of time-variable strong RFI were not removed
entirely. The most effective way to remove these was by
using a separate time-flagging step.  This was done with the
CASA {\tt flagdata} task (again within CARACal) where we average in
bins of 100 channels, and use the {\tt tfcrop} algorithm to fit the
data in the time dimension and remove outliers that deviate more than
5 times the standard deviation from the fit.  After merging with the
initial pipeline flags, these flags are then extended such that if
more than half of the visibilities at a certain observing time are flagged, all data at
that time are flagged. Similarly for the frequency channels: if more
than half of the data in a frequency channel are  flagged,
then all data at that frequency are flagged.  These final flagging
steps complete the reduction of a single-track
observation and the individual measurement sets are now ready to be 
combined and for the full-depth data cubes to be created. 

\subsection{Standard resolutions\label{sec:sens}}

MeerKAT is capable of producing high-quality images over a large range
in angular resolution (Fig.\ \ref{fig:flat}), resulting in mapping of
compact, high-column density sources, as well as extended, low-column
density features (Fig.~\ref{fig:surveys}).  It is difficult to capture
this wealth of structure in a single data cube at a single
resolution. For MHONGOOSE, we have therefore defined six standard
resolutions for our data products, which cover the angular resolution range of
MeerKAT as indicated in Fig.\ \ref{fig:surveys}.

These different resolutions are achieved by changing the robust
weighting parameter used in creating the data cubes, and, for the two
lowest resolutions, some additional tapering. We find that these six
standard combinations give a comprehensive overview of the \HI
morphology and kinematics of the sample galaxies.

\begin{figure}
  \resizebox{\hsize}{!}{\includegraphics{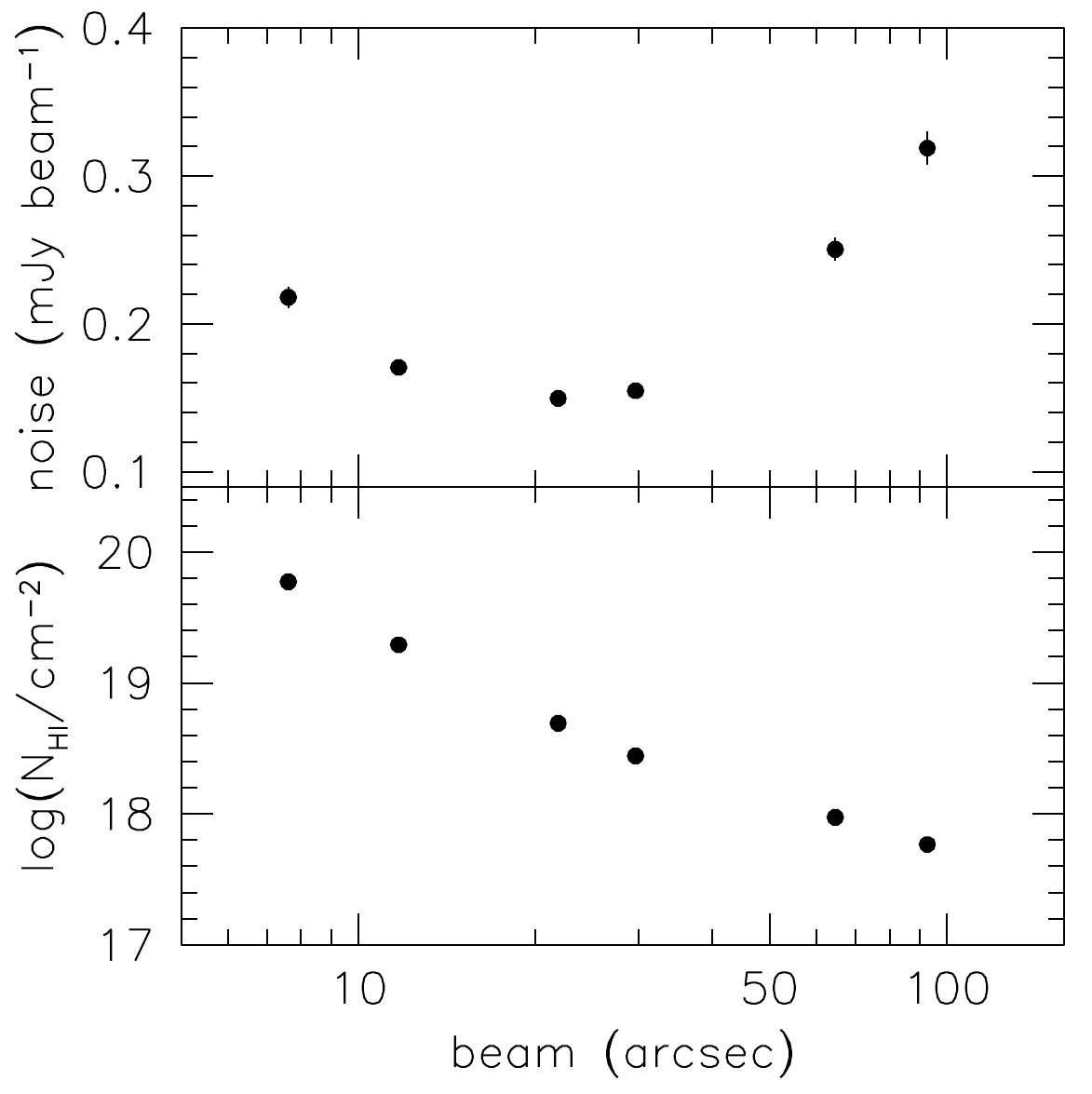}}
  \caption{Sensitivities of the full-depth data cubes. The top panel
    shows the average noise per 1.4 \kms channel for the six standard
    resolutions, as averaged over full-depth observations of ten
    galaxies (cf.\ Fig.~\ref{fig:flat}). Error bars show the rms difference in noise levels between the ten
    galaxies, but are generally comparable to or smaller
    than the symbol size. The bottom panel shows the $3\sigma$, 16
    \kms column density sensitivities. These are the same points as
    shown in Fig.~\ref{fig:surveys}. }
  \label{fig:noisevalues}
\end{figure}

Table~\ref{tab:resolutions} and Fig.~\ref{fig:noisevalues} show the average noise per channel and beam size as derived from
the full-depth standard cubes of ten galaxies available at the time of writing.  We also list the column density sensitivities where we
give the values for $1\sigma$ over a single channel, as well as
the $3\sigma$ over 16 \kms values used in Fig.~\ref{fig:surveys}.
Finally, we also list the $3\sigma$ \HI mass detection limit for an unresolved source, assuming
a distance of 10 Mpc (the median distance of the galaxies in the
sample) and a velocity width of 50 \kms.

The highest resolution of $\sim 8''$ is achieved using a robustness
parameter of zero. Decreasing the robustness parameter even further to
negative values resulted in too small a decrease of the beam size to justify the loss of sensitivity.
This loss is due to the relatively small number of
longest baselines (cf.\ Table~\ref{tab:baseline}). Emphasising these through weighting does therefore
not result in an improvement in image quality
at these
resolutions. For the purpose of MHONGOOSE \HI imaging, a robustness
parameter of zero therefore yields the highest usable resolution.

On the opposite end of the resolution range, tapering beyond $\sim 90''$
does not yield additional gains due to the absence of sufficiently
short baselines (cf.\ Table \ref{tab:baseline}). This can be seen by
the upturn in the noise curve at large resolutions in
Fig.\ \ref{fig:flat} and the top panel in Fig.~\ref{fig:noisevalues}.

\begin{table*}
\small
\caption{Standard resolutions}
\label{tab:st_res}
\centering
\begin{tabular}{l r r r r r  r c c c c}
  \hline
  \hline
  \noalign{\vskip 2pt}
  label & robust &taper &       pixel &$b_{\rm maj}$& $b_{\rm min}$ & $b_{\rm PA}$ & noise & $\log N_{\matHI}^{1\sigma,1{\rm ch}}$ &  $\log N_{\matHI}^{3\sigma,16{\rm km/s}}$& $\log{M_{\matHI}}$ \\
& value & ($''$) & ($''$) & ($''$) & ($''$) & $(^{\circ})$& (mJy beam$^{-1}$)& (cm$^{-2})$& (cm$^{-2}$)& $(M_{\odot})$\\
(1) & (2) & (3) & (4) & (5) & (6) & (7) & (8) & (9) & (10) &(11) \\
\hline
  \noalign{\vskip 2pt}
{\tt r10\_t90} & 1.0 &  90&     30 &    94.2 & 91.2 & 47&       0.318 $\pm$ 0.011&  16.76 & 17.77 & 6.05\\
{\tt r05\_t60} &0.5 &   60 &    20 &    65.3 & 64.0 & 92  &     0.250 $\pm$ 0.009&  16.97 & 17.98 & 5.95\\
{\tt r15\_t00} &1.5 &   0 &     7 &     34.4 & 25.4 & 135 &     0.154 $\pm$ 0.004&  17.44 & 18.44 & 5.74\\
{\tt r10\_t00} &1.0 &   0 &     5 &     26.4 & 18.2 & 136 &     0.150 $\pm$ 0.004&  17.69 & 18.69 & 5.73\\
{\tt r05\_t00} &0.5 &   0 &     3 &     14.1 & 9.7 & 137 &  0.171 $\pm$ 0.005&  18.29 & 19.29 & 5.78\\
{\tt r00\_t00} &0.0 &   0 &     2 &     8.2 & 7.1 & 142  &      0.219 $\pm$ 0.007&  18.77 & 19.77 & 5.89\\
 \hline
\end{tabular}\\
\tablefoot{(1) Label used to refer to this resolution. (2) Robustness weighting value. (3) Taper used. "0" mean no taper. (4) Pixel size of the data cubes and moment maps. (5) Average major axis beam size. (6) Average minor axis beam size. (7) Average position angle of major axis of beam. (8) Noise per channel. (9) $1\sigma$, one-channel column density sensitivity. (10) $3\sigma$, 16 \kms column density sensitivity. (11) \HI mass sensitivity for a $3 \sigma$, 50 \kms unresolved source at 10 Mpc.}

\label{tab:resolutions}
\end{table*}

\subsection{Creating the full-depth cubes\label{sec:fullcube}}

For the full-depth observations, we combine the ten individual 5.5h observations and create data cubes measuring
$1.5\degree \times 1.5\degree$ with a velocity range from $-500$ \kms
to $+500$ \kms with respect to the central velocity of the target
galaxy. We choose to image an area larger than the usual primary beam of
$\sim 1\degree$ due to the high sensitivity of the observations, which
enable \HI detections at a large distance from the pointing center.

The ten input observations were first all corrected from a topocentric
spectral coordinate grid to a common heliocentric spectral grid. 
We adopt the radio velocity definition where all channels have a constant frequency spacing and velocity is defined as $\varv_{\rm radio} = c(\nu_0 - \nu)/\nu_0$, where $\nu_0$ is the rest-frequency and $\nu$ the observed frequency.
The
ten measurement sets were then all simultaneously input into
\wsclean\ \citep{Offringa.2014} to create and deconvolve the final data cubes.  So-called
``clean masks'' were used to  indicate areas with emission where
deconvolution was necessary. These masks were created automatically
using a sequence of cubes at each of the standard resolutions listed in Table~\ref{tab:resolutions}.

Starting at the lowest resolution {\tt r10\_t90}, we created an
initial deconvolved cube ({\tt cube\_0}) using the auto-masking option
in \wsclean, where the mask ({\tt mask\_0}) was defined to include values of $5\sigma$
and higher. We then used SoFiA-2 \citep{Westmeier.2021} outside of
CARACal to create a new mask ({\tt mask\_1}) based on the
deconvolved {\tt cube\_0}.

Here, and for all subsequent SoFiA-2 runs, we used spatial kernels of 0
and 3 pixels, and velocity kernels of 0, 3, 7 and 15 channels. Using
the smooth-and-clip (S+C) method \citep{Serra.2012}, we used a threshold of 4$\sigma$ to define the mask
with spectral noise scaling enabled.  To define sources, valid pixels
were linked across a maximum spatial and spectral distance of,
respectively, 3 pixels or 3 channels. For the detected sources we then
imposed a spatial minimum size equal to the current beam size rounded
up to the nearest integer number of pixels. For the spectral minimum
size, we used 10 channels (14 \kms). Sources with negative flux density values
were not retained.  These SoFiA-2 input parameters were extensively
investigated to ensure that they optimally captured the emission
present in the cubes.

The resulting {\tt mask\_1} was used to create and deconvolve a new
{\tt cube\_1}. A new run of SoFiA-2 on this cube resulted in a {\tt
  mask\_2}, which was in turn used to create {\tt cube\_2}, which is
the final deconvolved {\tt r10\_t90} cube.  Within the masks, emission
was deconvolved down to $0.5\sigma$. The gain value for the major
iterations in \wsclean\ was set to 0.95 to ensure at least one cycle
of inversion of the clean model to visibility space.

For subsequent, higher spatial resolutions, the final mask of the
previous, lower resolution run was used as the initial mask. That is, for
the {\tt r05\_t60} cubes, the final {\tt mask\_2} of the {\tt
  r10\_t90} run was regridded and used as the initial {\tt mask\_0} of
the {\tt r05\_t60} runs.

The procedure is then as above where a {\tt cube\_1} is produced and SoFiA-2
is run on that cube to build a {\tt mask\_1}.  This is then used to create and deconvolve the
final {\tt cube\_2}.

This is then repeated in sequence for each of the resolutions, where the final mask of a given resolution is used as the initial mask of the next higher resolution.
This process is a fully automated part of our reduction pipeline. It
delivers high-quality deconvolved cubes almost entirely unsupervised.

\begin{figure}
  \resizebox{\hsize}{!}{\includegraphics{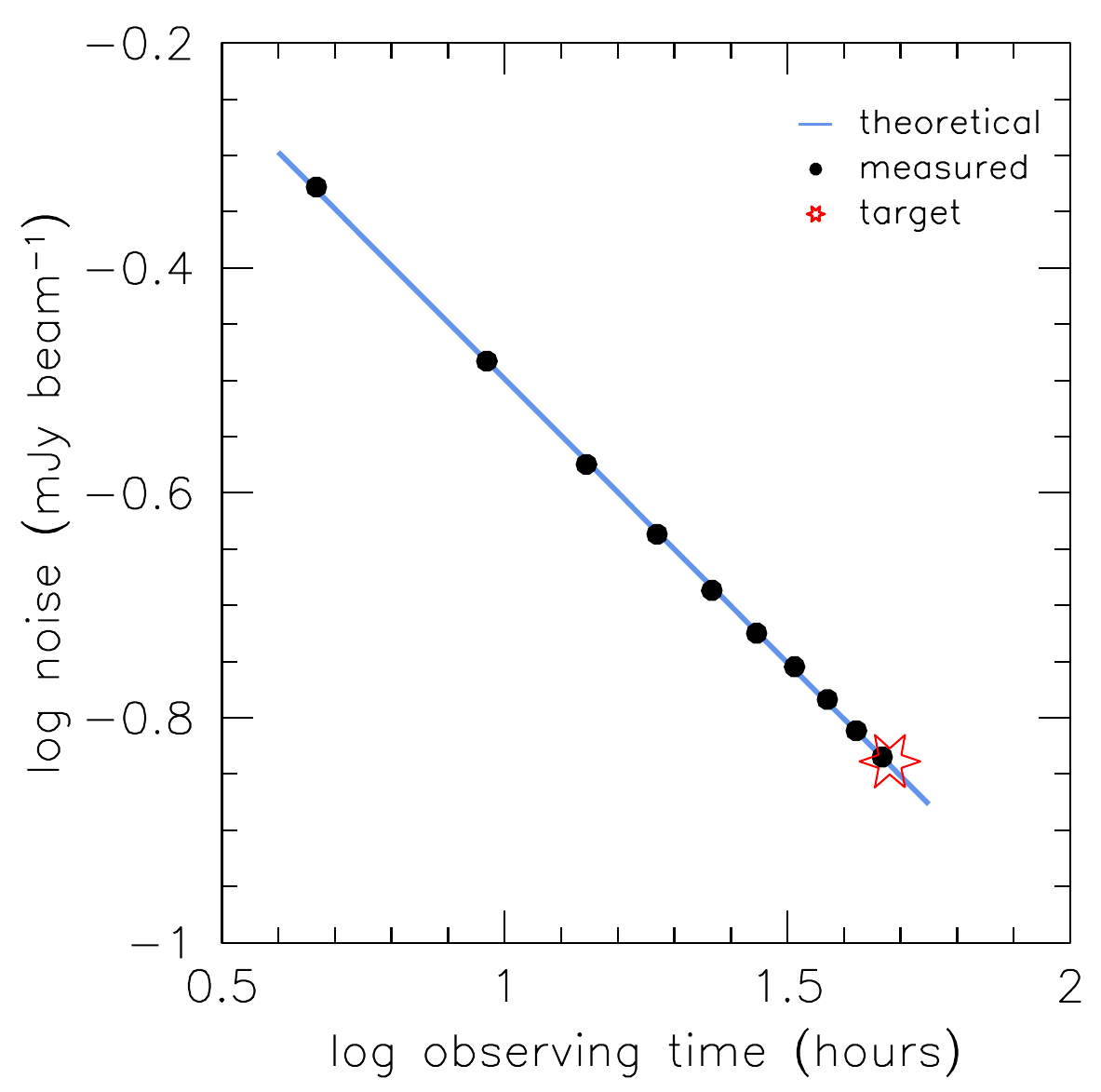}}
  \caption{Natural-weighted noise values measured over a 1.4 \kms
    channel as a function of integration time for the observations of
    ESO 300-G014 (J0309--41). The black filled points indicate the measured noise
    level. The blue line shows the expected theoretical noise levels with a slope of $-0.5$. A linear fit to the observed noise levels gives 
    a slope of $-0.504$. The red star
    indicates the target sensitivity from \citet{deBlok.2016}. }
  \label{fig:noisetime}
\end{figure}

As expected, the noise in the data cubes keeps decreasing as the
square root of the integration time as each of the ten observations is
added in turn. This is illustrated in Fig.~\ref{fig:noisetime} where
we show the noise per channel measured in natural-weighted
data cubes of  ESO 300-G014 (J0309--41), created by combining an increasing number of
tracks. We used natural weighting, instead of one of the standard
resolutions, to circumvent uncertainties due to changes in the noise
because of weighting and tapering and to enable an unambiguous comparison with the expected theoretical noise. 
The latter is calculated based on the number of visibilities and the properties of the MeerKAT antennas (13.5m dishes and
$T_{\rm sys}/\eta = 20.5$ K) and is also shown in Fig.~\ref{fig:noisetime}. The theoretical values have (by
definition) a slope of $-0.5$. The measured values have a slope of
$-0.504$.  The very small offset between the lines is equivalent to an
offset in $T_{\rm sys}/\eta$ of $\sim 0.4$ K, in the sense that MeerKAT
performs slightly better than our assumptions (though such a small
offset is likely to be within the variations between different observations).

The measured noise level thus meets the requirements presented
in the MHONGOOSE white paper \citep{deBlok.2016}, where a noise level of 0.074 mJy beam$^{-1}$ per 5 \kms channel after 48 hours of
on-source is specified. Scaling this to the channel width and
observing time that were eventually used, we see an almost exact
match, as indicated in Fig.~\ref{fig:noisetime}, showing that the full-depth MHONGOOSE data are of the required sensitivity to 
deliver the envisaged survey science.

\subsection{Creating full-depth moment maps\label{sec:full}}

We also create zeroth-, first- and
second-moment maps showing the integrated \HI distribution, the
intensity-weighted mean velocity and the velocity width, repectively.

The second-moment map is commonly referred to as a velocity
dispersion map. However, strictly speaking, second-moment
values only represent a physically meaningful velocity dispersion in
cases where the \HI emission line profile is well approximated by a 
single Gaussian\footnote{We note that here we adopt the informal definition that the second moment is equal to the dispersion. Mathematically, the second moment is defined as the variance.}.
In all other cases (non-Gaussian profiles or multiple separate components) they simply 
quantify the
spread of \HI velocities along each line of sight.

We used SoFiA-2 to find the signal in the cubes and create the
masks. The S/N in the cubes is high, and we found that we only needed
a limited number of spatial kernels to isolate the signal. In fact,
using spatial kernels larger than $\sim 1$--$2$ beam sizes had
the effect of introducing additional noise into the moment maps at the edges of the disks
and these larger kernels were therefore avoided. We used cubes that were not corrected for the
primary beam. This correction was applied to the zeroth-moment maps afterwards.

The SoFiA parameters we eventually used were spatial kernels of 0 and
4 pixels and velocity kernels of 0, 9 and 25 channels using a source-finding
S+C threshold of $4\sigma$. Sources were linked over a maximum of 5
pixels spatially and 8 channels spectrally. Minimum sizes of 4 spatial
pixels and 15 channels were imposed.

We then used the reliability parameter, as implemented in SoFiA, to further separate 
signal from the noise. We did find, however, that this parameter was
less critical as almost all sources were well separated from the noise
and an integrated S/N cut was usually sufficient to identify the
significant sources. In our case, we mostly used a reliability value
of 0.8 and an integrated S/N value of 5.0.

The masks as defined here are stricter than the masks used during deconvolution. 
This is due to the different purpose that they serve. The deconvolution masks were created 
to include all emission at many scales, even if this leads to including some noise in the masks. 
For the moment maps, instead,  it is important that as little noise as possible is included in the 
masks to avoid affecting the moment values. 

\subsection{Examples of full-depth moment maps\label{sec:fullmom}}

A full scientific analysis of the full-depth data will be presented in
forthcoming papers.  For this reason, and as data are still being
collected, we restrict ourselves here to presenting a short overview
of the full-depth data for two of the galaxies that were completed
early in the survey. We refer to \citet{Healy.2024} and
\citet{Maccagni.2024} for in-depth discussions. The purpose of
this overview is to give an indication of the high quality of the
data.

\begin{figure*}[htb!]
  \resizebox{0.95\hsize}{!}{\includegraphics{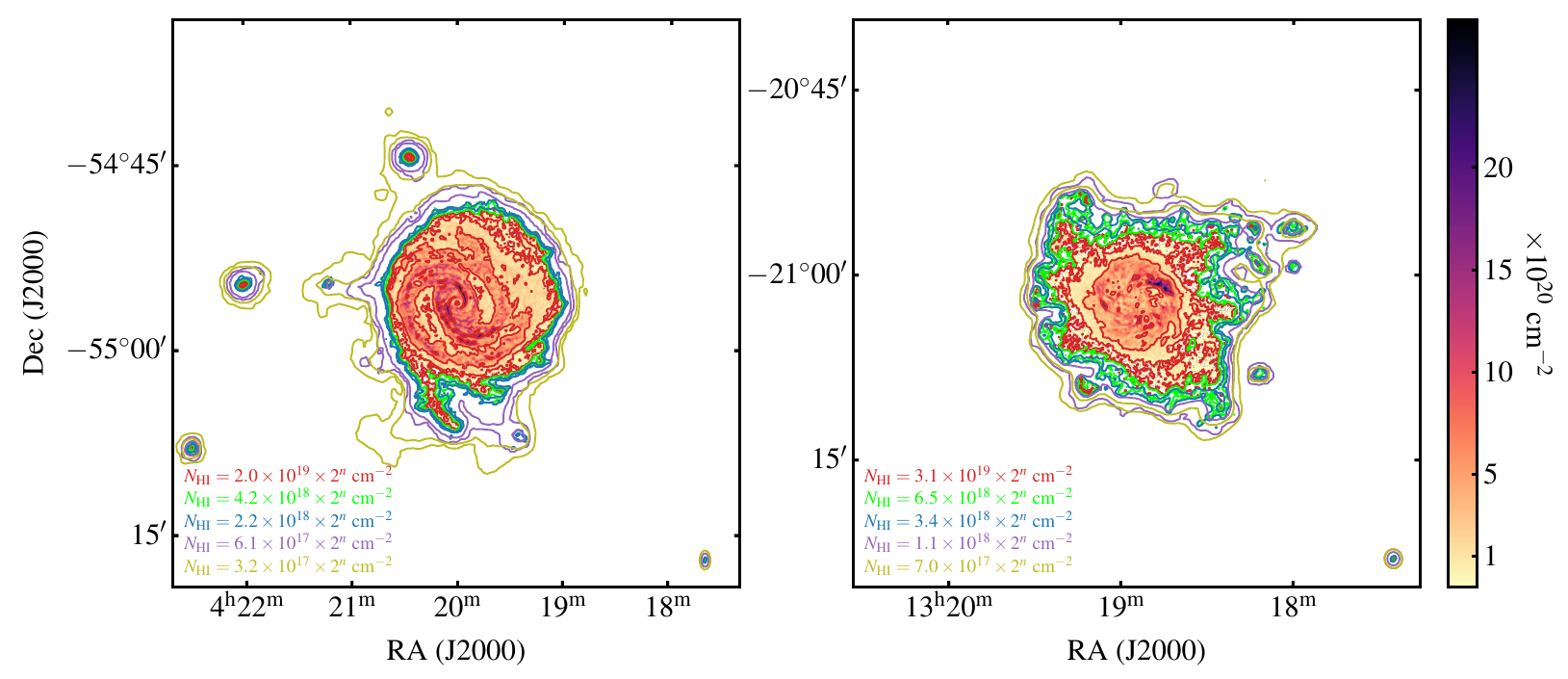}}
  \caption{Multi-resolution zeroth-moment maps of NGC 1566 (left panel) and NGC 5068 (right panel) based on the full-depth data. For each resolution, contours of $2^n$ times the S/N = 3 value  are shown (as listed in the legend). The resolutions shown are {\tt r05\_t00} (red), {\tt r10\_t00} (green), {\tt r15\_t00} (blue), {\tt r05\_t60} (purple), {\tt r10\_t90} (olive). For each resolution two contours are shown, i.e., $n=0,1$, except for the highest resolution {\tt r05\_t00} (red contour), where $n = 0, 1, 2, ...$ The {\tt r05\_t00} moment map is shown in the background as a false color image, with the column density levels indicated by the color bar on the right. The beam sizes (colored according to the respective resolution) are shown in the bottom-right corner.}
  \label{fig:multires}
\end{figure*}

Figure \ref{fig:multires} shows multi-resolution maps of the \HI
distribution in NGC 1566 (J0419--54) and NGC 5068 (J1318--21). Here we
show for five of the six standard resolutions the contour at the
column density level that corresponds to S/N = 3 in the zeroth-moment
map at that resolution.

During the initial reduction procedure, the velocity channels were not Hanning-smoothed and 
binned in velocity every two input channels (cf.\ Sect.~\ref{sec:reduction}). 
The channels in the current data cubes are therefore independent, meaning that 
the noise is expected to behave as $\sqrt{N}$ when summing $N$ channels.

We therefore can construct a noise map with
pixel values defined as $\sigma_{\rm chan}\, \delta v\, \sqrt{N}$,
where $\sigma_{\rm chan}$ is the average noise value measured in empty
channel maps, $\delta v$ is the velocity width of a channel (in this
case 1.4 \kms) and $N$ is the number of independent channels contributing to each
pixel in the zeroth moment map. 

Taking the ratio of the
zeroth-moment map and this noise map then gives a S/N map. We derive
the median column density value of all pixels with $2.75 < {\rm S/N} <
3.25$ and adopt this as the column density corresponding to S/N = 3 (also known as the ``pseudo-3$\sigma$'' column density; see \citealt{Verheijen.2001, Kregel.2004})

We plot contours of multiples of this value, thus clearly capturing in a single figure  
the detailed, compact high-column density distribution at the
highest resolutions, and then transitioning into the diffuse, extended distribution of
the low-column density \HI at the lowest resolutions.  Note the extended
and irregular distribution of \HI in the outer parts of NGC 1566, as
well as the clumpy outer \HI in NGC 5068. At column densities of $\sim
10^{20}$ cm$^{-2}$ (corresponding with the edge of the main
star-forming disks) very little of this irregular outer \HI would be
detected and both galaxies would give the impression of regular, symmetric
disks; cf.\ the observations of NGC 1566 in \citet{Elagali.2019} which concentrated on the high-column density disk. As
we show below, the outer \HI components contain only a small
fraction of the total \HI mass of the galaxies, but they dramatically
change the apparent morphology.  As mentioned, more
extensive analyses of these and other sample galaxies will be
presented in future papers, but Fig.~\ref{fig:multires} already shows the power of these deep observations.

\begin{figure*}[htb!]
  \centering
  {\resizebox{0.95\hsize}{!}{\includegraphics{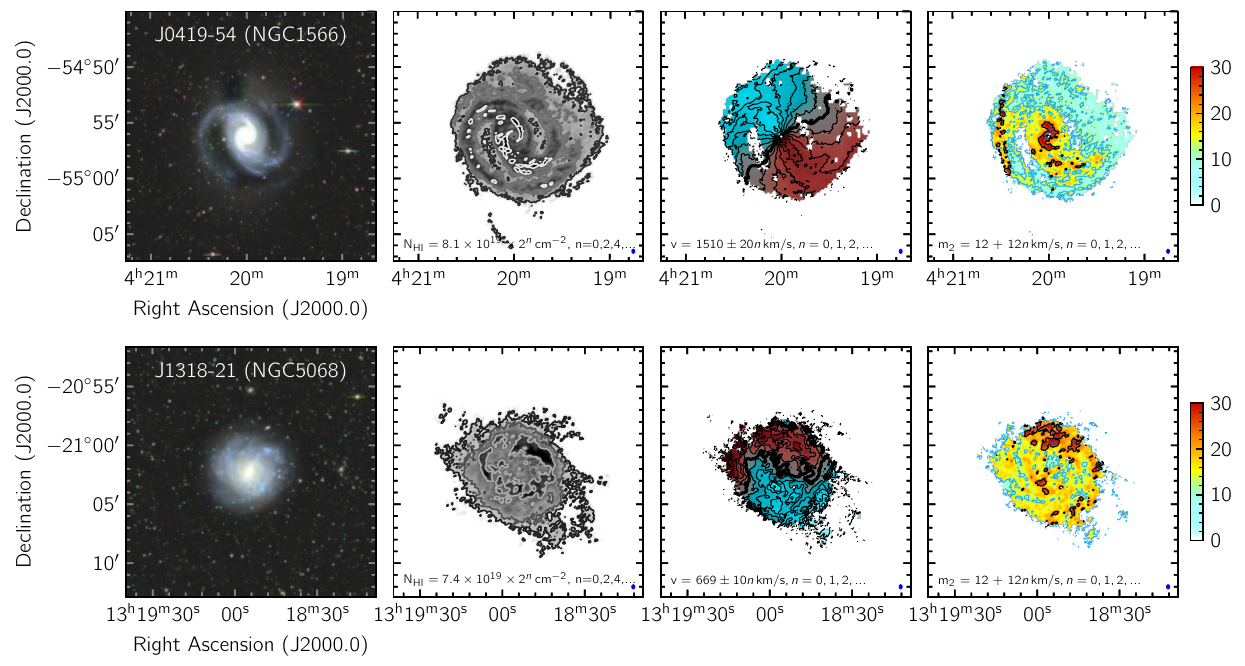}}}
  \caption{Example single-track moment maps using the {\tt r05\_t00} resolution for two MHONGOOSE galaxies. Top row: NGC 1566 (J0419--54); bottom row: NGC 5068 (J1318--21). From left to right: \emph{(i):} Combined \emph{grz}-color image from DECaLS. \emph{(ii):} primary-beam corrected zeroth-moment or integrated \HI intensity map. Contours as indicated in the figure. The lowest contour represents S/N = 3, with subsequent contour levels increasing by a factor of two. \emph{(iii):} First-moment map or intensity-weighted velocity field. Red colors indicate the receding side, blue colors the approaching side. The central velocity (listed in Tab.~\ref{tab:sample}) is indicated by the thick contour. Other contours are spaced by 10 or 20 \kms, as indicated in the Figure. \emph{(iv):} Second-moment map: colors show the range from 0 (light-blue) to 30 (red) \kms. The lowest contour shows the 12 \kms level, and subsequent contours are spaced by 12 \kms. The 24 \kms contour is shown in black. For both the first- and second-moment maps, pixels corresponding to  values below the S/N = 3 column density in the zeroth-moment map were blanked. See Appendix A for a more extensive description.} 
  \label{fig:sample}
\end{figure*}

\subsection{Single-track moment maps\label{sec:single}}

Observations are still ongoing at the time of writing, and
full-depth observations are not yet available for the entire
sample. However, every galaxy in the sample has by now been observed
for at least one 5.5h track.  The results presented in the remainder of this paper are based on the single track data, unless otherwise stated. 
To give an overview of the sample and pursue initial sample-wide science, 
we created single-track moment maps at two of the standard resolutions 
{\tt r05\_t00} and
{\tt r15\_t00} (beam size of $\sim 12''$ and $\sim 30''$, respectively) in an identical manner as for the full-depth data.  We
have limited this to two resolutions as the differences between the
standard resolutions are not as pronounced here due to the ten times
shorter integration time. Average beam sizes for these single-track
observations are identical to those listed in Table
\ref{tab:resolutions}, while column density and mass limits are 
a factor $\sqrt{10}$ less (cf.\ Fig.~\ref{fig:noisetime}).
These moment maps give a
comprehensive overview of the entire MHONGOOSE sample as observed by
MeerKAT.

In Fig.~\ref{fig:sample} we show the {\tt r05\_t00} single-track moment maps for NGC 1566 (J0419--54) and NGC 5068 (J1318--21) which were also discussed in Sect.~\ref{sec:fullmom}.
In the Figure, we show the Dark Energy Camera Legacy Survey (DECaLS)\footnote{\tt http://www.legacysurvey.org} (\citealt{Dey.2019}) optical image, the zeroth-moment (integrated \HI intensity) map, the first-moment (intensity-weighted
mean velocity field)\footnote{The color map used for the velocity field  is based on the  perception-based color maps  available as part of the CosmosCanvas project: {\tt https://ascl.net/code/v/3580}} and finally, the second-moment map.  For both the first- and second-moment maps, pixels corresponding to  values below the S/N = 3 column density in the zeroth-moment map were blanked. A comparison between Figs.~\ref{fig:multires} and \ref{fig:sample} for these galaxies clearly show the improvement in sensitivity that we can expect from the full survey.

Moment maps for the {\tt r05\_t00} resolution of the complete sample are given in Appendix A and we refer to this Appendix for a more extensive description.

\section{Global \HI profiles}

The global \HI profile or integrated \HI spectrum of a galaxy shows the
spatially integrated \HI flux density as a function of velocity or frequency. It is often used in scaling relation studies, such as the
Tully-Fisher relation \citep{Tully.1977} and also in studies involving the total \HI
masses of galaxies. Global \HI profiles are often measured using
single-dish telescopes.  These measure the total power, and therefore
the total \HI mass. Global \HI
profiles measured with synthesis radio telescope arrays are less straight-forward to interpret.  
Depending on the baseline distribution of the array and the extent of the source, these may suffer
from the so-called ``missing spacing'' problem, which is caused by a lack of short baselines which prevents measurements of flux on larger angular scales. For  galaxies  
that have an angular size similar or larger than the scales probed by the shortest baselines of the array this can lead to a severe
underestimate of the total \HI mass. In these cases, a comparison with
the single-dish \HI global profile is often used to determine
whether any \HI has been missed in the synthesis observations. For MeerKAT, with a shortest baseline of 29m, the largest scale at which signal can be detected is $\sim 20'$.

Comparisons between the single-dish \HI profiles and
those derived from synthesis observations have been used to constrain the 
existence of low-column density \HI in
galaxies (e.g., \citealt{Pingel.2018, Kamphuis.2022}). The underlying assumption 
is that any flux not
detected in the synthesis observations must be extended and/or of low
column density.

MeerKAT has been specifically designed to have a compact core with
many short baselines, and one may expect that any missing-spacing
issue will be less severe. The MHONGOOSE data, both the single-track
and full-depth observations, offer a chance to systematically compare
the measured masses with single-dish masses for a significant number of
our sample galaxies. Below we first compare the single-track and full-depth profiles to gauge whether the increased observing time led to the detection of additional \HI. This is then followed by a comparison with the single-dish measurements.

\subsection{Comparing the single-track and full-depth profiles\label{sec:fluxden}}

A first step is to compare the \HI global profiles of the single-track
data and full-depth data which differ by a factor of ten in observing
time. This obviously means an increase in S/N for the
full-depth data, but an interesting question is whether this can be
used to deduce the existence of low-column density \HI.

\begin{figure}
  \resizebox{\hsize}{!}{\includegraphics{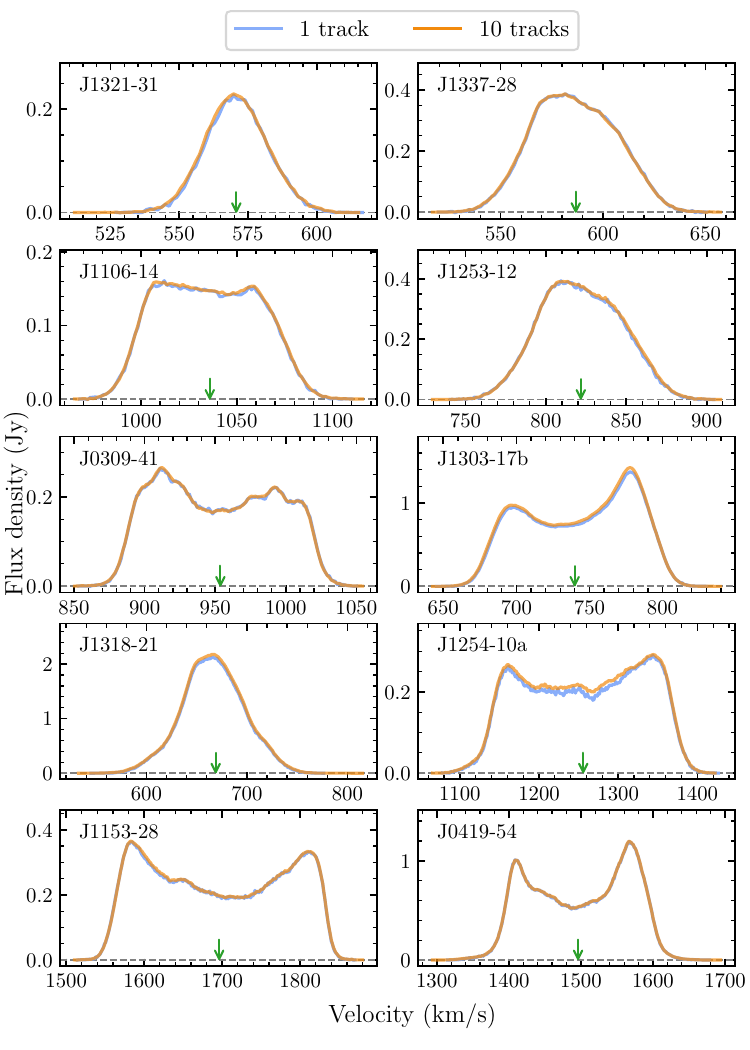}}
  \caption{Single-track (blue) and full-depth (orange) global \HI
    profiles for ten of the MHONGOOSE galaxies as identified in the
    top left of each panel. The green arrows indicate the central velocities. We show only the velocity ranges where signal was present in the mask.}
  \label{fig:deepshallowcomp}
\end{figure}

We compare the global profiles and total \HI masses for ten of the
MHONGOOSE galaxies for which the full-depth global profiles were
determined. The global profiles are derived by measuring the flux in each
channel, applying the masks that were used to create the moment
maps.  The fluxes are corrected for primary beam attenuation. The global
profiles are compared in Fig.~\ref{fig:deepshallowcomp} with the corresponding 
flux densities (as
listed in Table~\ref{tab:deepshallowcomp}).
The full-depth \HI masses are on average $\sim 2\%$ higher
than the single-track \HI masses. Such differences may sound insignificant, 
but they make a big difference when it comes to deducing the
existence of a low-column density component.

If we assume that for each galaxy, the ``extra'' detected \HI 
is due to a previously undetected low-column density \HI component that is 
spread evenly over the \HI disk, 
then we can use the additional mass and the area of the disk to calculate the column density.
This results in column density components of a few times $10^{17}$
cm$^{-2}$ to a few times $10^{18}$ cm$^{-2}$, as listed in Table~\ref{tab:deepshallowcomp}. This is consistent with the increase in column density 
detection limits going 
from  single-track to full-depth observations.
These numbers highlight that to deduce the presence or absence of low column density gas based
purely on global profiles, an accuracy of a few percent or better is
needed. 
Such an accuracy is much higher than the typical
absolute flux calibration uncertainty of \HI observations. This 
is compounded when comparing with results from different telescopes.

\begin{table*}
\small
\centering
\caption{\HI mass comparison between full-depth and single-track MeerKAT observations}
\begin{tabular}{l r r r r r c}
\hline
\hline
Galaxy    &   $S_f$     & $S_s$  & $S_f/S_s$ & $\log{M_{\matHI,f}/M_{\odot}}$  & $\log{M_{\matHI,s}/M_{\odot}}$& $\log{\Delta N_{\matHI}/{\rm cm}^{-2}}$\\
& (Jy \kms) & (Jy \kms) &  & &&\\
(1) &  (2) &  (3) &  (4) & (5) & (6) & (7)\\ 
\hline
J1321--31   &   6.45    & 6.23   & 1.036 & 7.615&       7.599 & 17.65\\
J1337--28   &   21.18   & 21.11  & 1.004 & 8.024&       8.023 & 17.21\\ 
J1106--14   &   12.28   & 12.10  & 1.014 &  8.748&      8.742 & 18.35\\
J1253--12   &   26.73   & 26.29  & 1.017 & 8.883&       8.875 & 17.98\\
J0309--41   &   27.61   & 27.52  & 1.003 & 8.889&       8.887 & 17.65\\
J1303--17b  &   114.23  & 110.47 & 1.034 &  8.987&      8.972 & 18.60\\
J1318--21   &   167.21  & 161.56 & 1.035 &  9.028&      9.013 & 18.29\\
J1254--10a  &   57.08   & 54.61  & 1.045 & 9.236&       9.216 & 18.79\\
J1153--28   &   71.96   & 71.22  & 1.010&  9.841&       9.836 & 18.56\\
J0419--54   &   164.90  & 163.98 & 1.006 &10.086&       10.084 &18.03\\
 \hline
\end{tabular}
\tablefoot{(1) Galaxy name. (2) Flux density $S_f$ from full-depth
  data. (3) Flux density $S_s$ from single-track data. (4) Ratio of full-depth and single-track flux $S_f/S_s$. (5) \HI mass from
  full-depth data. (6) \HI mass from single-track data. (7) Column density of the
  additional \HI in the
  full-depth data calculated as ratio of additional \HI and \HI disk area as derived from the zeroth-moment maps.}
\label{tab:deepshallowcomp}
\end{table*}

\subsection{Comparison with single-dish profiles\label{sec:singledish}}

Above, we show that the longer integration times of the full-depth
data allow us to detect a small amount of additional \HI. However,
this internal comparison does not address the ``missing spacing''
issue. A further question is therefore whether MeerKAT is able to give
us an accurate measurement of the total \HI mass of our galaxies.

As the MHONGOOSE sample is ultimately selected from HIPASS,  
single-dish Parkes observations are available with a beam size of $15'$ and flux
measurements for
all but one of the sample galaxies are given in the HIPASS Bright Galaxy Catalog (
BGC; \citealt{Koribalski.2004}). The one galaxy that is not in the BGC (J0429--27) falls 
below the  flux limit of that catalog. Instead we use the flux density value as listed in the 
\citet{Meyer.2004} HIPASS Catalogue.
In addition, 15 of
our galaxies have been observed with the GBT with a beam size of $9'$
\citep{Sorgho.2019, Sardone.2021}.

We can therefore test
whether the MHONGOOSE observations manage to recover the total flux.
As the differences between the flux densities derived from the
single-track and full-depth data are only small (as discussed above),
we use the single-track flux densities here, as this maximizes the
number of galaxies where a comparison can be made. 

The single-track flux densities are listed in
Tab.~\ref{tab:shallowsinglecomp}, with the velocity widths given in
Tab.~\ref{tab:sample}.  These are based on the {\tt r15\_t00} data
using masks as described in Sect.~\ref{sec:single} and
\ref{sec:fluxden}. The 20\% velocity width $W_{20}$ is defined as
  the difference between the velocities where the profile reaches 20\%
  of its peak value. The central velocity is defined to be halfway
between these velocities.

\begin{figure*}
  \centering
  \resizebox{0.85\hsize}{!}{\includegraphics{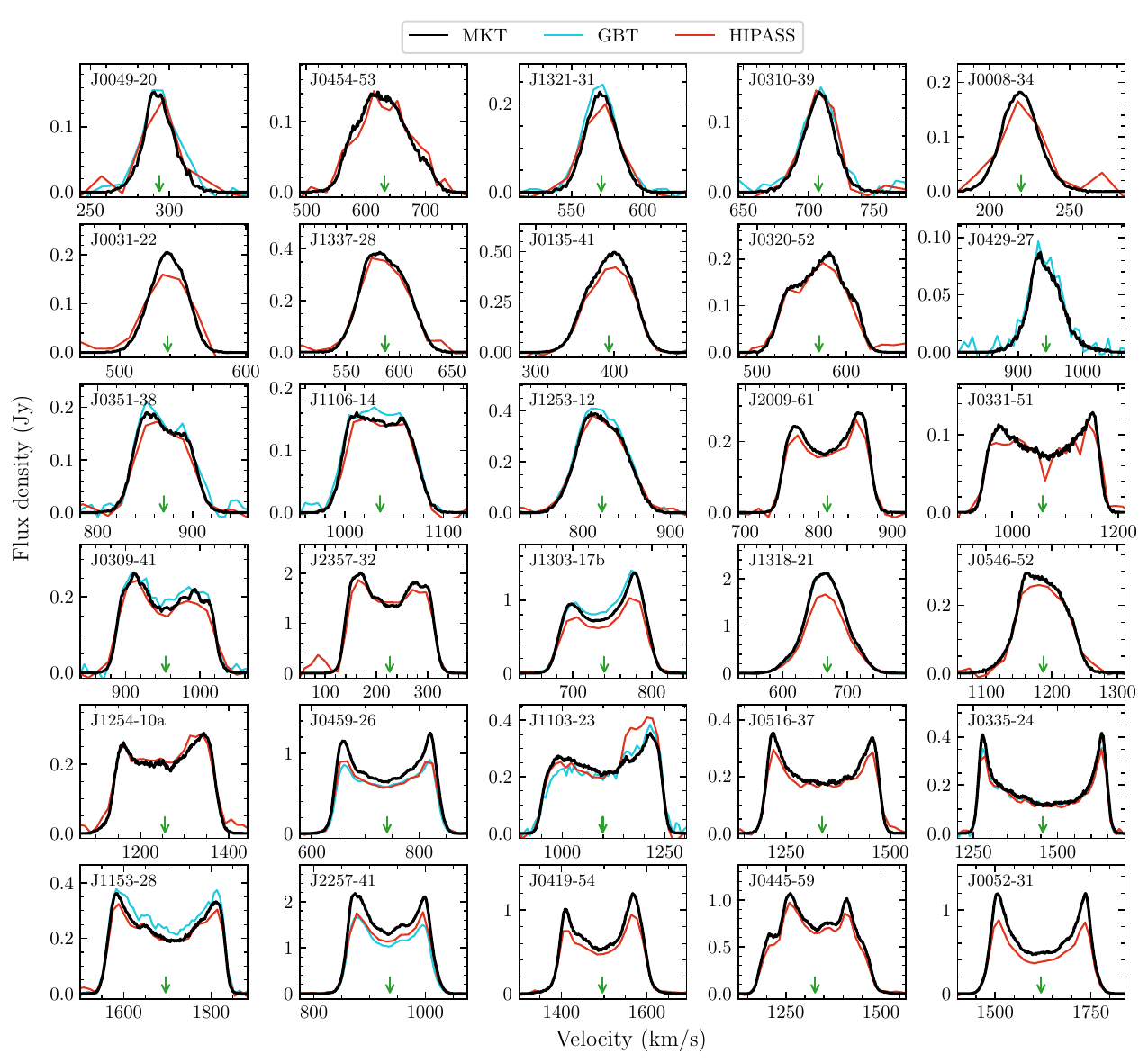}}
  \caption{Comparison of the MeerKAT and single-dish profiles. MeerKAT {\tt r15\_t00} profiles are shown in black, GBT profiles \citep{Sardone.2021} in cyan, and HIPASS profiles \citep{Koribalski.2004} in red. The green arrows indicate the central velocties as derived from the MeerKAT data.}
  \label{fig:allprofcomp}
\end{figure*}

We additionally list the HIPASS and GBT flux densities in Tab.~\ref{tab:shallowsinglecomp} for those galaxies where at least one
single-dish measurement is available.  We have remeasured
the GBT flux densities from the original profiles, as the values
listed in \citet{Sardone.2021} represent cumulative flux densities
integrated over the entire respective areas mapped by the GBT, rather than direct
integration of the global profiles as discussed here. 
 The \HI profiles themselves are shown and compared in Fig.~\ref{fig:allprofcomp}. Where necessary we have shifted the 
literature spectra to the same radio velocity scale as the MeerKAT profiles. Any ``peaks'' detected in the single-dish data outside the velocity range of the target galaxies (e.g., the HIPASS profile of J0008--34 at $v \sim 270$ \kms in Fig.~\ref{fig:allprofcomp}) are due to noise.

\begin{figure}
  \centering
  \resizebox{0.9\hsize}{!}{\includegraphics{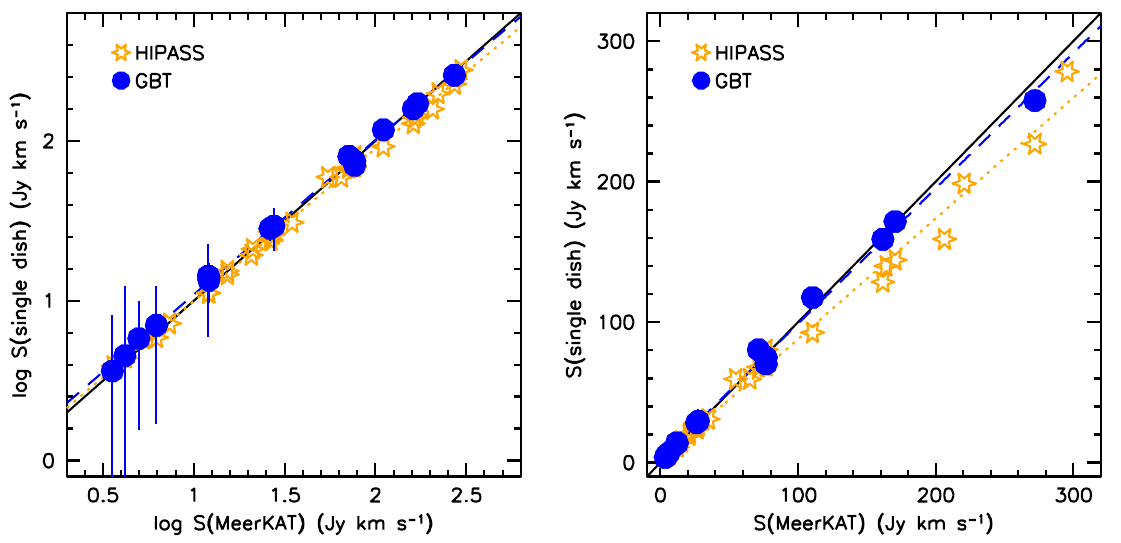}}
  \caption{Comparison of the single-track MeerKAT flux densities with
    single-dish flux densities. Stars (orange) indicate HIPASS values,
    filled circles (blue) GBT values. The full black line indicates a
    unity slope.  The left panel compares the flux densities on a
    logarithmic scale, the right panel shows the same data points on a
    linear scale.  The dashed blue line is a fit to the GBT
    points. The fit to the logarithmic GBT values has a slope of
    0.968, the linear values give a fit with a slope of 0.954. The
    dotted orange line is a fit to the HIPASS data. Here the
    logarithmic slope is 0.964, the linear fit gives a slope of
    0.860.}
  \label{fig:singledishcomp}
\end{figure}

In Fig.~\ref{fig:singledishcomp} we compare the MHONGOOSE, HIPASS and
GBT flux densities. The HIPASS values are systematically lower than
the MeerKAT and GBT values.  One explanation could be that  some of the \HI has 
been subtracted during data processing by the running bandpass correction method
\citep{Barnes.2001}.  The flux densities of \HI-bright galaxies are
also known to be affected by the gridding used in the HIPASS pipeline
\citep{Barnes.2001, Koribalski.2004}.

The agreement with the remeasured GBT flux densities is good. The
ratio of the GBT and MeerKAT flux densities is $1.058 \pm 0.080$, so
the GBT on average detects $\sim 6\%$  more flux, but this is
within the scatter of $\sim 8\%$. A least-squares fit to the linear values, as shown
in Fig.\ref{fig:singledishcomp}, has a slope of 0.964, but this slope
is mainly determined by NGC 7424 (J2257--41), the galaxy with the
largest flux density, where we find that the GBT actually detects less
\HI than MeerKAT. Leaving that point out we end up with a slope of
0.997, again indicating good agreement between the MeerKAT and GBT fluxes. This
is especially so when one considers the uncertainty in the MeerKAT
flux calibration of $\sim 10\%$ \citep{Serra.2023} and a most likely
similar uncertainty in the GBT flux calibration.
The good agreement is consistent with the fact that most sample galaxies are significantly less extended than the $\sim 20'$  maximum angular scale that the MeerKAT shortest baseline of 29m is sensitive to.

\begin{table}
\small
\centering
\caption{MeerKAT and single-dish \HI mass comparison}
\begin{tabular}{l r r r}
\hline
\hline
Galaxy    &   $S$ (MKT) & $S$ (HIPASS) & $S$ (GBT)  \\
& (Jy \kms) & (Jy \kms) &   (Jy \kms) \\
(1) & (2) & (3) & (4) \\
\hline
J0008--34       &5.3    &5.8    &-\\
J0031--22       &7.3    &7.2    &-\\
J0049--20       &3.6    &3.9    &3.6\\
J0052--31       &206.3  &159.1  &-\\
J0135--41       &34.7   &30.9   &-\\
J0309--41       &27.5   &26.1   &29.4\\
J0310--39       &4.2    &4.4    &4.5\\
J0320--52       &15.3   &14.6   &-\\
J0331--51       &20.7   &19.4   &-\\
J0335--24       &76.6   &68.8   &70.2\\
J0351--38       &12.0   &11.2   &13.47\\
J0419--54       &164.0  &140.0  &-\\
J0429--27       &5.0    &5.9        &5.8\\
J0445--59       &220.7  &198.8  &-\\
J0454--53       &15.3   &15.4   &-\\
J0459--26       &170.4  &144.3  &171.7\\
J0516--37       &64.9   &59.6   &-\\
J0546--52       &26.3   &24.3   &-\\
J1103--23       &76.6   &80.4   &75.0\\
J1106--14       &12.1   &11.2   &14.2\\
J1153--28       &71.2   &68.1   &80.3\\
J1253--12       &26.3   &27.2   &28.3\\
J1254--10a      &54.6   &59.4   &-\\
J1303--17b      &110.5  &92.6   &117.5\\
J1318--21       &161.6  &128.6  &159.0\\
J1321--31       &6.2    &5.9    &7.0\\
J1337--28       &21.1   &21.1   &-\\
J2009--61       &26.6   &23.9   &-\\
J2257--41       &271.9  &227.0  &257.8\\
J2357--32       &295.5  &278.5  &-\\
 \hline
\end{tabular}
\label{tab:shallowsinglecomp}
\tablefoot{(1) Galaxy identification. (2) Single-track MHONGOOSE flux density. (3) HIPASS flux density from \citet{Koribalski.2004}, except for J0429--27 where the flux density is from \citet{Meyer.2004}. (4) GBT flux density rederived from \citet{Sardone.2021}.}

\end{table}

\section{Comparing the moment maps}

\subsection{Pixel-pixel density histograms of individual galaxies}

The moment maps presented in Fig.~\ref{fig:sample1}
give an overview of the morphology and kinematics of the \HI
in the MHONGOOSE galaxies.  In terms of column density sensitivity these
single-track observations approach the sensitivity of the HALOGAS
survey to within a factor of two (cf.\ Fig.~\ref{fig:surveys}, shifting the MHONGOOSE line up by 0.5 dex).

An efficient method to further summarize this information is
pixel-by-pixel scatter plots of the various moment values: for each location
(pixel) the value of one moment is plotted against that of a different
moment at that same location.  A number of previous studies have used
these  plots to study properties of galaxies. For example,
\citet{deBlok.2018} used the distribution of the zeroth- and
second-moment values of the \HI distribution in the M81 triplet to
find that each of the three triplet galaxies occupied unique positions
in that particular parameter space, and addressed the question of
whether the M81 triplet could be the equivalent of a Lyman-$\alpha$
absorber (cf.\ their Fig.\ 20). \citet{Maccagni.2021} presented optical line emission data
of Fornax A and used the distribution of zeroth- and first-moment
values to distinguish between rotating and outflowing gas. \citet{Sun.2018, Sun.2020} 
studied the zeroth and second moments of the CO distribution in the Physics at High Angular resolution 
in Nearby GalaxieS (PHANGS) galaxies. 

The single-track MHONGOOSE observations give us the opportunity to do
a similar analysis for 30 galaxies in \HI, with all of them observed and reduced in an
identical, homogeneous manner. Here we explore the relation between
column density and velocity width (zeroth and second moment).

We use the {\tt r15\_t00} data of all of our galaxies and plot the
zeroth- and second-moment map pixels of each of our galaxies against
each other.  The $\sim 30''$ resolution data gives a good column density sensitivity
while not greatly sacrificing angular resolution
(cf.\ Fig.~\ref{fig:noisevalues}). We realise that 
the individual pixels are not independent of each other. However, here we only consider 
the relative distributions at a single resolution. Taking the size of the beam into 
account would only change the absolute scaling.

Rather than using scatter plots where subtle differences are often
difficult to make out, we derive a two-dimensional density histogram
covering the column density-velocity width parameter space with a $75
\times 75$  grid.

The zeroth-moment values are corrected for inclination and converted
to surface densities by multiplying them by $\cos i$, where we assume
the inclinations given in Table \ref{tab:sample}. We found that this
simple correction was sufficient for our purposes, with conclusions
insensitive to the exact inclinations values. 

\begin{figure*}[h]
  \centering
  \resizebox{0.85\hsize}{!}{\includegraphics{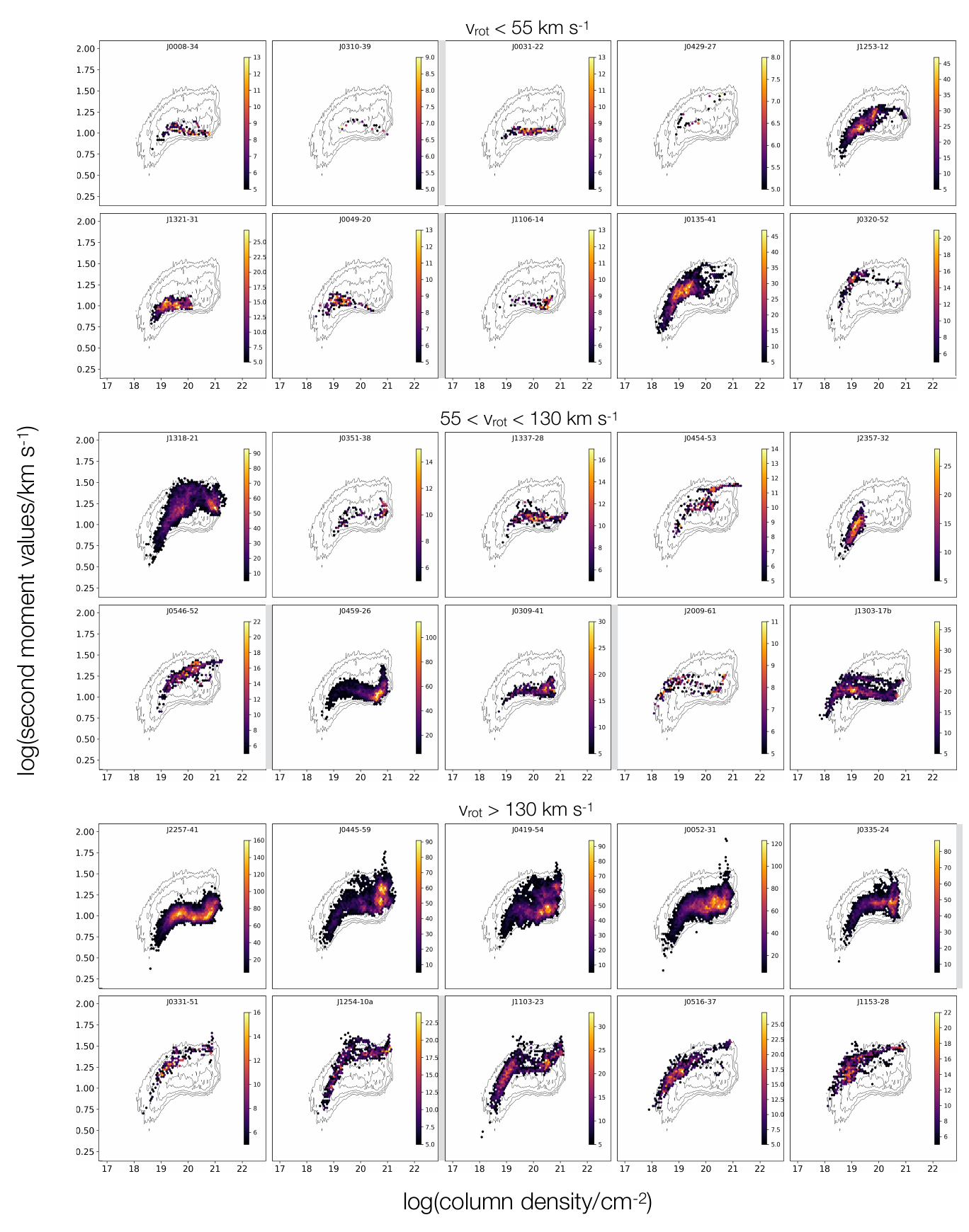}}
  \caption{Two-dimensional density histograms showing the distribution
    of inclination-corrected column densities  and second moment values for the MHONGOOSE
    galaxies. Each histogram consists of $75 \times 75$ hexagonal
    bins, and only bins containing more than 4 pixels are shown. The
    color bar on the right in  each panel indicates the number of pixels per
    bin. The background contours indicate the total distribution of
    the entire sample as shown in Fig.~\ref{fig:m0m2total}. Galaxies are divided in three bins in rotation velocity. Within each bin, galaxies are ordered by increasing inclination, with the most face-on galaxies in the top-left, and the most edge-on in the bottom-right. The two light-gray bars between the panels in each bin indicate inclinations of 50\degree\ $(b/a=0.67)$ and 70\degree\ $(b/a=0.33)$, respectively.}
  \label{fig:m0m2panel}
\end{figure*}

The panels in Fig.~\ref{fig:m0m2panel} show, for each galaxy, the 
number of pixels having values corresponding to any given combination of column density 
(horizontal axis) and velocity spread (vertical axis). This is shown as 
a two-dimensional density-histogram  with each bin having the value given 
by the color scale for that panel. Bins are only shown if
there are a minimum of four moment pixels falling in them. 
The contours show a similar density histogram but in this case derived from all the 
single-track moment maps of all  galaxies combined.  The juxtaposition of the two 
distributions allows for a  quick comparison of the distribution for  a 
particular galaxy with that of the entire sample.
This total distribution is further discussed in Sect.\ \ref{sec:totalpix}.

To highlight any differences between galaxies of different masses, we divide the galaxies in three equal-number bins 
in rotation velocity, as shown in Fig.~\ref{fig:m0m2panel}. The rotation velocities are calculated from the $W_{20}$ 
and $i$ values given in Tab.~\ref{tab:sample}. Within each rotation velocity bin we show the galaxies in order of 
increasing inclination. 

We see a large spread in the typical second-moment values for each
galaxy. Compare for example the high values found in J1318-21 with the
low ones in J0351-38. While our galaxies have a range in distance,
this is not the cause of the spread in these typical values (which
would be due to the beam measuring a larger number of kpc for more
distant galaxies, and thus potentially capturing a larger range in
velocities). The main determining factor seems to be the SFR. We
discuss this further below.

In the highest-rotation velocity bin we can make out a number of interesting trends. The lowest inclination galaxies in 
general cover most of the parameter space as traced by the entire sample. The only exception is J2257--41 which 
has low second-moment values. Moving to the high-inclination galaxies we see that they exclusively cover the 
upper part of the diagram. This is not a beam-smearing effect, but is rather caused by the longer line-of-sight through 
the disk, resulting in increased linewidths. We refer to Appendix B for a more in-depth discussion 
of the impact of beam smearing and inclination. 

The face-on massive galaxies show significantly increased second-moment values at the highest column densities. While some of 
this will be due to beam smearing in combination with a steeply rising inner rotation curve, part of it will be intrinsic to the galaxies.
Most of these high values occur in the central parts of
the galaxies and are likely related to the intense star formation in these portions of the galaxy. They mark the presence of  winds or outflows powered by star formation or an AGN.
The most massive galaxies in the sample, NGC 1566 (J0419--54), NGC 1672 (J0445--59), 
NGC 1371 (J0335--24) and NGC 289 
(J0052--31), are known Seyfert galaxies. 

The lowest-rotation velocity galaxies do not show these extreme second-moment values, but rather cling to the bottom of 
the distribution for the lower inclinations, though we do see larger second-moment values towards the more edge-on galaxies. 

The situation for the intermediate rotation velocity galaxies is complex. We do not see a clear trend of the distribution 
changing with inclination. An extreme example is NGC 5068 (J1318--21) which shows the highest second-moment values in this bin, 
yet is the most face-on galaxy. The complex situation in this galaxy is described in more detail in \citet{Healy.2024}. 
Another complex distribution can be seen for NGC 1705 (J0454--53). This is a starbursting dwarf galaxy, and the 
parallel track of high second-moment values are indications of the gas outflows in this galaxy (cf.\ \citealt{Meurer.1998}). 
A few more of these intermediate-rotation velocity galaxies show multiple velocity
width values at a given surface density. In many cases the high
second-moment values are linked to locations where there are multiple
components along the line of sight due to, e.g., spiral arms, the
presence of extra-planar gas, or long lines of sight through the
disk.



An in-depth analysis of these density histograms, with the addition of
information on the distribution of  molecular gas, such as presented in \citet{Sun.2018, Sun.2020},
will give much information on the conditions for star
formation in nearby galaxies and the link with the properties of the
atomic and molecular gas. Such a quantitative analysis is, however, beyond
the scope of this paper.

\subsection{Total pixel-by-pixel density histogram\label{sec:totalpix}}

\begin{figure}
  \resizebox{0.99\hsize}{!}{\includegraphics{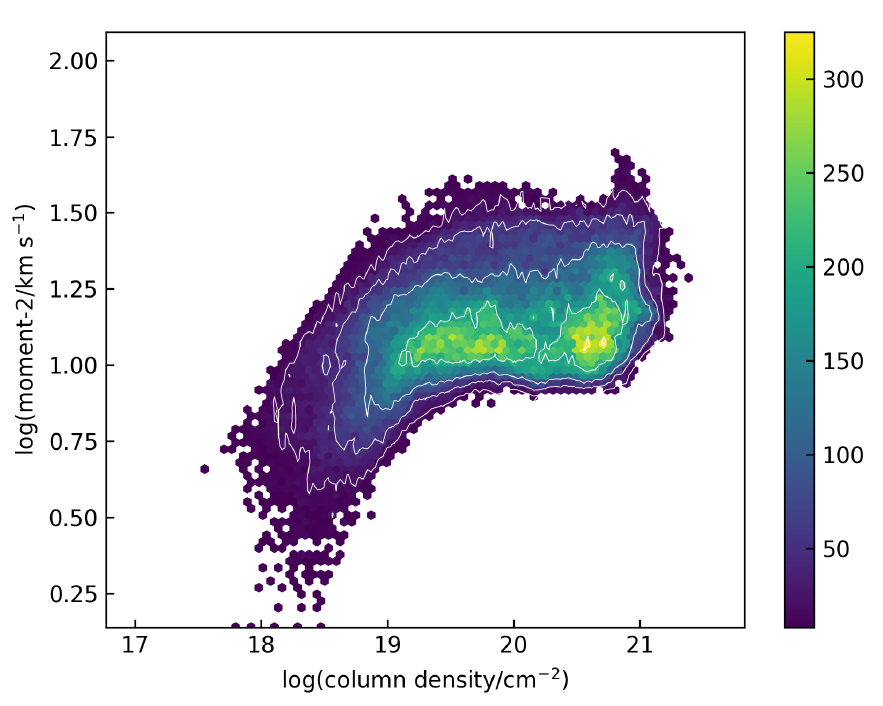}}
  \caption{Two-dimensional density histograms showing the
    distribution of zeroth and second moment values for the entire
    MHONGOOSE sample.  The histogram consists of $75 \times 75$
    hexagonal bins, and only bins containing more than 8 pixels are
    shown. The color bar indicates the number of
    pixels with the \HI column density and velocity widths indicated by the horizontal and vertical axes, respectively.}
  \label{fig:m0m2total}
\end{figure}

We now consider the surface density-velocity width density histogram
of the entire sample.  Here we again use the same $75 \times
75$ grid to sample the displayed parameter range, but only
display bins containing more than eight sample values. This integrated
histogram is shown in Fig.~\ref{fig:m0m2total} and contains several
interesting features. Some of these are known from the literature, but
others have not been explicitly noted before.

\begin{figure}
  \resizebox{0.99\hsize}{!}{\includegraphics{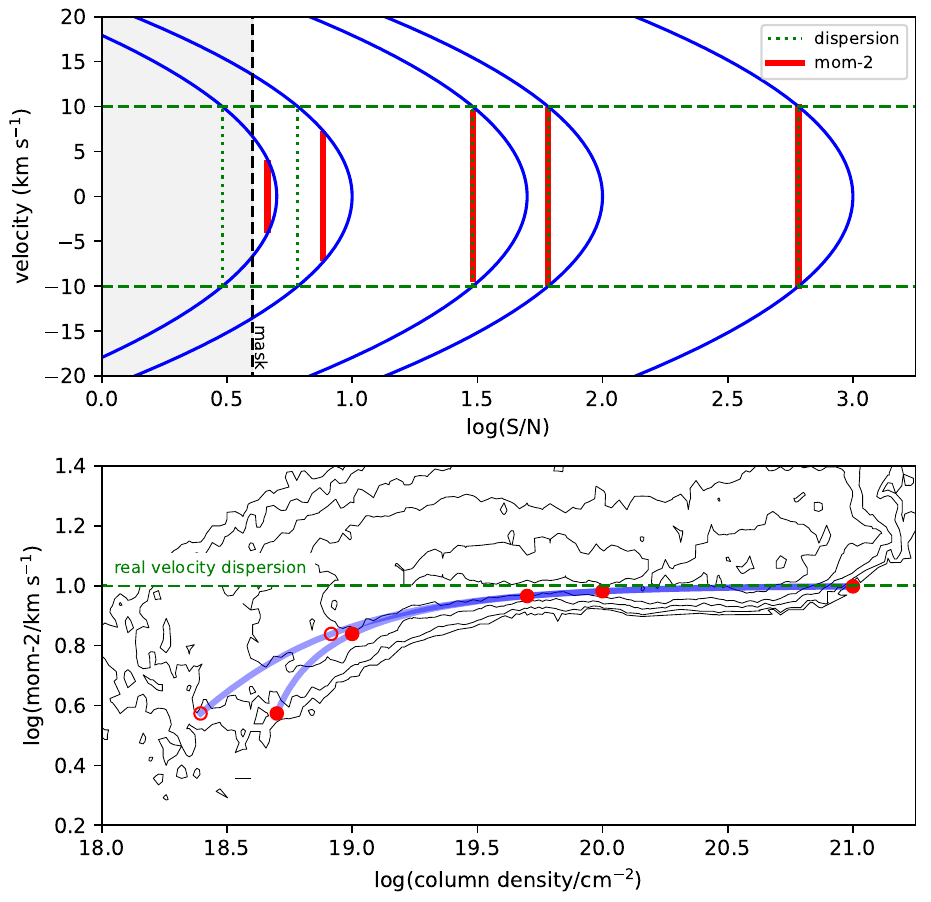}}
  \caption{Illustration of the change in zeroth- and second-moment values due to masking. 
  The top panel shows a number of Gaussians, here plotted as a function of velocity as given by the vertical axis, while the horizontal axis indicates the peak value 
  of the Gaussian in terms of S/N. From right to left the peak values are 1000, 100, 50, 10 and 
  5 times the S/N. The dashed vertical black line indicates the 4 times S/N level where masking occurs. 
  Each profile has a velocity dispersion of 10 \kms as indicated by the horizontal dashed green 
  lines, and, for each Gaussian separately, the dotted vertical green lines. The thick red lines 
  indicate the measured second-moment values for each Gaussian, assuming they are masked at 4 times 
  the S/N. The bottom panel shows the change in
  zeroth and second moment values. Here we have lined up the positions of the peaks 
  in the top panel with those of the points in the bottom panel by assuming that a 
  column density of $10^{21}$ cm$^{-2}$ corresponds to a S/N of 1000. This is 
  consistent with the data presented in Fig.\ \ref{fig:m0m2total} that are shown here as contours.
  The curve extending to the left (open circles) takes into account that the total 
  flux of the Gaussian decreases due to
    masking, the right-hand curve (closed circles) ignores this. For other assumed velocity dispersions, the shape of the curve does not change but will shift up (larger dispersions) or down (lower dispersions).}
  \label{fig:gauss}
\end{figure}

At the highest column densities we see a steep drop-off in the density
distribution, despite this being the part of the diagram with the
highest S/N. This drop-off occurs at surface density values
of $\log(N_{\matHI}/{\rm cm}^{-2}) \sim 21.1$, or $\sim 10$ \msunpc, and thus implies
an upper limit to the \HI column density (at this resolution).  This value agrees well with
the upper limit of $\approx 9$ \msunpc found by \citet{Bigiel.2008}
for the THINGS sample at similar resolutions (also see \citealt{Leroy.2008}; for a more detailed observational view, see \citealt{Schruba.2017}). A comparison with molecular
gas column density values shows that this upper limit corresponds with a
transition from an atomic to a molecular ISM (cf.\ \citealt{Krumholz.2009, McKee.2010, Sternberg.2014}) 

Turning to the second-moment values, we see that at surface densities
$\log(N_{\matHI}/{\rm cm}^{-2}) > 19.5$ there is a well-defined lower
limit to the second-moment values. This limit occurs at $\sim 8.5$
\kms. Given the 1.4 \kms velocity resolution of the data
(corresponding to the lower limit of the plotted second-moment range)
and the high S/N of the profiles in this
part of the diagram, this cannot be a velocity resolution artefact but
must indicate an absence of \HI profiles with second-moment values
less than $\sim 8.5$ \kms. To first order, the ISM is considered to 
consist of a cool neutral medium (CNM) with a temperature of a few
tens of K and a warm neutral medium (WNM) with temperatures around
$\sim 8000$K . The CNM line profiles have a line width of $\sim 1$ \kms
while WNM line width are $\sim 8$ \kms (in the absence of any other
processes that can broaden line-widths) \citep{Braun.1997,Wong.2002, Wolfire.2003}.  The lower limit found in our
data does not indicate an absence of CNM, but merely shows that the
\HI emission, and therefore the line width, is dominated by the WNM,
where the latter has a lower limit to its observed velocity dispersion as shown
in Fig.~\ref{fig:m0m2total}.

As is evident from Fig.~\ref{fig:m0m2panel}, many of the galaxies show
distinct local density concentrations in the histograms. We also see this in
the total density histogram, suggesting this is a general feature of
our sample. Figure~\ref{fig:m0m2total} shows a  local maximum in the combined distribution
centered on $\log(N_{\matHI}/{\rm cm}^{-2}) \sim 20.6$ and another (less
concentrated) one around $\log(N_{\matHI}/{\rm cm}^{-2}) \sim
19.5$. Comparison with the integrated \HI maps as well as SINGG
H$\alpha$ and GALEX UV images shows that the profiles corresponding to
the high-column density concentration are generally found in the star-forming disks of the galaxies.   The profiles contributing to the
low-column density concentration are generally found in the
outer disk or in inner disk regions with little to no detected star formation as seen 
in the H$\alpha$ and FUV images.  The high-density clump in Fig.~\ref{fig:m0m2total}
also shows a ``tail'' at high velocity dispersion, which is not visible in the low density clump, possibly induced by feedback from star formation.

Perhaps the most interesting is the clear division between these two
concentrations around $\log(N_{\matHI}/{\rm cm}^{-2}) \sim 20.3$
($N_{\matHI} \sim 2 \cdot 10^{20} {\rm cm}^{-2}$). 
If \HI complexes gradually collapse either under their own gravity or 
triggered by an external event, we would expect the observed \HI 
column density to continuously increase, and therefore expect to see 
a continuous \HI column density distribution in Fig.~\ref{fig:m0m2total}. 
If we assume that collapsing \HI complexes move from left to right 
in the Figure (i.e., from low- to high-column density), then the presence of the division points to a short timescale for this column density increase.

A possible explanation for the presence of this division is related to the formation of the CNM.  This was put in a physical context in
\citet{Kanekar.2011}. Based on measurements of the spin temperature in
a large number of Galactic \HI clouds they identify a rapid change in
spin temperature at $\sim 2 \cdot 10^{20}$ cm$^{-2}$. They associate
this with a threshold \HI surface density for formation of the
CNM. Above this $\sim 2 \cdot 10^{20}$ cm$^{-2}$ threshold, the \HI can
sufficiently self-shield against UV photons to enable formation of a
cool component. Below this threshold the UV photons can heat and
ionize the gas, preventing the formation of a CNM. Note that this threshold also
corresponds to that used to define a Damped Lyman-$\alpha$ system.

Following the interpretation by \citet{Kanekar.2011}, these data show
that the existence of a CNM, which is more closely associated with star
formation, leaves an imprint on the moment maps, including a
rapid transition to higher column densities once this component
starts forming.  Figure~\ref{fig:m0m2total} thus shows, in a single
plot, the various transitions occurring towards star formation.
At the lowest column densities, between $\sim 10^{19}$ and $\sim 10^{20}$ cm$^{-2}$, the gas is dominated by the WNM, and
there is little or no star formation (\citealt{Bigiel.2010, Serra.2012, Maccagni.2017}). At $\sim 2 \cdot 10^{20}$ cm$^{-2}$ the
gas is sufficiently dense that a significant CNM can rapidly
develop. The approximate column density limits of the high-column density concentration correspond closely to the values given in \citet{Bigiel.2010} ($\sim 3-10$~\msunpc or $\log(N_\matHI/{\rm cm}^{-2}) \sim 20.5-21.1$) where star formation is efficient 

We expect that the CNM
will still be embedded in a WNM (at the resolution of our
observations), where the WNM is the source of the observed
second-moment values. We therefore do not see the lower CNM velocity
dispersions reflected in the plot, but studies of the individual profiles should provide clear evidence for its presence.  
The gas turns completely molecular at the column
density cut-off discussed above.

It will be interesting to repeat this analysis once all full-depth data are available. The increase in S/N will allow exploration at different angular resolutions, but also, perhaps more crucially, at identical physical resolutions.

\subsection{Masking and moments\label{sec:mask}}

One of the distinct features in Figs.~\ref{fig:m0m2panel} and \ref{fig:m0m2total}
 is the tail of decreasing second-moment values
towards low surface densities. Naively, one might presume that this
could be a consequence of the low-column density gas being mostly in
the outer disk where there is less energy available to heat and stir
up the gas, leading to a low velocity dispersion. The true cause is
more prosaic, however, as we show here that it is caused by the
masking procedure used to create the moment maps.

The SoFiA-2 moment map creation steps described above are a sophisticated combination 
of source finding and masking at different spatial and velocity scales.  
The essentials of this can, fortunately, already be captured by considering
an idealized Gaussian \HI profile with constant velocity dispersion and variable 
normalisation or peak value  (corresponding to different column densities).

We consider a data cube with a noise per channel $\sigma_{\rm ch}$ in
which we embed Gaussian line profiles with a fixed velocity dispersion
and a peak value expressed as some multiple of $\sigma_{\rm ch}$.
Analogous to our SoFiA moment map construction, we assume that we mask
the data at the $4\sigma_{\rm ch}$ level. For a Gaussian with a peak
close to the noise level most of the profile will be masked, while a
high S/N profile will be much less affected.

We consider the second moment of the remaining $>4 \sigma_{\rm
  ch}$ profile \emph{after masking} as a function of the
S/N of the profile. Without masking (and for noise-free data), this would 
equal the
velocity dispersion of the Gaussian. With masking, it decreases as the
peak value decreases.

A second effect is that the area
underneath the masked Gaussian (the zeroth-moment value or integrated
flux) also decreases after masking due to removal of the high- and
low-velocity wings of the profile.  This also causes a relatively
larger decrease in column density as the peak flux density declines.

In Fig.~\ref{fig:gauss}, we plot the curve showing the calculated
changes in the measured zeroth- and second-moment values for a single
fixed-velocity dispersion Gaussian as a function of its peak value and
superimpose this on the surface density-velocity width density
histogram. The absolute vertical position of the curve is determined by the input
velocity dispersion of the Gaussian. The horizontal position of the curve is 
fixed by identifying the typical S/N for a given column density. In
Fig.~\ref{fig:gauss} we show the resulting changes for a Gaussian
masked at $4\sigma_{\rm ch}$ with a velocity dispersion of 10 \kms,
and a S/N of 1000 for a column density of $10^{21}$ cm$^{-2}$.  This
is approximately correct for the data shown here, as deduced from the
S/N maps discussed in Sect.~\ref{sec:full} and \ref{sec:single}.  We
show two models in Fig.~\ref{fig:gauss}, one where
we ignore the decrease in flux due to masking, and one where we do take
this into account. The latter extends to lower column densities.

For the assumptions made here, the shape of this relation is
fixed. Changing the input velocity dispersion merely shifts it up and
down, while increasing the S/N of the underlying data would shift it
horizontally. The tail towards lower column densities is therefore an unavoidable byproduct of the masking
procedure. Longer integrations (i.e., a decreasing noise)  would push the tail to lower column
densities. 
A similar analysis using full-depth cubes would result in the tail starting at column densities that are $\sqrt{10}$, or 0.5 dex lower in column density.

Note how well the down-turn in the tail is described by the simple
single Gaussian model. We can deduce for the masking procedure
described here that at S/N values between $\sim 50$ and $\sim 10$,
the second moment values start to underestimate the velocity
dispersion. At a ${\rm S/N} \sim 5$, the second moment underestimates
the velocity dispersion by a factor of $\sim 2.5$. Column densities,
though not as affected, can nevertheless be underestimated by factors of
$\sim 1.6$ at these lower S/N values.
Multiple components along the line of sight will not significantly influence the 
lower bounds of the distribution as, at a given column density, these will be found 
at higher second-moment values. Nevertheless, fitting multiple Gaussians to the 
velocity profiles could further elucidate (or better: circumvent) the effects of masking 
in the presence of multiple profiles.

In short, in our specific case of masking at $4\sigma$, moment values
below a S/N $\sim 20$ should be used with care, as these will introduce significant biases with respect to the true column density or velocity dispersion.
Different masking levels and methods produce different S/N level
limits, but the general conclusion remains and will always be relevant
when masking is used. As noted, deeper data cannot totally remove this
this issue, but will shift these effects towards lower column
densities.

\section{Other galaxies in the MHONGOOSE fields}

\begin{figure*}
  \resizebox{\hsize}{!}{\includegraphics{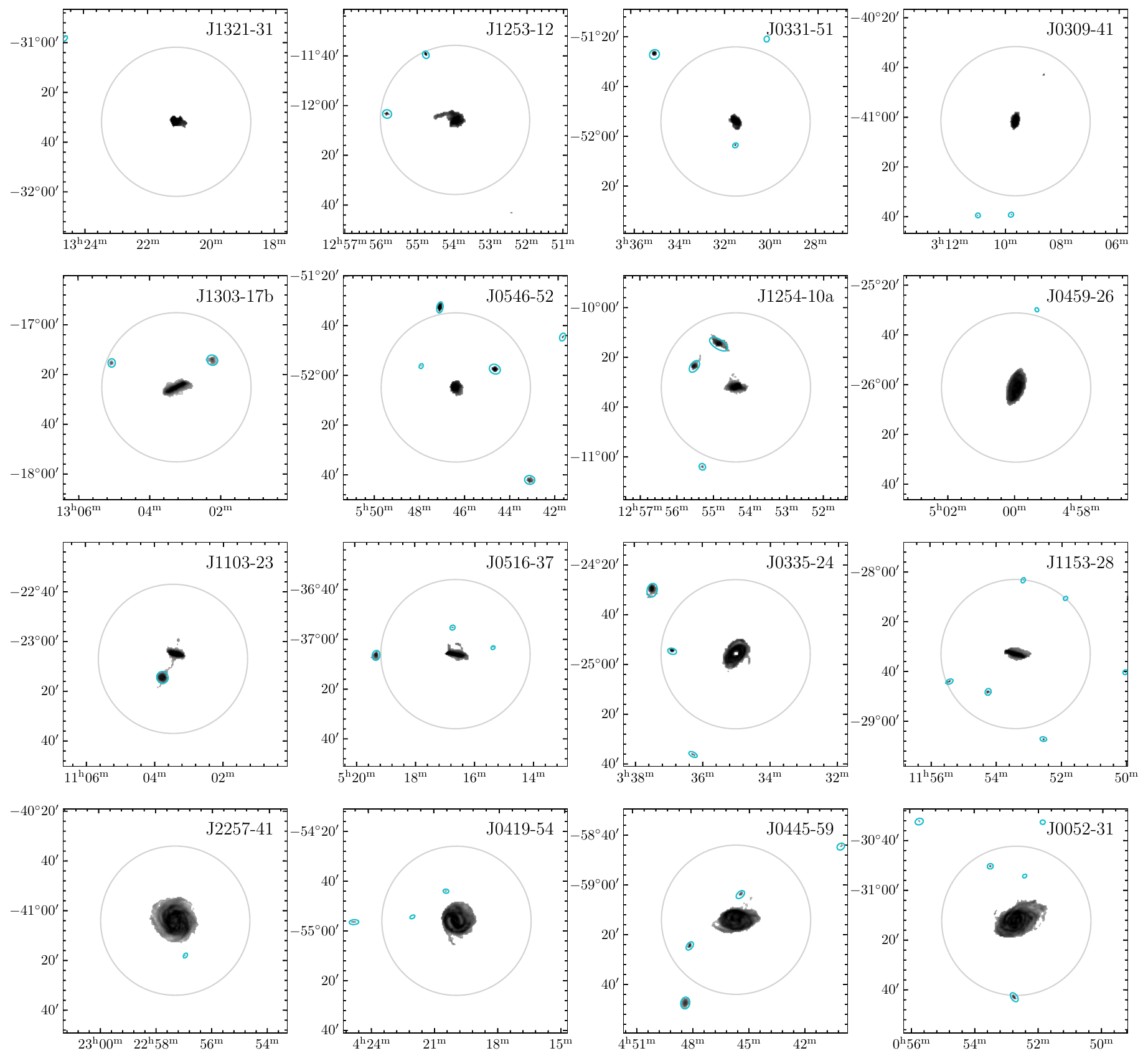}} \caption{Overview
    of the zeroth-moment maps of the fields where additional galaxies
    are present. The HIPASS name of the target galaxy (center) is
    identified in the top-right corner of the zeroth-moment maps. The
    additional galaxies are indicated by blue circles and their
    properties are listed in Table \ref{tab:companions}. A small
    number of galaxies are not indicated as they partially fall
    outside of the field of view shown here. The large dotted circles
    indicate a diameter of 1 degree, equivalent to the FWHM of the
    MeerKAT primary beam. The maps have been corrected for primary
    beam attenuation. }
  \label{fig:fields}
\end{figure*}

The sensitivity and large field of view of MeerKAT result in the
detection of additional companion and field galaxies. Here we give a
first inventory of the galaxies detected in the 1.5\degree\ by 1.5\degree\ 
single-track fields in a velocity range of $-500$ to $+500$ \kms
with respect to the central velocity of the target galaxies.  
Detections are based on the single-track zeroth-moment maps. 
A list of these additional galaxies is given in Table \ref{tab:companions}.
For the fields where additional galaxies are present, Fig.~\ref{fig:fields} shows the
positions of these galaxies. 
We expect an even larger number of galaxies to be detected in the full-depth data once available.
While the search volume is defined with respect to our target galaxies, this does not imply that every one of these additional galaxies is also physically associated with those target galaxies. Most of them will be, but it is possible that (especially at the edges of the volume probed) we are picking up some foreground or background galaxies.

The Table reports a variety of integrated \HI properties, as well as the identification of the optical 
counterparts. These identifications were made by  consulting the SIMBAD and NED data bases.
We find that half of the additional \HI detections can be definitively associated 
with an optical counterpart through a match both in position (generally to within one or two beams) 
and in velocity (generally to within a few tens of \kms). Most of the other
half can be reliably associated with an optical counterpart through a position match only, because no previously measured velocity is noted in either the NED or SIMBAD databases, and our redshifts are 
the first ones measured for these objects.
Finally, we find four galaxies (indicated with the MKT prefix in Table~\ref{tab:companions}) that have 
no previous identification, but where an optical counterpart is clearly visible in DECaLS images. 

Three additional galaxies visible in the moment maps are partially outside of the volume of the data cubes used here. These are LEDA 583148 in the 
J0309--41 field, ESO 486--21 in the J0459--26 field, and MKT J125225.4-124304.1 in the J1253--12 field. 
For these three galaxies we list the position and approximate radial velocity in 
Table~\ref{tab:companions}. 
We refrain from presenting their \HI line widths and fluxes because the large uncertainty due to the incomplete coverage by the data used here. 
The last of these three galaxies is a fifth, previously uncataloged galaxy.
All of the additional galaxies listed here can thus be associated with an optical counterpart; we have not found any ``dark'' galaxies.

We note that one of our 
lowest \HI mass galaxies, UGCA~15 (J0049--20), is itself a companion to the larger galaxy NGC 247. This is the only case where a ``companion'' galaxy is 
more massive than the target galaxy and we do not include NGC 247 in the following analysis.

\begin{figure*}
\centering
  \includegraphics[width=0.99\linewidth]{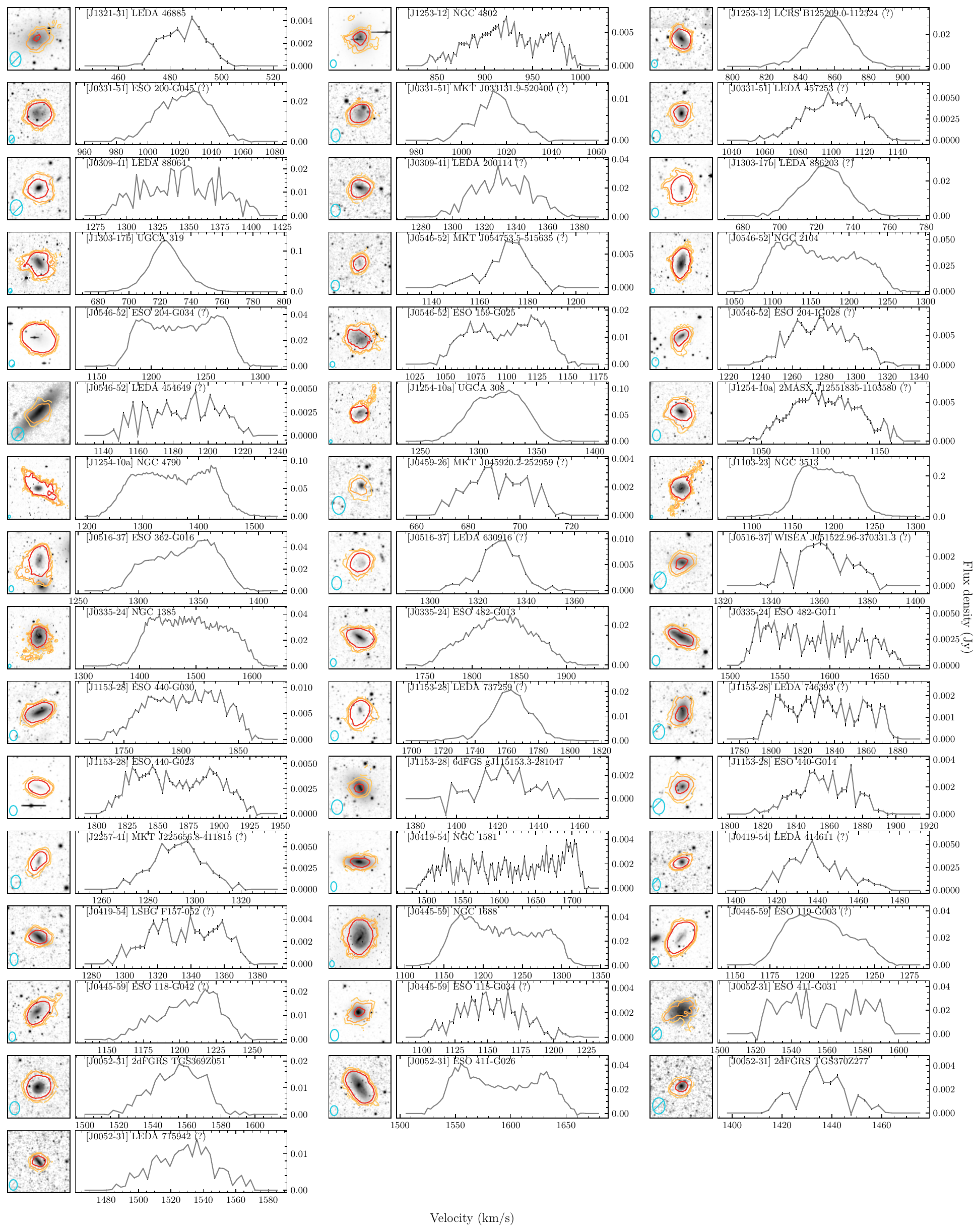}
  \caption{Grayscale cutouts from DECaLS DR10 $g$-band centred on the additional 
  \HI detections. Contours indicate \HI column densities at S/N = 3, 5, 10 (from orange to red). Spectra have been binned 
  by a factor two with a resulting  channel width of 2.8 \kms. These spectra were used 
  to derive the parameters listed in Table \ref{tab:companions}. All spectra are based on {\tt r15\_t00} data, 
  except for  J0546-52 LEDA 454649 which is based on {\tt r05\_t00} data.}
  \label{fig:dwarfspectra}
\end{figure*}

 Figure~\ref{fig:dwarfspectra} shows the integrated \HI 
maps overlaid on optical DECaLS images along with the global \HI profiles (corrected for the primary beam) of the additional galaxies.

\begin{figure*}
  \centering
  \sidecaption
  \resizebox{12cm}{!}{\includegraphics{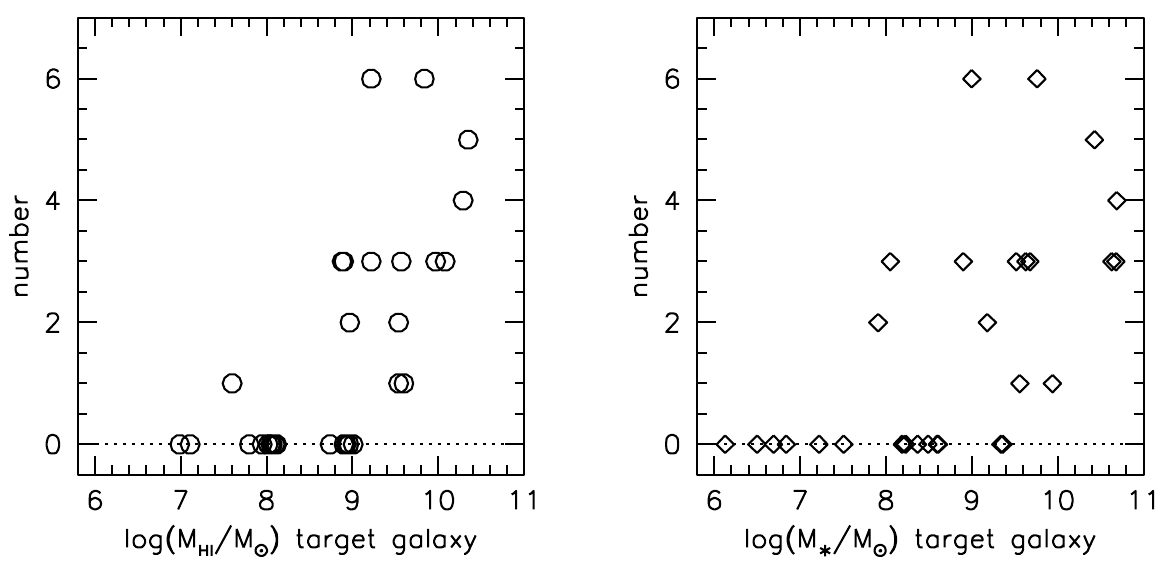}}
  \caption{Number of additional galaxies in a 1.5\degree\ $\times$ 1.5\degree\ area around the target galaxies, within a velocity range
    of $-500$ to $+500$ \kms with respect to the central velocity of
    the target galaxy. This number is plotted against the \HI mass of the target galaxy in the left panel and against the stellar mass in the right panel. Note the sudden increase at $M_{\matHI}  \sim 10^9$ \msun\ (left) and $M_\star \sim 10^8$ \msun\ (right).}
  \label{fig:dwarfnumbers}
\end{figure*}

The number of additional galaxies detected varies significantly
from field to field.  Some target galaxies have no additional galaxy in
the field, while others boast five or six companions.
We can get a first indication of the variety of the environments
around the target galaxies by plotting the number of additional
galaxies per field as a function of the \HI mass and the stellar mass of the target
galaxy as shown in Fig.~\ref{fig:dwarfnumbers}. It is striking that we see only 
one additional galaxy until just below $M_{\matHI} \sim 10^9$~\msun. Above that mass the number of additional galaxies suddenly
increases. A similar steep increase can be seen at a stellar mass $M_\star \sim 10^8$ \msun. 

We can compare this with a similar observational result found by
\citet{Zhu.2023}. In a census of gaseous satellites around spiral
galaxies, they identify companion galaxies around local spiral
galaxies identified in the ALFALFA survey, and find that the number starts to
increase at a halo mass of $\log(M_{\rm halo}/M_{\odot}) \sim
11.5$. Using their Table 1 for a first-order conversion of their halo
masses to \HI masses, we find that this sudden increase occurs at an
\HI mass just below $10^9$ \msun, strikingly close to where we find an
increase in our sample.

Applying a more restrictive definition of a companion galaxy, analogous to
that used in \citet{Zhu.2023}, does not substantially change our
results.  For example, imposing a cut of $\pm 300$ \kms with respect
to the systemic velocity, and a cut in projected separation to the companion of $200$ kpc (the
approximate value of the virial radius $R_{200}$ for the upper mass
range of the sample) does not change the observed sudden increase.

It is interesting to note that
that many of these gas-rich companion galaxies are found projected within the virial radius, contrary to what we see in the Local Group \citep{Spekkens.2014}, but consistent with recent SAGA results \citep{Mao.2021} with many star forming galaxies projected within the virial radii.

A more complete census of the companion galaxy population will be
possible when the full-depth data for the full sample are available and
we can probe the companion population to the fullest extent possible for our survey. As with the galaxies presented here, these satellites and companions will be resolved spatially and spectrally, meaning that in-depth studies of the dark matter content and baryon fractions can be made down to an \HI mass
limit of $\sim 10^6$ \msun (cf.~Table \ref{tab:resolutions}), a regime previously only accessible in the Local Group. A first discovery of such a low-mass galaxy in the MHONGOOSE observations is presented in \citet{Maccagni.2024}.

\begin{table*}
\small
\centering
\caption{Additional galaxies in the field}
\begin{tabular}{l l r r r r r r}
\hline
\hline
Field     &     Galaxy  & $\alpha$ (J2000.0) & $\delta$ (J2000.0) & $V_{\rm central}$       & $W_{20}$& $W_{50}$ & $S_{\matHI}$ \\
 & & & & (\kms) & (\kms) & (\kms) &   (Jy \kms) \\
\hline
\noalign{\vskip 2pt}
J1321-31 & \textbf{LEDA 46885} & 13 24 35.7  & -30 58 18.3  & 484.0 & 25.7 & 18.4 & 0.075 \\
J1253-12 & \textbf{NGC 4802} & 12 55 49.7  & -12 03 25.3    & 923.8 & 143.4 & 94.7 & 0.531 \\
J1253-12 & LCRS B125209.0-112324 & 12 54 45.9  & -11 39 39.6  &857.7 & 38.6 & 22.1 & 0.892 \\
J0331-51 & ESO 200-G045 & 03 35 00.8  & -51 27 15.0 & 1023.4 & 52.4 & 38.6 & 1.001 \\
J0331-51 & {\it MKT J033131.9-520400} & 03 31 31.9  & -52 04 00.0 & 1014.1 & 28.5 & 11.9 & 0.225 \\
J0331-51 & LEDA 457253 & 03 30 11.3  & -51 21 14.4 & 1102.3 & 56.0 & 37.7 & 0.212 \\
J0309-41 & \textbf{LEDA 88064} & 03 10 59.2  & -41 39 40.7 & 1343.0 & 107.5 & 66.2 & 1.227 \\
J0309-41 & LEDA 200114 & 03 09 48.0  & -41 39 25.5 & 1330.3 & 59.7 & 38.6 & 1.225 \\
J1303-17b & LEDA 886203 & 13 05 03.9  & -17 15 24.1 & 723.9 & 44.1 & 27.8 & 0.872 \\
J1303-17b & \textbf{UGCA 319} & 13 02 14.5  & -17 14 17.8 & 724.3 & 34.0 & 22.1 & 3.300 \\
J0546-52 & {\it MKT J054753.5-515635} & 05 47 53.5  & -51 56 35.9 & 1171.6 & 29.4 & 13.8 & 0.130 \\
J0546-52 & \textbf{NGC 2104} & 05 47 04.4  & -51 33 02.4 & 1162.6 & 167.2 & 146.1 & 5.868 \\
J0546-52 & ESO 204-G034 & 05 44 41.5  & -51 57 48.7 & 1227.4 & 104.8 & 93.7 & 3.294 \\
J0546-52 & \textbf{ESO 159-G025} & 05 43 07.2  & -52 42 13.9 & 1097.5 & 94.7 & 76.3 & 1.156 \\
J0546-52 & ESO 204-IG028 & 05 41 45.9  & -51 44 35.8 & 1277.0 & 63.4 & 49.6 & 0.350 \\
J0546-52 & LEDA 454649 & 05 47 30.7  & -51 35 45.7 & 1181.9 & 64.3 & 56.1 & 0.171 \\
J1254-10a & \textbf{UGCA 308} & 12 55 31.1  & -10 23 39.8 & 1317.4 & 74.4 & 55.1 & 5.590 \\
J1254-10a & 2MASX J12551835-1103580 & 12 55 18.2  & -11 04 02.3 & 1104.8 & 108.4 & 68.0 & 0.430 \\
J1254-10a & \textbf{NGC 4790} & 12 54 51.5  & -10 14 43.4 & 1353.6 & 216.9 & 182.0 & 14.917 \\
J0459-26 & {\it MKT J045920.2-252959} & 04 59 20.2  & -25 29 59.8 & 692.1 & 40.4 & 31.2 & 0.087 \\
J1103-23 & \textbf{NGC 3513} & 11 03 46.1  & -23 14 31.4 & 1190.9 & 98.3 & 80.0 & 20.664 \\
J0516-37 & \textbf{ESO 362-G016} & 05 19 18.8  & -37 06 32.1 & 1334.6 & 91.0 & 75.4 & 3.029 \\
J0516-37 & LEDA 630916 & 05 16 44.7  & -36 55 28.9 & 1329.4 & 24.8 & 13.8 & 0.184 \\
J0516-37 & WISEA J051522.96-370331.3 & 05 15 23.1  & -37 03 31.1 & 1360.8 & 38.6 & 31.2 & 0.074 \\
J0335-24 & \textbf{NGC 1385} & 03 37 28.8  & -24 30 12.7 & 1492.4 & 214.1 & 186.5 & 6.488 \\
J0335-24 & \textbf{ESO 482-G013} & 03 36 53.8  & -24 54 46.2 & 1830.1 & 127.7 & 79.9 & 2.121 \\
J0335-24 & \textbf{ESO 482-G011} & 03 36 17.3  & -25 36 18.6 & 1578.5 & 148.9 & 130.5 & 0.379 \\
J1153-28 & \textbf{ESO 440-G030} & 11 55 25.7  & -28 44 10.1 & 1804.1 & 110.3 & 91.9 & 0.763 \\
J1153-28 & LEDA 737259 & 11 54 14.8  & -28 48 20.2 & 1763.8 & 38.6 & 24.8 & 0.609 \\
J1153-28 & LEDA 746393 & 11 53 10.5  & -28 03 28.6 & 1833.1 & 79.0 & 74.4 & 0.113 \\
J1153-28 & \textbf{ESO 440-G023} & 11 52 33.3  & -29 07 20.2 & 1865.8 & 110.3 & 85.5 & 0.351 \\
J1153-28 & \textbf{6dFGS gJ115153.3-281047} & 11 51 53.3  & -28 10 44.4 & 1424.1 & 52.4 & 21.1 & 0.063 \\
J1153-28 & \textbf{ESO 440-G014} & 11 50 03.4  & -28 40 17.9 & 1855.8 & 54.2 & 28.5 & 0.113 \\
J2257-41 & {\it MKT J225656.8-411815} & 22 56 56.8  & -41 18 15.9 & 1292.8 & 40.4 & 20.2 & 0.148 \\
J0419-54 & \textbf{NGC 1581} & 04 24 44.4  & -54 56 31.2 & 1632.8 & 216.9 & 190.2 & 0.387 \\
J0419-54 & LEDA 414611 & 04 22 00.9  & -54 54 41.5 & 1437.4 & 43.2 & 22.1 & 0.131 \\
J0419-54 & LSBG F157-052 & 04 20 26.8  & -54 44 25.1 & 1340.2 & 71.7 & 52.4 & 0.192 \\
J0445-59 & \textbf{NGC 1688} & 04 48 23.4  & -59 47 50.4 & 1225.5 & 160.8 & 143.4 & 4.775 \\
J0445-59 & ESO 119-G003 & 04 48 07.4  & -59 24 54.6 & 1211.4 & 77.2 & 63.4 & 2.155 \\
J0445-59 & ESO 118-G042 & 04 45 27.6  & -59 04 21.4 & 1207.9 & 71.7 & 34.9 & 0.922 \\
J0445-59 & ESO 118-G034 & 04 40 16.6  & -58 44 37.4 & 1153.1 & 85.5 & 51.4 & 0.171 \\
J0052-31 & \textbf{ESO 411-G031} & 00 55 42.5  & -30 32 23.7 & 1558.7 & 68.9 & 63.4 & 1.524 \\
J0052-31 & \textbf{2dFGRS TGS369Z051} & 00 53 30.1  & -30 50 29.0 & 1555.6 & 51.5 & 30.3 & 0.665 \\
J0052-31 & \textbf{ESO 411-G026} & 00 52 45.2  & -31 43 08.4 & 1590.7 & 113.0 & 98.3 & 3.037 \\
J0052-31 & \textbf{2dFGRS TGS370Z277} & 00 52 25.7  & -30 54 25.7 & 1436.0 & 35.8 & 13.8 & 0.078 \\
J0052-31 & LEDA 715942 & 00 51 52.0  & -30 32 46.0 & 1528.4 & 72.6 & 35.8 & 0.542 \\
\hline
\noalign{\vskip 2pt}
\multicolumn{2}{l}{\it Galaxies with uncertain parameters} &&&&&&\\
J0309--41 & LEDA 583148 & 03 08 38.0  & -40 43 29.5   & (1412.1\rlap{)}  & --  & --  & -- \\
J0459--26 & \textbf{ESO 486-21}  & 05 03 16.2  & -25 25 18.8  &  (845.8\rlap{)}  & --  & --  & -- \\
J1253--12 & MKT J125225.4-124304.1        & 12 52 25.4  &  -12 43 04.1  & (401.7\rlap{)} &  --  & --   & --    \\
 \hline
\end{tabular}
\label{tab:companions}
\tablefoot{Galaxies in {\bf bold} font match their optical counterparts both in position and velocity. Galaxies in a normal font match the cataloged optical counterpart in position, but with no prior cataloged redshift. For these galaxies, our data give the first redshift measurements. For the galaxies in {\it italic} font, there is a spatial match with a previously uncataloged optical counterpart. The ``Galaxies with uncertain parameters'' are only partially included in the data presented here. The central velocities listed are approximate.}
\end{table*}

\section{Summary}

We present a description of the MHONGOOSE survey, a large, deep, high-resolution
\HI survey on MeerKAT of 30 nearby gas-rich spiral and dwarf
galaxies. The total survey comprises 1650h of observing time, with
55h allocated per galaxy. The main science goal of MHONGOOSE is to
characterize the low-column density \HI in these galaxies, linking it
with various cold-gas-accretion scenarios.  Other science goals
include detailed explorations of the link between \HI  and
star formation, and the
kinematics of the galaxies, constraints on the presence and distribution of dark matter, as well as the environments of the targets. The
sample covers  approximately three orders of magnitude in \HI mass, from low-mass
$M_\matHI \sim 10^7$ \msun as typically found in dwarf galaxies to $M_\matHI \sim 10^{10}$ \msun as found in large spiral
galaxies.
In this paper, we present the survey, the sample, and a summary
of the observations. In addition we describe the data reduction
procedure and data products.

Each galaxy will eventually be observed for 55h in ten tracks of 5.5h each, and we produce
full-depth data cubes at a velocity resolution of 1.4~\kms and a range
in angular resolutions from $\sim 8''$ as the highest resolution to
$\sim 90''$ as our lowest resolution. The resulting column density
sensitivity ($3\sigma$ over 16 \kms) ranges from $4.3 \cdot 10^{19}$
cm$^{-2}$ at the highest resolution to $4.2 \cdot 10^{17}$ cm$^{-2}$
at the lowest resolution. The $3\sigma$ and 50 \kms mass-detection limit
is $5.4 \cdot 10^5$ \msun at 10 Mpc, which is the median distance of the
sample.

We present first results based on the full-depth data
of a limited number of galaxies and single-track observations of the
complete sample (with the latter observations comprising 10\% of the survey data). 
Comparing full-depth, single-track, and single-dish
data, we show that MeerKAT is excellent at recovering the total fluxes
of our sample galaxies. We also show that the full-depth data detect on average $\sim 2\%$ more \HI than the single-track data. While this increase may seem modest, the additional \HI mass it represents is sufficient to allow the 
existence of a $\sim 10^{18}$ cm$^{-2}$ low-column-density \HI 
component in the disks of our galaxies.  This 
demonstrates the difficulty in investigating the
presence of low-column-density \HI using total \HI mass measurements.

We also investigated the link between the local \HI surface density and
the velocity ``dispersion'' in all of our sample galaxies. We recover
the well-known upper limit on \HI surface density (measured on kpc scales), and find a lower
limit to the measured second-moment value of $\sim 8.5$ \kms. We find
that the \HI surface density has two distinct locations in a surface
density--velocity dispersion diagram: one at $\sim 5 \cdot 10^{20}$
cm$^{-2}$, representing the gas in the star-forming disks, and one at
$\sim 5 \cdot 10^{19}$ cm$^{-2}$ representing gas not participating in
star formation. The two regions are distinct and separated at an \HI
surface density of $2 \cdot 10^{20}$ cm$^{-2}$. Previous studies (e.g.,\ \citealt{Kanekar.2011}) showed that, at this surface density, a rapid change in spin temperature
occurs, and that this is the column density where the formation of a cool and
dense component of the ISM starts.

We give an overview of all additional galaxies detected in the single-track data. These are defined as the sources detected in \HI that were not targeted by the survey but were nevertheless found within the 
$1.5\degree \times 1.5\degree$ area and spanning 1000 \kms in velocity range around the target galaxies.
We find that half of these 49 galaxies have  cataloged optical counterparts that coincide in position and velocity. Just under half have cataloged optical counterparts that coincide in position but that had until now no previously measured redshifts. Five of our additional galaxies have optical counterparts that have not been previously cataloged. We have not found any \HI detections that do not have optical counterparts.

Studies of the full-depth data will be presented in future
papers. However, based on the results shown here, it is
already clear that the MHONGOOSE data are of excellent quality, showing
great promise for future work on the low-column-density \HI in nearby
galaxies.

\bigskip
\noindent {\bf Data:} The single-track moment maps presented in this paper are available for download at the MHONGOOSE website ({\tt https://mhongoose.astron.nl}) or through a DOI ({\tt https://zenodo.org/doi/10.5281/zenodo.10907079}).

\begin{acknowledgements}

It is a pleasure to thank SARAO staff past and present for their help in making MHONGOOSE a reality.
The MeerKAT telescope is operated by the South African Radio Astronomy Observatory, which is a facility
of the National Research Foundation, an agency of the Department of Science and Innovation.
This work has received funding from the European Research Council
(ERC) under the European Union’s Horizon 2020 research and innovation
programme (grant agreement No. 882793 ``MeerGas'').

We thank SURF, and particularly Dr.\ Raymond Oonk for the support in the early stages of the data reduction.
KAO is supported by a Royal Society Dorothy Hodgkin Fellowship, by STFC through grant ST/T000244/1, 
and by the European Research Council (ERC) through Advanced Investigator 
grant to C.S. Frenk, DMIDAS (GA 786910).
DJP, NZ and SK are supported through the South African Research Chairs Initiative of
the Department of Science and Technology and National Research Foundation (Grant number 77825).
EA and AB thank the CNES for financial support.
LC acknowledges funding from the Chilean Agencia Nacional de Investigaci\'on y Desarrollo (ANID) through Fondo Nacional de Desarrollo Cient\'ifico y Tecnol\'ogico (FONDECYT) Regular Project 1210992.
The work at AIRUB is supported by the German Federal Ministry of Education and Research (BMBF) Verbundforschung grant 05A20PC4 (Verbundprojekt D-MeerKAT-II).
KS acknowledges support from the Natural Sciences and Engineering Research Council of Canada (NSERC).
LVM and AS acknowledge financial support from the grant CEX2021-001131-S funded by MCIN/AEI/10.13039/501100011033 and from the grant PID2021-123930OB-C21 funded by MCIN/AEI/10.13039/501100011033, by ``ERDF A way of making Europe'' and by the European Union.
BN acknowledges financial support from the Severo Ochoa grant CEX2021-001131-S funded by MCIN/AEI/10.13039/501100011033.
This research has made use of the NASA/IPAC Extragalactic Database (NED), which is funded by the National Aeronautics and Space Administration and operated by the California Institute of Technology. This research has made use of the SIMBAD database,
operated at CDS, Strasbourg, France.

The Legacy Surveys consist of three individual and complementary projects: the Dark Energy Camera Legacy Survey (DECaLS; Proposal ID \#2014B-0404; PIs: David Schlegel and Arjun Dey), the Beijing-Arizona Sky Survey (BASS; NOAO Prop. ID \#2015A-0801; PIs: Zhou Xu and Xiaohui Fan), and the Mayall z-band Legacy Survey (MzLS; Prop. ID \#2016A-0453; PI: Arjun Dey). DECaLS, BASS and MzLS together include data obtained, respectively, at the Blanco telescope, Cerro Tololo Inter-American Observatory, NSF’s NOIRLab; the Bok telescope, Steward Observatory, University of Arizona; and the Mayall telescope, Kitt Peak National Observatory, NOIRLab. Pipeline processing and analyses of the data were supported by NOIRLab and the Lawrence Berkeley National Laboratory (LBNL). The Legacy Surveys project is honored to be permitted to conduct astronomical research on Iolkam Du’ag (Kitt Peak), a mountain with particular significance to the Tohono O’odham Nation.

NOIRLab is operated by the Association of Universities for Research in Astronomy (AURA) under a cooperative agreement with the National Science Foundation. LBNL is managed by the Regents of the University of California under contract to the U.S. Department of Energy.

This project used data obtained with the Dark Energy Camera (DECam), which was constructed by the Dark Energy Survey (DES) collaboration. Funding for the DES Projects has been provided by the U.S. Department of Energy, the U.S. National Science Foundation, the Ministry of Science and Education of Spain, the Science and Technology Facilities Council of the United Kingdom, the Higher Education Funding Council for England, the National Center for Supercomputing Applications at the University of Illinois at Urbana-Champaign, the Kavli Institute of Cosmological Physics at the University of Chicago, Center for Cosmology and Astro-Particle Physics at the Ohio State University, the Mitchell Institute for Fundamental Physics and Astronomy at Texas A\&M University, Financiadora de Estudos e Projetos, Fundacao Carlos Chagas Filho de Amparo, Financiadora de Estudos e Projetos, Fundacao Carlos Chagas Filho de Amparo a Pesquisa do Estado do Rio de Janeiro, Conselho Nacional de Desenvolvimento Cientifico e Tecnologico and the Ministerio da Ciencia, Tecnologia e Inovacao, the Deutsche Forschungsgemeinschaft and the Collaborating Institutions in the Dark Energy Survey. The Collaborating Institutions are Argonne National Laboratory, the University of California at Santa Cruz, the University of Cambridge, Centro de Investigaciones Energeticas, Medioambientales y Tecnologicas-Madrid, the University of Chicago, University College London, the DES-Brazil Consortium, the University of Edinburgh, the Eidgenossische Technische Hochschule (ETH) Zurich, Fermi National Accelerator Laboratory, the University of Illinois at Urbana-Champaign, the Institut de Ciencies de l’Espai (IEEC/CSIC), the Institut de Fisica d’Altes Energies, Lawrence Berkeley National Laboratory, the Ludwig Maximilians Universitat Munchen and the associated Excellence Cluster Universe, the University of Michigan, NSF’s NOIRLab, the University of Nottingham, the Ohio State University, the University of Pennsylvania, the University of Portsmouth, SLAC National Accelerator Laboratory, Stanford University, the University of Sussex, and Texas A\&M University.

BASS is a key project of the Telescope Access Program (TAP), which has been funded by the National Astronomical Observatories of China, the Chinese Academy of Sciences (the Strategic Priority Research Program “The Emergence of Cosmological Structures” Grant \# XDB09000000), and the Special Fund for Astronomy from the Ministry of Finance. The BASS is also supported by the External Cooperation Program of Chinese Academy of Sciences (Grant \# 114A11KYSB20160057), and Chinese National Natural Science Foundation (Grant \# 12120101003, \# 11433005).

The Legacy Survey team makes use of data products from the Near-Earth Object Wide-field Infrared Survey Explorer (NEOWISE), which is a project of the Jet Propulsion Laboratory/California Institute of Technology. NEOWISE is funded by the National Aeronautics and Space Administration.

The Legacy Surveys imaging of the DESI footprint is supported by the Director, Office of Science, Office of High Energy Physics of the U.S. Department of Energy under Contract No. DE-AC02-05CH1123, by the National Energy Research Scientific Computing Center, a DOE Office of Science User Facility under the same contract; and by the U.S. National Science Foundation, Division of Astronomical Sciences under Contract No. AST-0950945 to NOAO.
\end{acknowledgements}

\bibliographystyle{aa} 
\bibliography{My_Library} 

\begin{appendix}

\section{Atlas of single-track moment maps}

In Fig.~\ref{fig:sample1} we show the {\tt r05\_t00} single-track moment maps of the entire sample.
A short description of the panels that are shown (from left to right) for each galaxy
follows.

The left-most panel shows a combined optical \emph{grz} color
image from DECaLS DR9 (except for J1303--17b where we use a \emph{giz} image from DR10). The  next panel shows the zeroth-moment or integrated \HI map.
The lowest column density contour shown
corresponds to a S/N of 3 (see Sect.~\ref{sec:fullmom}
for a description).  This value is listed at the bottom of the
zeroth-moment panel.  Subsequent contours then increase by a factor
$2^n$, where $n=0,2,4, ...$. 

In the third panel we show the first-moment map or intensity-weighted velocity field. 
Here we indicate the central velocity (as listed in Table~\ref{tab:sample}) with a
thick contour. The central velocity is defined from the integrated \HI profile as the velocity halfway between the velocities where the flux density equals 20\% of the peak value (see Sect.~\ref{sec:singledish}).
Other contours are spaced by 10 or 20 \kms with respect to
this velocity, as listed in the panel. The color map indicates
whether velocities are receding (red) or approaching
(blue). 

The second-moment maps are shown using the same intensity scale for
all galaxies. For all galaxies, the color map shows the same velocity range from 0 (light-blue) to 30 (red) \kms. Contours start at 12 \kms, and increase in steps of 12
\kms. For reference, the 24 \kms contour is shown in black. Both the first- and second-moment maps have been masked such that only
values with ${\rm S/N} > 3$ in the zeroth-moment map are shown.
While we do not present a full analysis of the first- and second-moment maps in this paper, we do caution that 
for the edge-on galaxies, these have to be interpreted with care because of the strong projection effects.

\newpage

\renewcommand{\thefigure}{A.\arabic{figure}}

\begin{figure*}
  \centering
  {\resizebox{0.73\hsize}{!}{\includegraphics[trim={4cm 0.7cm 5.0cm 0.83cm},clip]{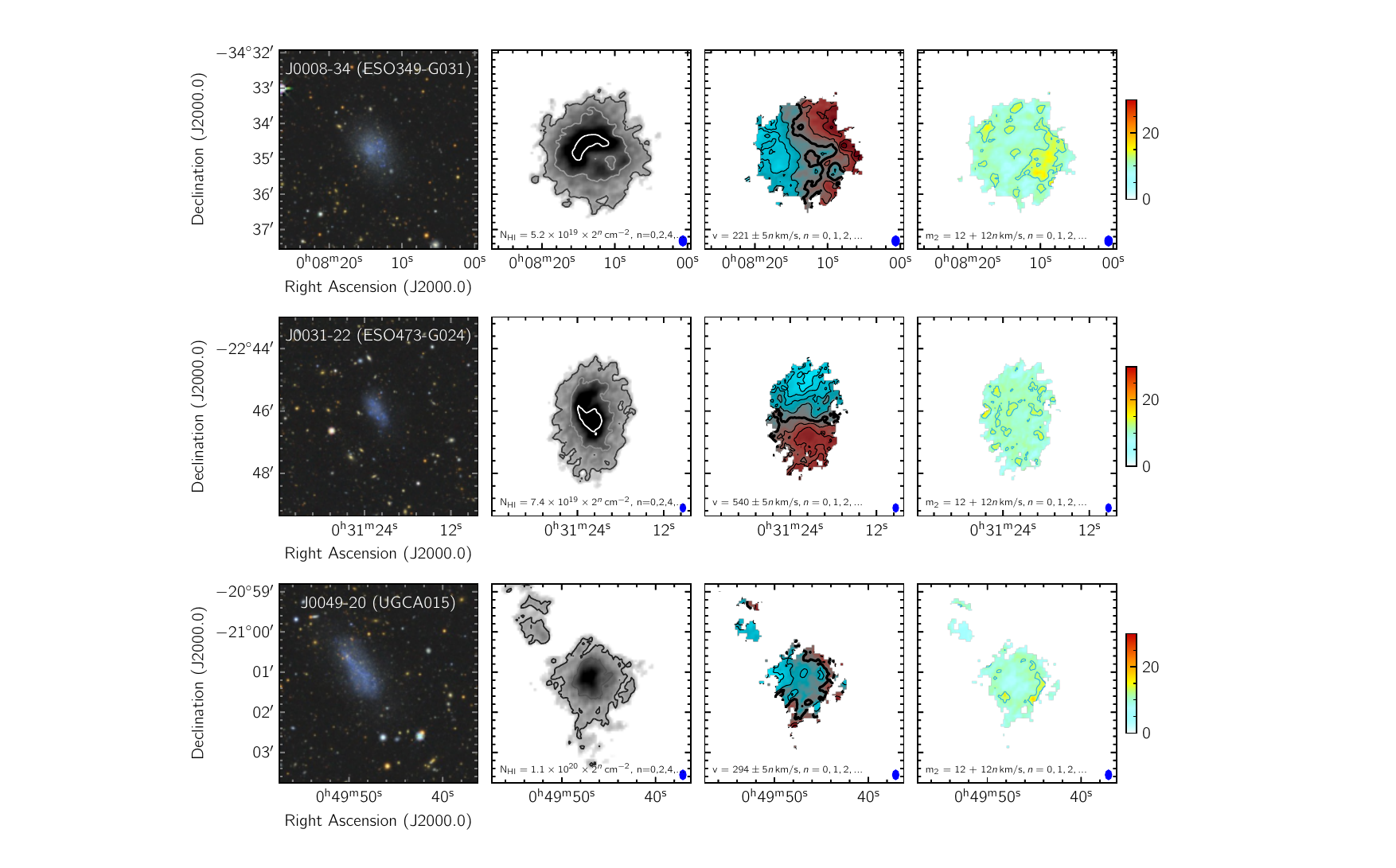}}}
  {\resizebox{0.73\hsize}{!}{\includegraphics[trim={4cm 0.7cm 5.0cm 0.83cm},clip]{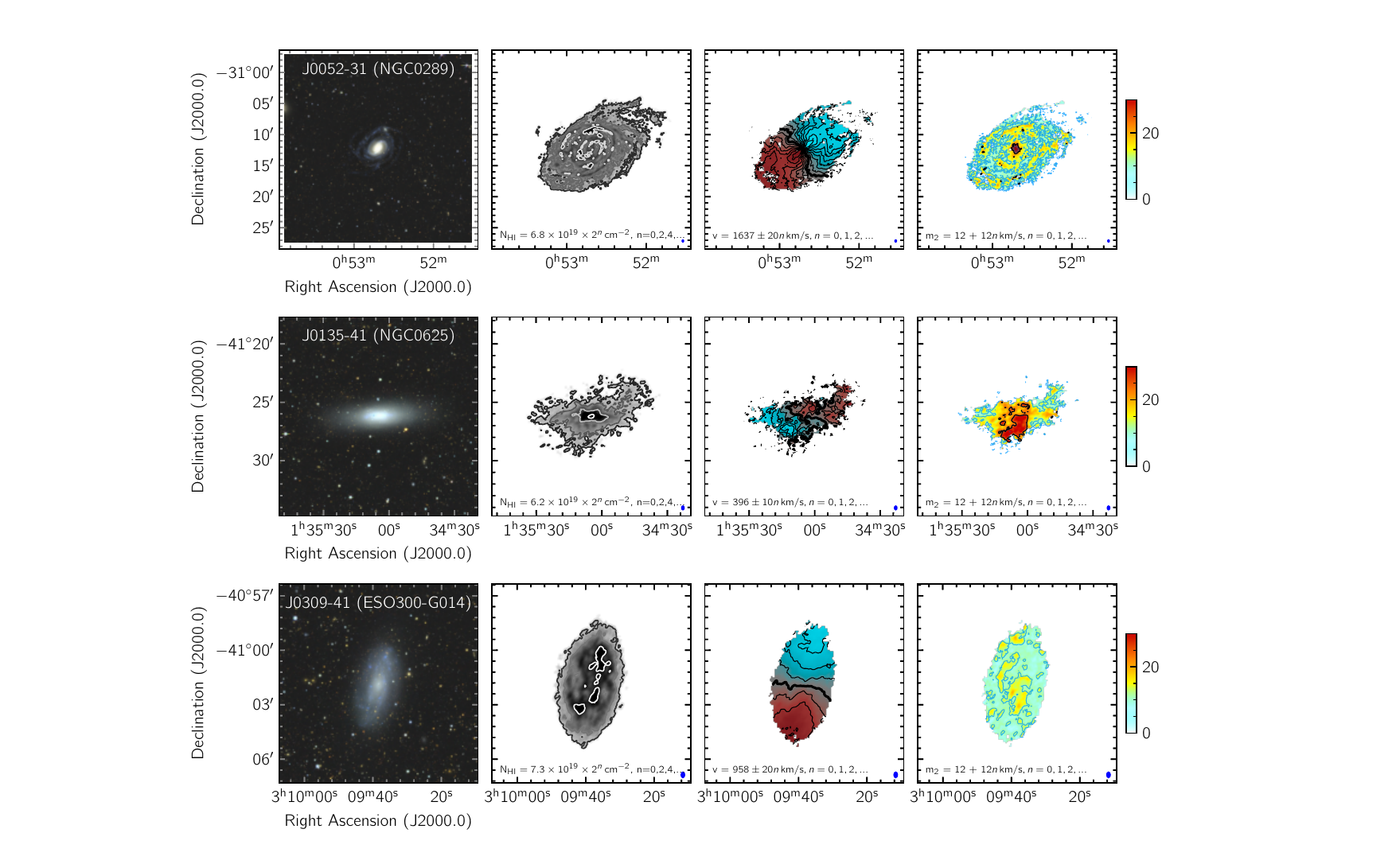}}}
  \caption{Single-track moment maps for the MHONGOOSE galaxies using the {\tt r05\_t00} resolution. From left to right: \emph{(i):} Combined \emph{grz}-color image from DECaLS. \emph{(ii):} primary-beam corrected zeroth-moment or integrated \HI intensity map. Contours as indicated in the Figure. The lowest contour represents S/N = 3, with subsequent contour levels increasing by factors of two. \emph{(iii):} First-moment map or intensity-weighted velocity field. Red colors indicate the receding side, blue colors the approaching side. The central velocity (listed in Tab.~\ref{tab:sample}) is indicated by the thick contour. Other contours are spaced by 10 or 20 \kms, as indicated in the Figure. \emph{(iv):} Second-moment map: the colors show the range from 0 (light-blue) to 30 (red) \kms. The lowest contour shows the 12 \kms level, and subsequent contours are spaced by 12 \kms. The 24 \kms contour is shown in black. See text for a more extensive description.}
  \label{fig:sample1}
\end{figure*}

\renewcommand{\thefigure}{A.\arabic{figure}. continued}
\addtocounter{figure}{-1}

\begin{figure*}
  \centering
  {\resizebox{0.73\hsize}{!}{\includegraphics[trim={4cm 0.7cm 5.0cm 0.83cm},clip]{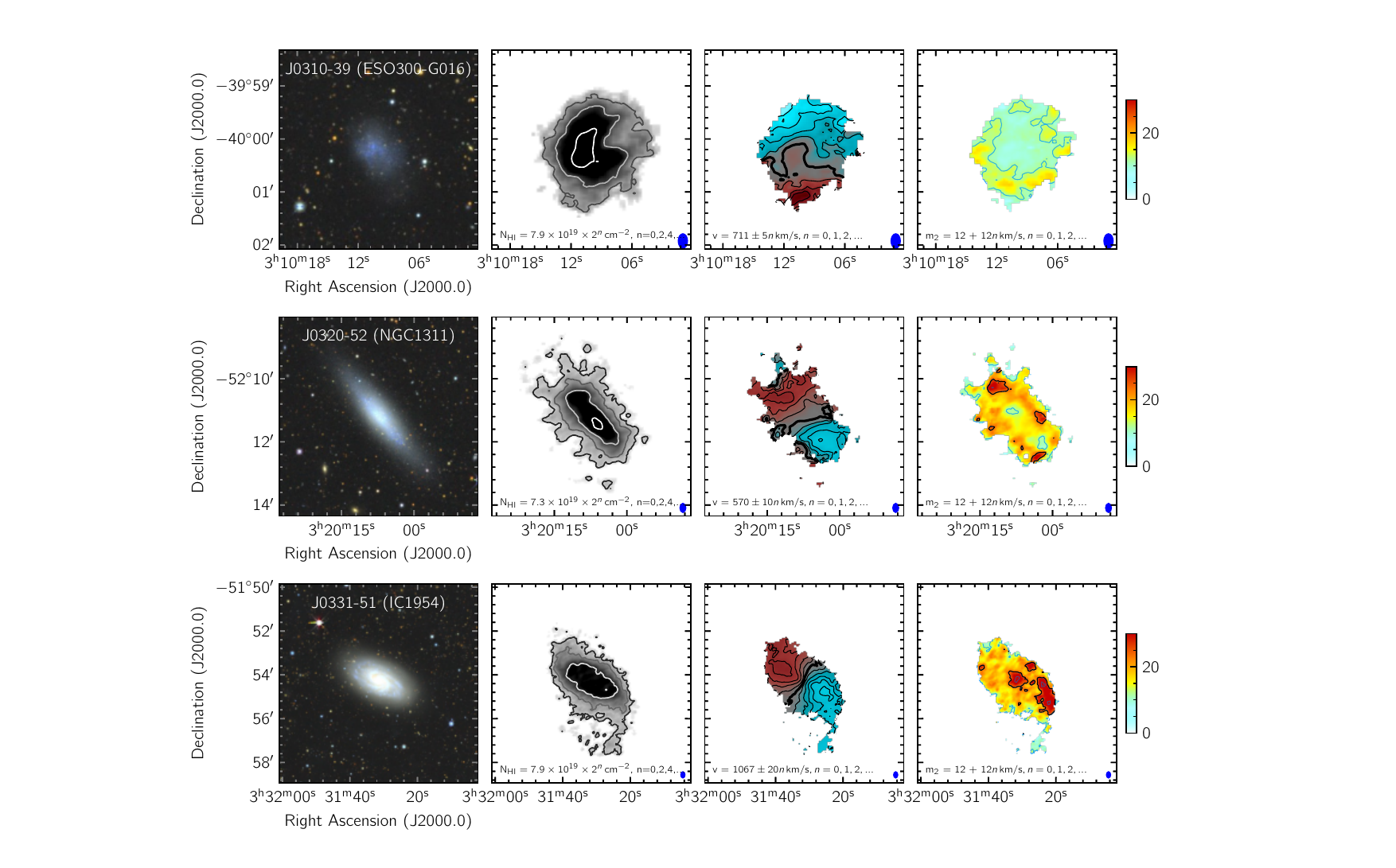}}}
  {\resizebox{0.73\hsize}{!}{\includegraphics[trim={4cm 0.7cm 5.0cm 0.83cm},clip]{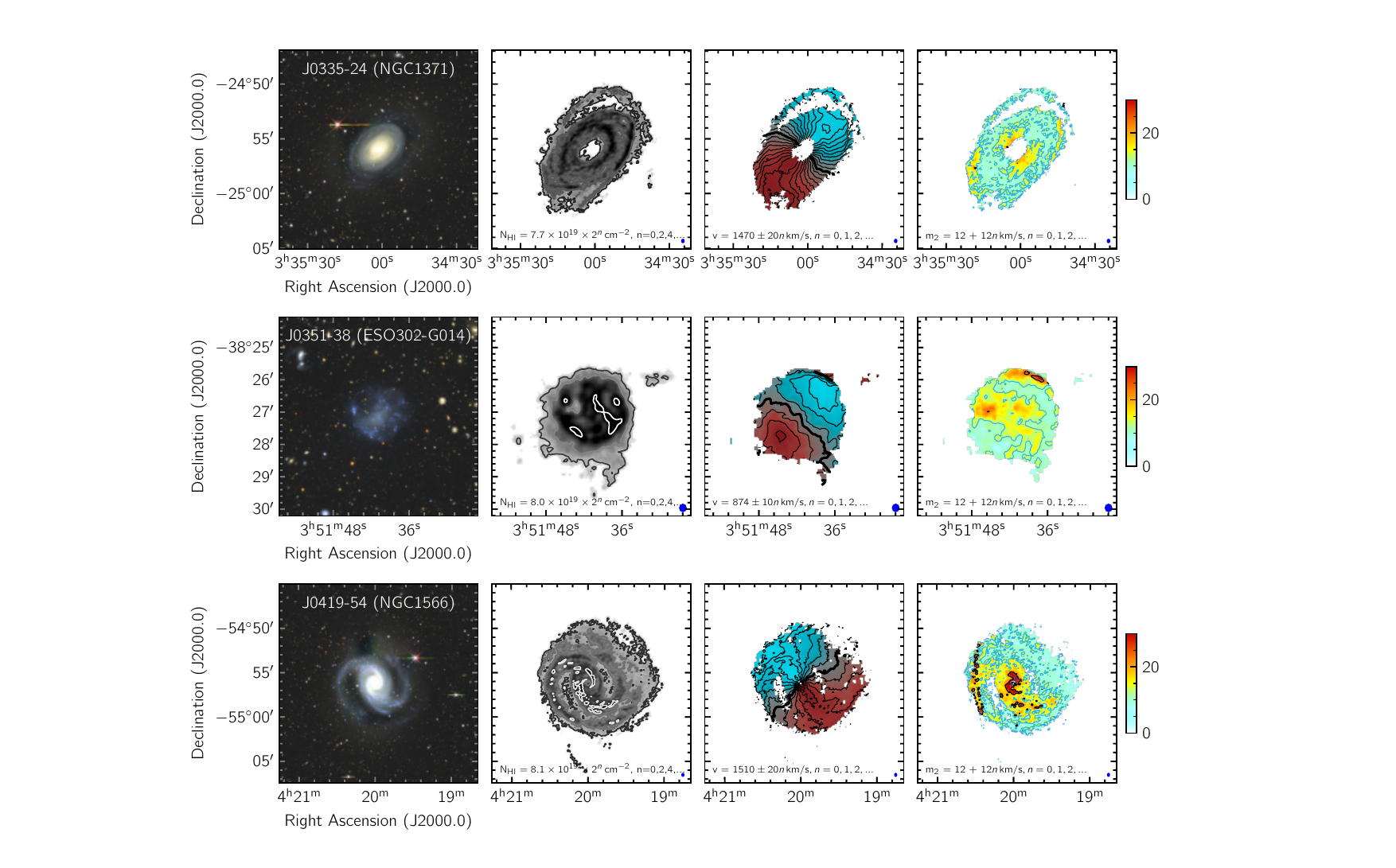}}}
  \caption{Single-track moment maps of MHONGOOSE galaxies.}
  \label{fig:sample2}
\end{figure*}

\addtocounter{figure}{-1}

\begin{figure*}
  \centering
  {\resizebox{0.73\hsize}{!}{\includegraphics[trim={4cm 0.7cm 5.0cm 0.83cm},clip]{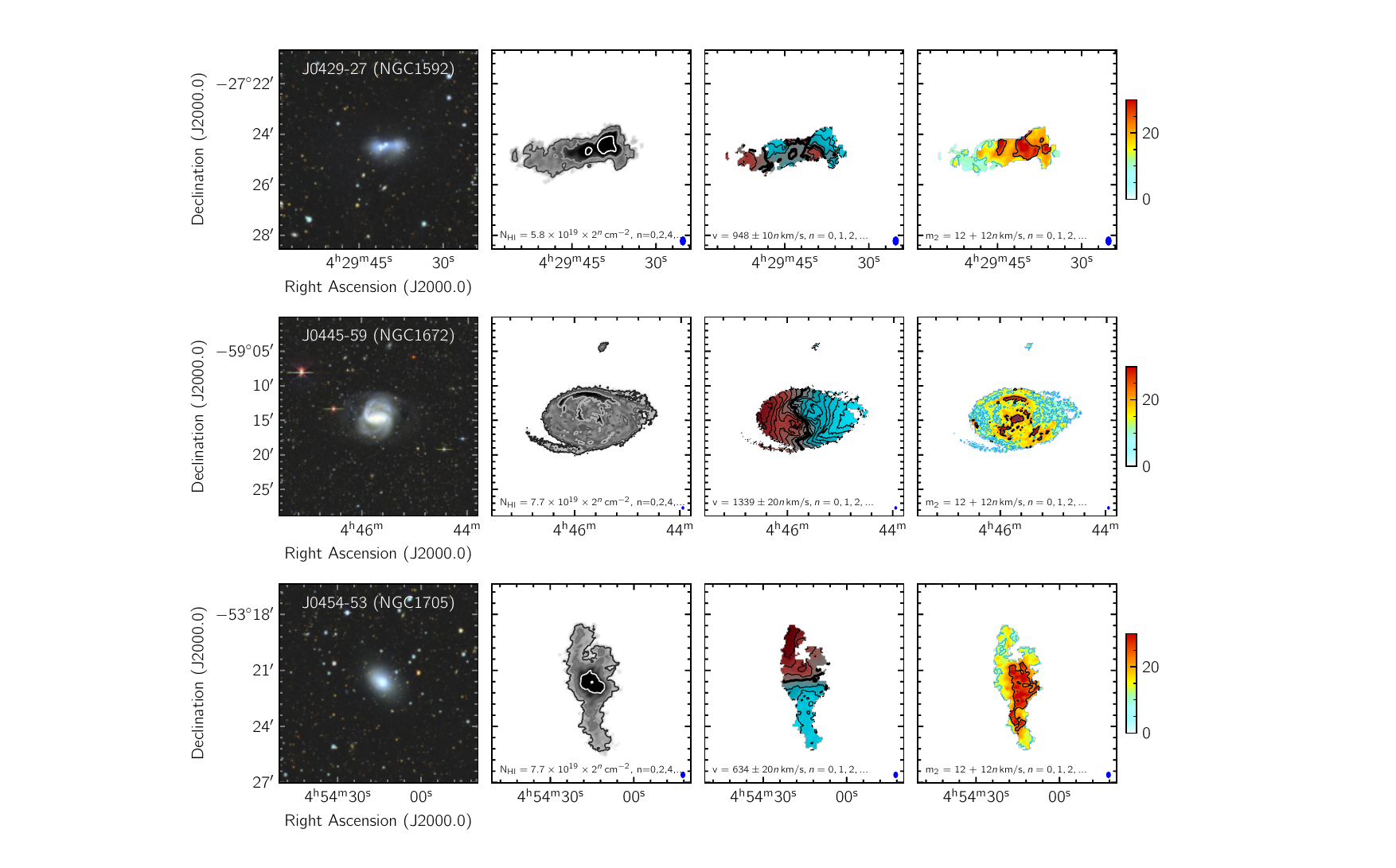}}}
  {\resizebox{0.73\hsize}{!}{\includegraphics[trim={4cm 0.7cm 5.0cm 0.83cm},clip]{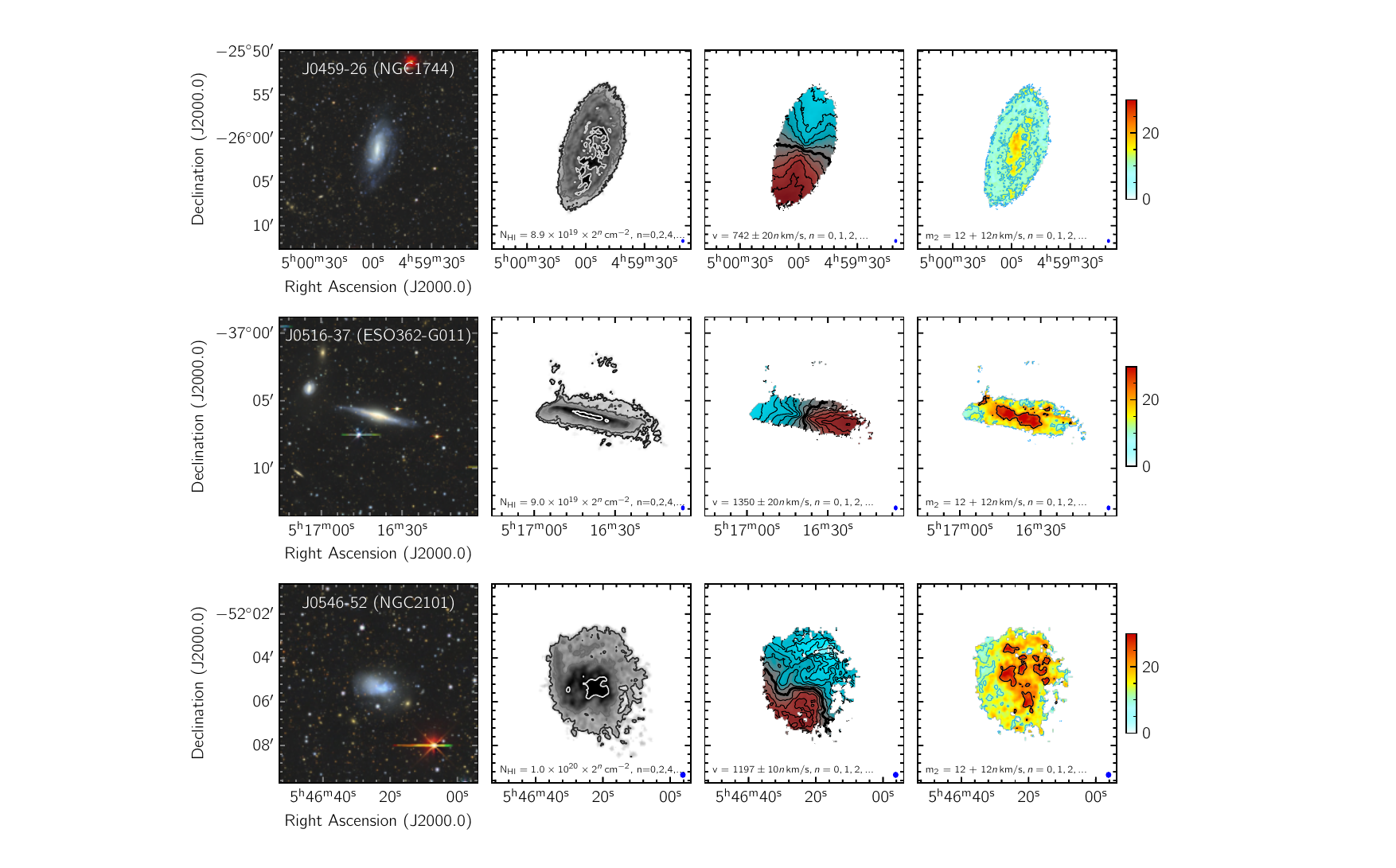}}}
  \caption{Single-track moment maps of MHONGOOSE galaxies.}
  \label{fig:sample3}
\end{figure*}

\addtocounter{figure}{-1}

\begin{figure*}
  \centering
  {\resizebox{0.73\hsize}{!}{\includegraphics[trim={4cm 0.7cm 5.0cm 0.83cm},clip]{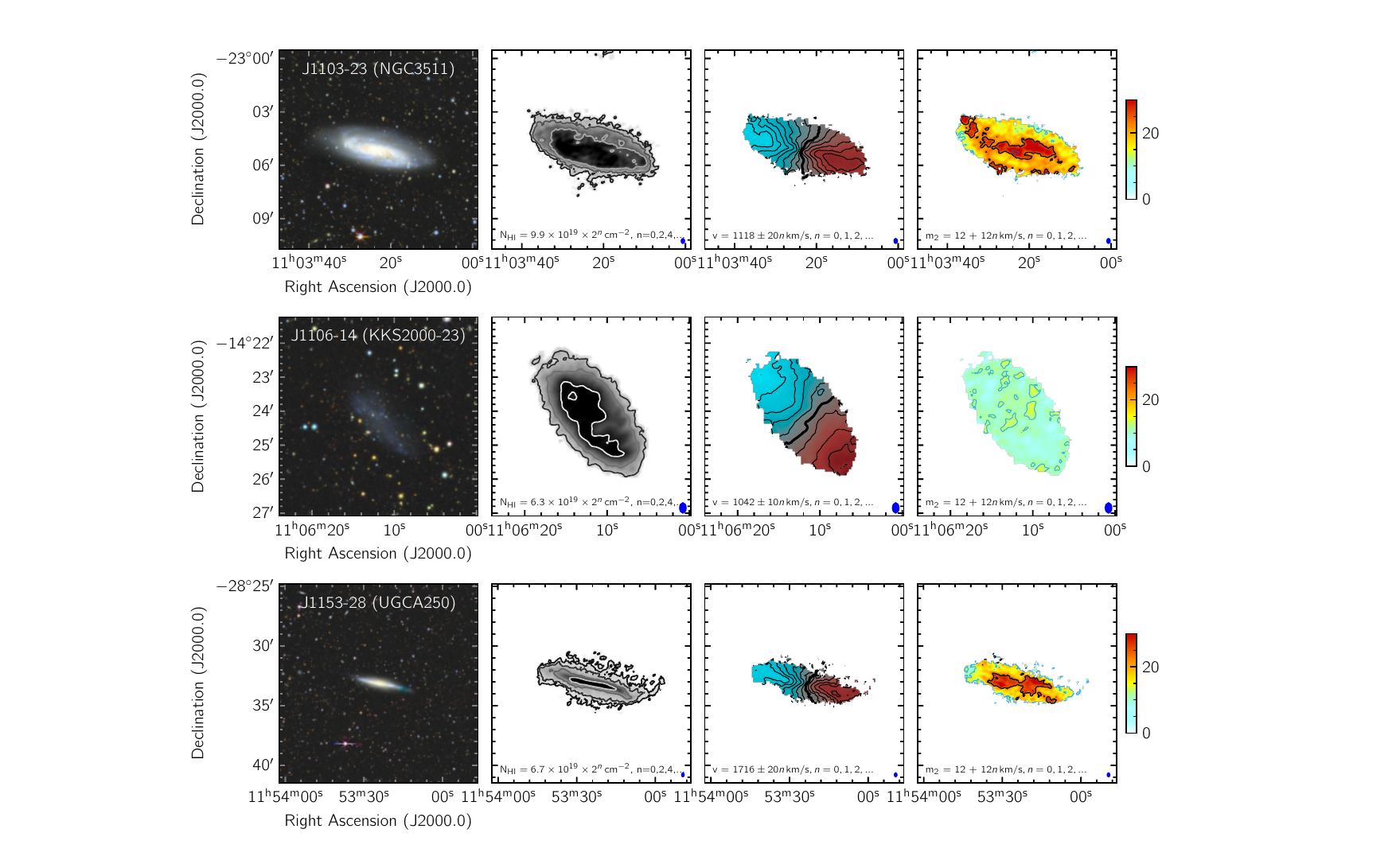}}}
  {\resizebox{0.73\hsize}{!}{\includegraphics[trim={4cm 0.7cm 5.0cm 0.83cm},clip]{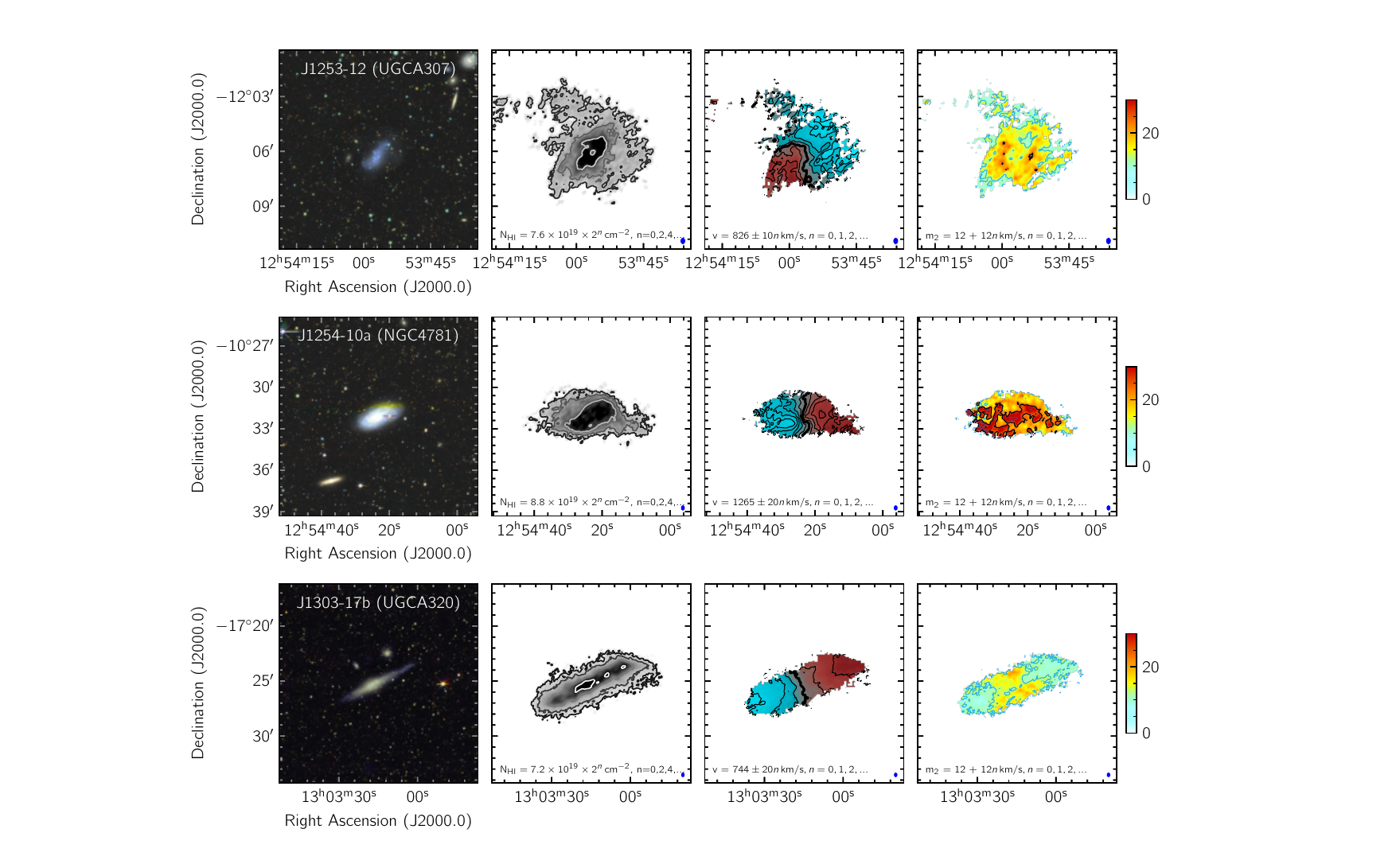}}}
  \caption{Single-track moment maps of MHONGOOSE galaxies.}
  \label{fig:sample4}
\end{figure*}

\addtocounter{figure}{-1}

\begin{figure*}
  \centering
  {\resizebox{0.73\hsize}{!}{\includegraphics[trim={4cm 0.7cm 5.0cm 0.83cm},clip]{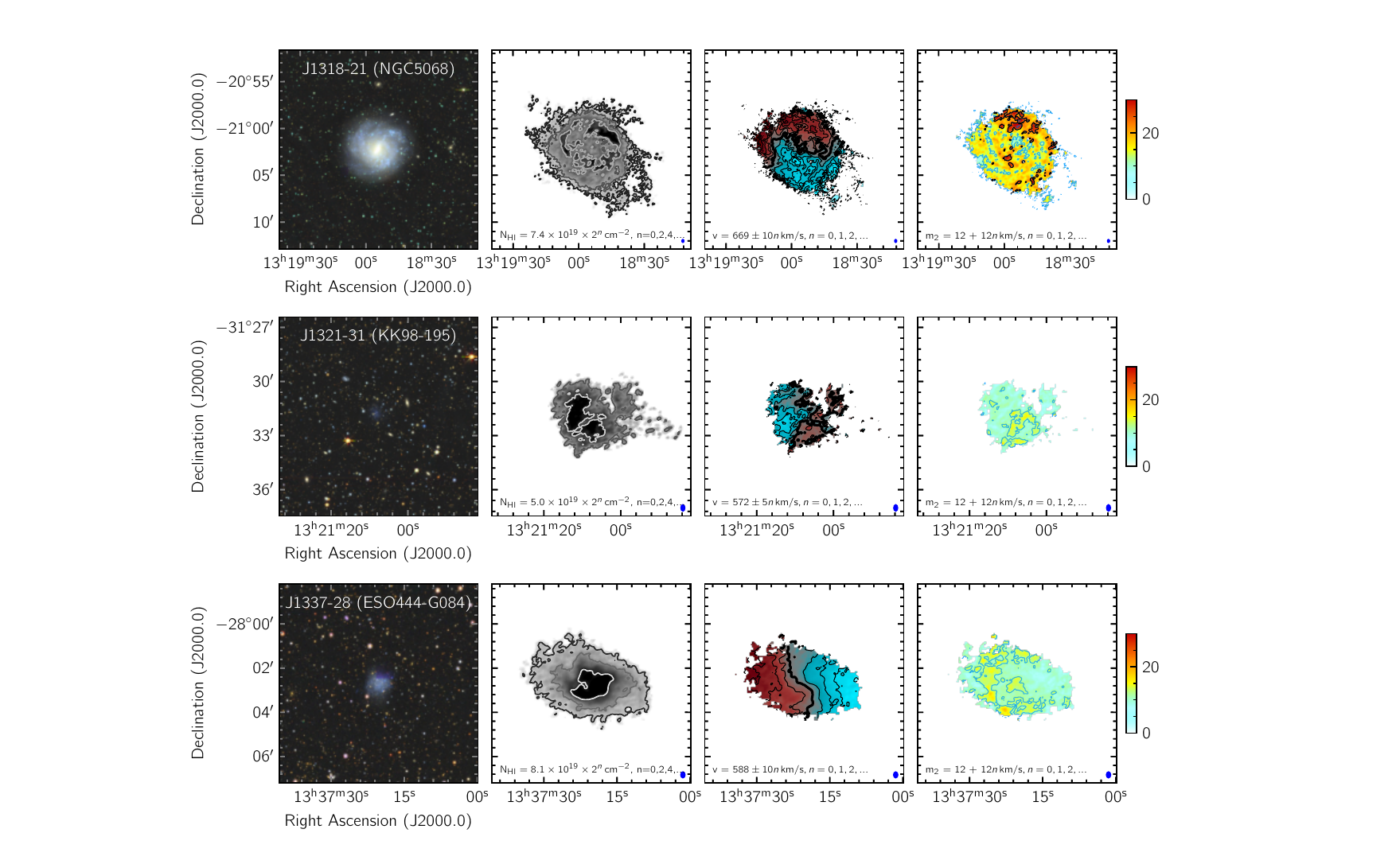}}}
  {\resizebox{0.73\hsize}{!}{\includegraphics[trim={4cm 0.7cm 5.0cm 0.83cm},clip]{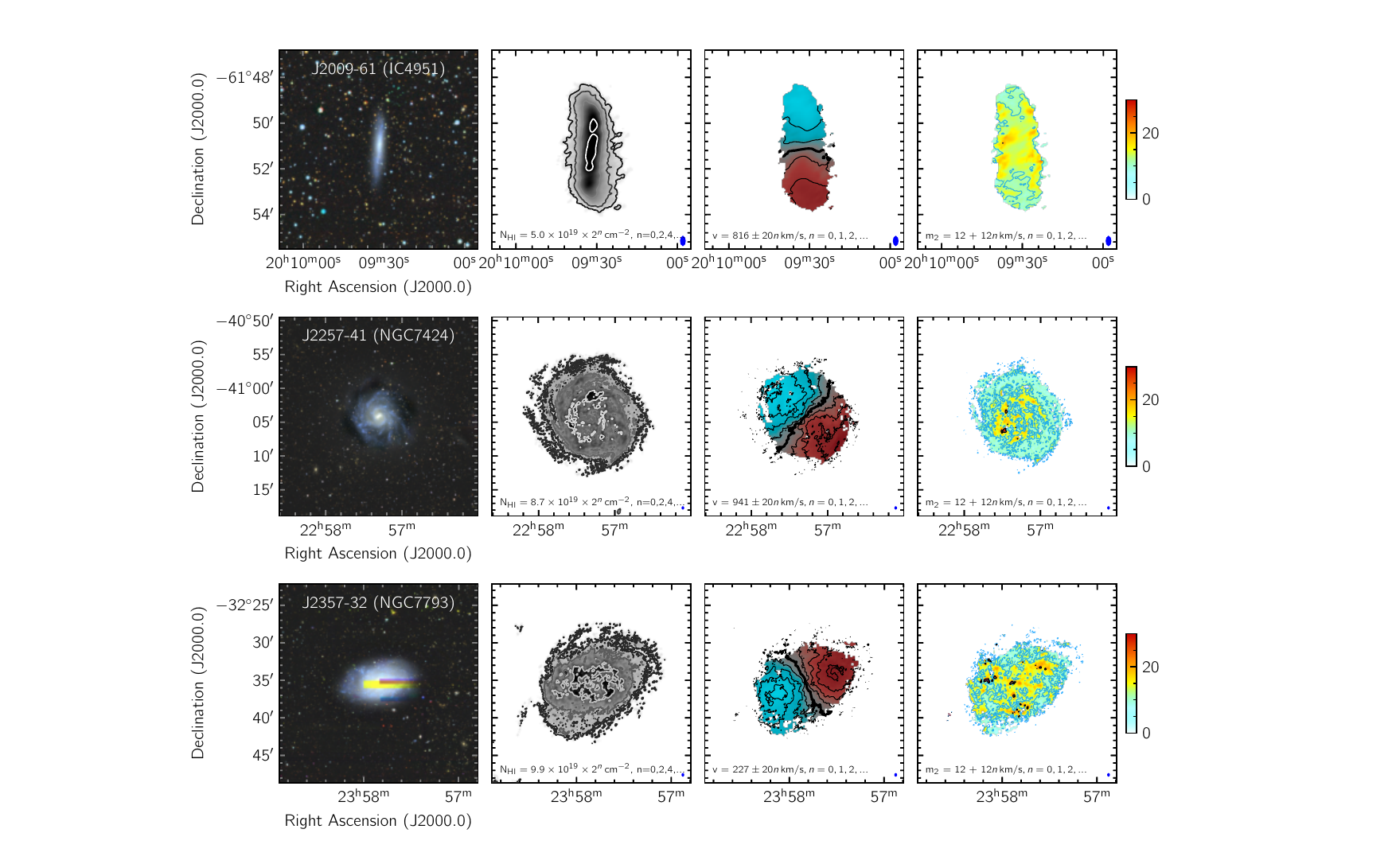}}}
  \caption{Single-track moment maps of MHONGOOSE galaxies.}
  \label{fig:sample5}
\end{figure*}

\FloatBarrier

\section{The effect of beam smearing and inclination on second-moment values}

The column density-second moment distributions shown in Figs.~\ref{fig:m0m2panel} and \ref{fig:m0m2total} were derived using the {\tt r15\_t00} standard resolution with a beam size of $\sim 30''$. This beam size, in combination with the angular sizes of the galaxies, could potentially lead to inflated second-moment values due to beam smearing (especially in the centers of the galaxies). In addition, long line-of-sight effects in the high-inclination galaxies could also contribute to higher second-moment values. Here we address  these potential issues and show that they do not significantly affect the distribution shown in Fig.~\ref{fig:m0m2total}.

We test the effect of beam smearing by creating a model galaxy similar to one of our sample galaxies, convolving it to the observed resolution and inserting it in a data cube with a noise level equal to that found for the observations. We then treat it exactly as the sample galaxies for the derivation of the moment maps and creation of the moment distributions.

We use as the basis of our model the rotation curve and \HI radial profile of NGC1566 as presented in \citet{Elagali.2019}. We approximate the rotation curve with a linear, steep rise from 0 to 180 \kms in the inner 5 kpc. At larger radii we assume a constant rotation velocity of 180 \kms (cf.\ Fig.~10 in \citealt{Elagali.2019}). For the \HI distribution we assume a central value of $\log(N_\matHI/\msunpc) = 1$, with $\log(N_\matHI)$ linearly decreasing with radius until $\log(N_\matHI/\msunpc) = -0.5$ at 50 kpc (cf.\ Fig.~12 in \citealt{Elagali.2019}). For the purposes of this exercise we extend the radial range of the model to 100 kpc by assuming that the rotation curve stays flat, and that the $\log(N_\matHI)$ values keep decreasing at the same rate as inside 50 kpc. We adopt a value of 30 degrees for the inclination (cf.\ Fig 10 in \citealt{Elagali.2019}). For the velocity dispersion of the \HI we use a constant value of 10 \kms. The \HI disk is assumed to be thin, with a scale height of 100 pc. 
We assume the same distance and systemic velocity as for the real NGC 1566 (17.6 Mpc and 1496 \kms, respectively; cf.~Tab.~\ref{tab:sample}).
With these parameters we create a model cube using the GALMOD section of the $^{3\rm{D}}$BAROLO software package \citep{DiTeodoro.2015a} with a beam size equal to that of the {\tt r15\_t00} standard resolution. We insert this model in a cube filled with an rms noise of $0.48$ mJy (again at the {\tt r15\_t00} resolution), consistent with the noise values typically found in the single-track cubes.

Zeroth- and second-moment maps of the model are then created with
SoFiA-2 in an identical manner as for the real galaxies. We plot the
resulting column density-second moment distribution in
Fig.~\ref{fig:beamsmear} (right panel), comparing it with the observed
NGC 1566 distribution (left panel). Several noteworthy features can be
seen. Firstly, in the model we see a plume towards high second-moment
values at the highest column densities. This is caused by beam
smearing due to the slope of the inner rotation curve. Secondly, on
the low-column density end of the distribution we see the familiar
tail towards low second-moment value due to masking (as discussed in
Sect.~\ref{sec:mask}). Finally, in the intermediate column density
range, we see the constant second-moment values. The limited
resolution has slightly increased the average value from its original
10 \kms, but only by $\sim 1$ \kms or less. These therefore are a good
reflection of the velocity dispersion in the model disk.

In summary, only the high second-moment values at the highest column densities (corresponding to the innermost part of the disk where the rotation curve rises) can be attributed to beam smearing. Higher second-moment values at lower column densities (i.e., further out in the disk) reflect higher velocity dispersions, or the presence of multiple components along the line of sight. 

We can compare the distribution shown in Fig.~\ref{fig:beamsmear}
(right panel) with the more complex distribution for the real NGC
1566 in Fig.~\ref{fig:beamsmear} (left panel). The latter shows higher
second-moment values over a larger range in column densities,
indicating that a significant fraction of these higher values are not
due to beam smearing. Inspection of Fig.~\ref{fig:m0m2panel} shows
that these are caused by high-second moment components in
galaxies with overall low second-moment values (e.g., J0419--54) as
well as galaxies that have overall higher second-moment values (e.g.,
J1318--21).

We can test this further by taking the inclinations of the galaxies
into account. In general, the broadening of profiles due to beam
smearing and length of the line of sight through the disk increases
dramatically towards high inclinations, with the effect strongest in
edge-on galaxies. There is therefore a possibility that
high-inclination galaxies contribute to the higher second-moment
values we observe at intermediate column densities (i.e., around $\sim
10^{20}$ cm$^{-2}$). In Fig.~\ref{fig:m0m2_edgeon} (left panel)
  we show the distribution from Fig.~\ref{fig:m0m2total}, which
  includes all galaxies in the sample, at all inclinations, including
  edge-ons. To remove the effect that high-inclination galaxies may
  have on the distribution, we show in the centre panel the
  distribution of galaxies with an inclination $i<60\degree$
  (16 galaxies) and in the right panel the distribution with
  $i < 70\degree$ (22 galaxies).  Comparison of these
distributions shows that omitting the high-inclination galaxies
mainly affects the width of the second-moment distribution in the
low-column density tail. The effects in the intermediate and
high-column density ranges are much less pronounced, and the
distribution remains essentially the same in these column density
regimes.

\renewcommand{\thefigure}{B.\arabic{figure}}

\begin{figure*}
  \resizebox{\hsize}{!}{\includegraphics{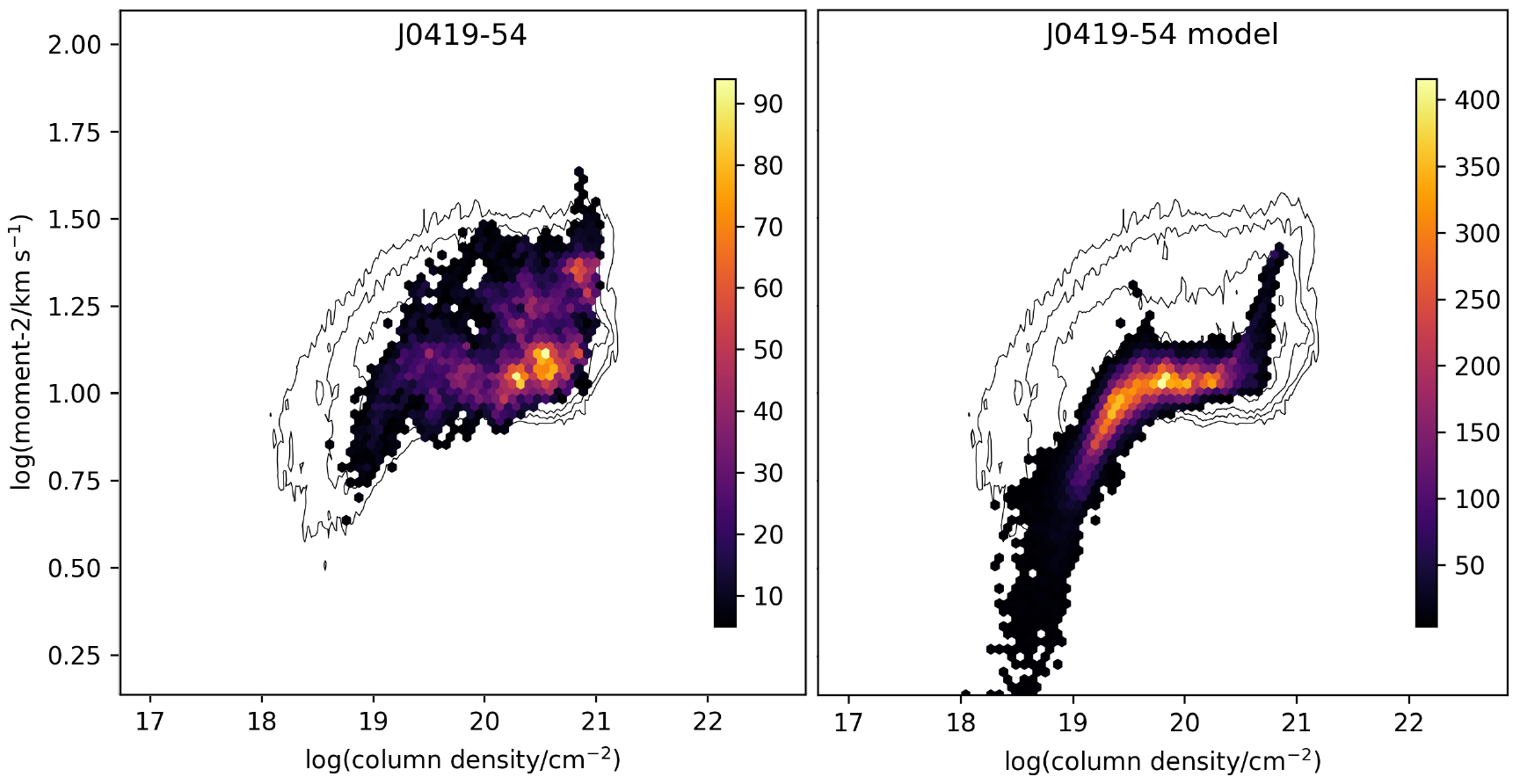}}
  \caption{Comparison of the column density-second moment distribution of a model galaxy built to resemble NGC 1566 with that of NGC 1566 itself. The observed distribution is shown in the left panel; the model in the right panel. For the model we use a single, thin \HI disk with a velocity dispersion of 10 \kms. The model galaxy is ``observed'' and treated just like the single-track MHONGOOSE galaxies. Note the plume due to beam smearing at the high column density end, and the masking tail at the low column density end. The model does not show the high second-moment values at intermediate column densities that are present in the observations.}
  \label{fig:beamsmear}
\end{figure*}

\begin{figure*}
  \resizebox{\hsize}{!}{\includegraphics{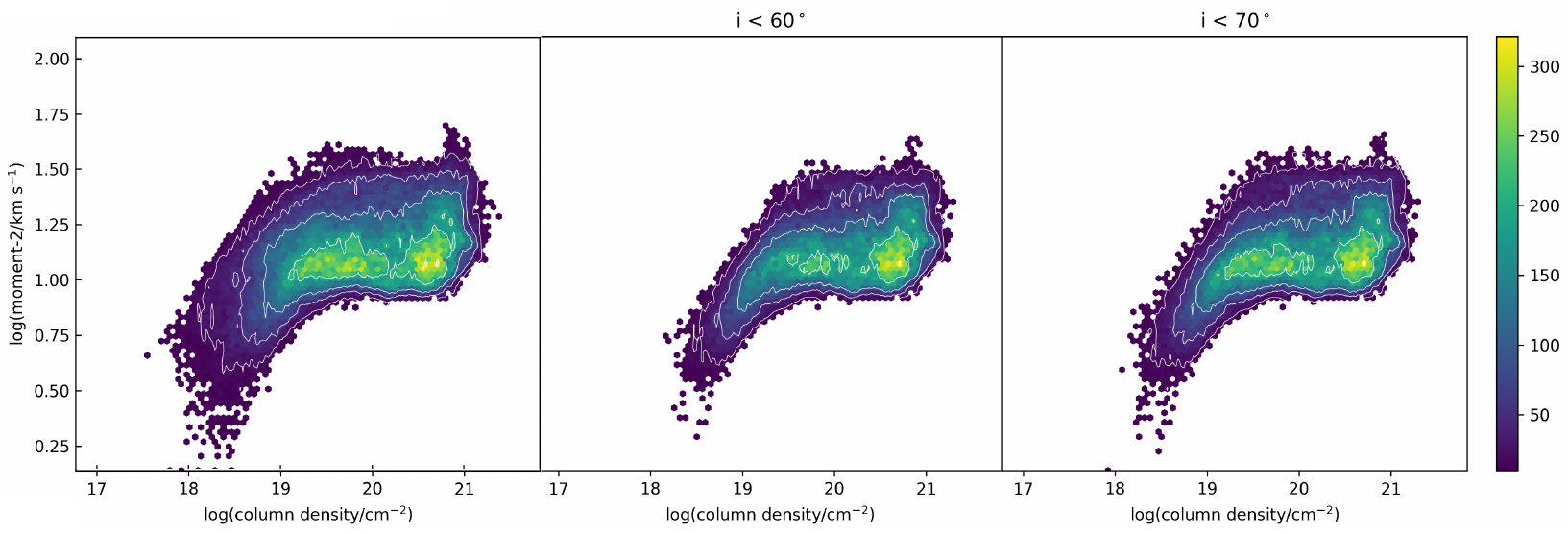}}
  \caption{Comparison of the column density-second moment distribution for a number of inclination limits. The left panel shows the distribution as shown in Fig.~\ref{fig:m0m2total} without any inclination limits. The centre panel shows the distribution when limited to galaxies with inclination $i<60\degree$. The right panel shows the same for $i<70\degree$.}
  \label{fig:m0m2_edgeon}
\end{figure*}

\end{appendix}

\end{document}